\pdfoutput=1 

\documentclass[
superscriptaddress,showpacs,twocolumn]{revtex4}
\usepackage{graphics}
\usepackage{epstopdf}
\usepackage{graphicx}
\usepackage{dcolumn}
\usepackage{bm}
\usepackage{amsmath}
\usepackage{color}
\usepackage[normalem]{ulem}

\newcommand{\be}{\begin{equation}}
\newcommand{\ee}{\end{equation}}
\newcommand{\ba}{\begin{eqnarray}}
\newcommand{\ea}{\end{eqnarray}}
\newcommand{\bfig}{\begin{figure}[t]\begin{centering}}
\newcommand{\efig}{\end{centering}\end{figure}}

\def\dm03{\hbox{$\Delta m^2_{03}$}}

\begin{document}

\title{Improving the neutrino mass hierarchy identification with
inelasticity measurement in PINGU and ORCA}

\author{Mathieu Ribordy}
\email{mathieu.ribordy@epfl.ch}
\affiliation{High Energy Physics Laboratory, \'Ecole Polytechnique F\'ed\'erale, CH-1015 Lausanne, 
Switzerland}

\author{A.\ Yu.\ Smirnov}
\email{smirnov@ictp.it}
\affiliation{The Abdus Salam International Centre for
Theoretical Physics, I-34100 Trieste, Italy}

\date{\today}

\begin{abstract} 

Multi-megaton scale under ice and underwater detectors of atmospheric neutrinos
with few GeV's energy threshold (PINGU, ORCA) open up new possibilities  
in the determination of neutrino properties, 
and in particular the neutrino mass hierarchy.  
With a dense array of optical modules it will be possible
to determine the inelasticity, $y$, of the charged current 
$\nu_\mu$ events in addition to the neutrino energy $E_\nu$ and 
the muon zenith angle $\theta_\mu$. The discovery potential of 
the detectors will substantially increase with the measurement of $y$. It will 
enable 
(i) a partial separation of the neutrino and antineutrino signals;
(ii)  a better reconstruction of the neutrino direction; 
(iii) the reduction of  the neutrino parameters degeneracy;    
(iv) a better control of  systematic uncertainties;  
(v) a better identification of the  $\nu_\mu$ events. 
It will improve the sensitivity to the CP-violation phase. The three dimensional, $(E_\nu, \theta_\mu, y )$,   
$\nu_\mu-$oscillograms with the kinematical as well as the experimental smearing are computed.  
We present the asymmetry distributions in the $E_\nu - \theta_\mu$ 
plane for different intervals of $y$ and study their properties. 
We show that the inelasticity information reduces  the 
effect of degeneracy 
of parameters by 30\%. With the inelasticity,
the total significance of establishing mass hierarchy may increase by 
$(20 - 50) \%$, thus effectively increasing the volume of the detector by factor $1.5 - 2$.  

\end{abstract}

\pacs{14.60.Pq
}          
\maketitle

\section{Introduction}

Multi-megaton scale atmospheric
neutrino detectors with few GeV's energy threshold
have an enormous and largely unexplored physics potential. 
These detectors are sensitive to the oscillatory patterns 
due to the 1-3 mixing in the neutrino energy - zenith angle 
($E_\nu - \theta_\nu$) plane.  
The patterns have several salient features, which include the 
MSW resonance peaks  due to oscillations in the mantle 
($E_\nu \sim  6$ GeV) and the core ($E_\nu \sim  4$ GeV) as well as  the parametric enhancement ridges  
at $E_\nu \sim (4 - 12)$ GeV which are realized for the 
core crossing neutrino trajectories 
(see~\cite{our1}  for detailed description  
and \cite{blennow} for recent review and references).
The patterns differ for neutrinos and antineutrinos   
and strongly depend on the type of neutrino  mass hierarchy. 
In particular, the indicated features appear in the neutrino 
channels in the case of normal mass hierarchy (NH) and in the 
antineutrino channels in the case of inverted hierarchy (IH)
(in the two neutrino approximation, inversion of the mass hierarchy is equivalent to 
switching the neutrino and antineutrino oscillation patterns).
This opens up a possibility to establish the neutrino mass hierarchy  
and also to measure the deviation of the 2-3 mixing from maximal as well as
the 1-3 mass splitting. Once the hierarchy is established,
one can consider a possibility to measure the CP-violation phase.

Multi-megaton detectors are expected to record 
of the order of $10^5$ events a year. Such a large statistics allows, 
in principle,  to compensate shortcomings related to flavor identification
of events and reconstruction of their energy and angular characteristics.
With so high statistics,  one can select some particular events in certain 
kinematical regions, which are most sensitive to a given
neutrino parameter, thus reducing effect of  degeneracy of parameters, {\it  etc}.

PINGU (Precision IceCube Next-Generation Upgrade)~\cite{pingu},  
the IceCube DeepCore~\cite{deepcore} augmented
with a denser instrumentation in its center,
and ORCA (Oscillation Research with
Cosmics in the Abyss)~\cite{orca} projects are possible future 
realizations of these Multi-megaton scale detectors.

A simplified estimation of the sensitivity to the  mass hierarchy
of the  DeepCore (DC)  experiment has been performed in \cite{Mena:2008rh}.
Due to the high energy threshold ($>10$ GeV), DC has a low sensitivity 
to the resonance pattern and therefore to the hierarchy. 
The sensitivity of DC to deviation of the 2-3 mixing from maximal
has been explored in \cite{fernandez}. 

The idea to send a neutrino beam from Fermilab to PINGU 
to determine the mass hierarchy has been elaborated in \cite{winter}.   

A possibility to use PINGU and the atmospheric neutrino flux 
for the identification of  the neutrino mass hierarchy (MH) 
and search for the CP violation effects    
was recently explored in~\cite{ARS}. 
The strategy is based on the measurement of the $E_\nu - \theta_\nu$ 
distribution of the sum of muon neutrino 
and antineutrino events.   
The smearing of the distribution over $\theta_\nu$ and $E_\nu$ 
has been performed that takes into account 
accuracy of reconstruction of the neutrino energy and direction.  

The estimator of discovery potential, $S$ (the hierarchy asymmetry), 
has been introduced \cite{ARS}, 
which allows one to make quick evaluation of sensitivities of the detector 
to neutrino mass hierarchy as well as to other parameters. 
For $ij$--bin  
in the reconstructed neutrino energy ($i$) and zenith angle ($j$), 
the  asymmetry is defined as 
\be
S_{\nu, ij} = \frac{N^{\rm{IH}}_{\nu,ij} - N^{\rm{NH}}_{\nu,ij}}{\sqrt{N^{\rm{NH}}_{\nu,ij}}}.  
\nonumber
\ee
Here $N^{\rm{NH}}_{\nu,ij}$ and   $N^{\rm{IH}}_{\nu,ij}$ are the numbers of events in the $ij$--bin 
for the normal and the inverted mass hierarchies correspondingly. 
The moduli of the asymmetry, $|S_{\nu, ij}|$,  gives the 
statistical significance for the identification of the mass 
hierarchy. The asymmetry  allows one  to explore in a transparent 
way the dependence of sensitivities 
on the experimental  energy and angular resolutions,   on degeneracies 
of parameters and on various systematic errors. 
In \cite{ARS} it was shown that the hierarchy can be established at $(3 - 10)\sigma$ 
level after 5 years of operation of PINGU depending on the energy 
and angular resolutions and on the size of the systematic error. 

A final answer concerning the sensitivity  should follow from the detailed Monte Carlo simulation
of the distributions of events. That should take into account
realistic parameters of the detectors after 
their geometries are determined.
Then, in a simple approach, the sensitivity can be obtained from the
fit of the simulated distributions with the distributions computed
for the cases of normal and inverted mass hierarchies. The fit can
be done using Poisson statistics without binning. 
Results of a ``toy'' Monte Carlo study
for large volume detectors have been presented in \cite{franco}.   
The physics potential of PINGU and ORCA was further explored in 
\cite{agarwalla} and \cite{franco}. 

As it was discussed in \cite{ARS},
several factors dilute the significance 
of the MH identification,   
although at the probability level the effect of inversion 
of the hierarchy is of the order 1. Indeed, 

(i) the hierarchy asymmetry has opposite
signs in different kinematical regions.
Therefore smearing over the angle and energy, $E_\nu - \theta_\nu$, 
leads to a substantial decrease of the
observable effect. The smearing originates
from finite energy and angular resolutions of the  detector 
(experimental smearing) and 
due to difference of the neutrino and muon directions
(kinematical smearing);

(ii) the hierarchy asymmetry has different signs for neutrinos and antineutrinos.  
Therefore summing up the neutrino and antineutrino signals leads to a 
partial cancellation of the  effect;

(iii) the presence of both $\nu_\mu$ and $\nu_{\rm e}$
flavors in the original atmospheric neutrino flux leads, in general,  
to a  suppression of oscillation effects.
The suppression becomes weaker at high energies, where the $\nu_e$ flux
is small;

(iv) current uncertainties of the oscillation parameters, 
such as $\Delta m_{32}^2$ and $\theta_{23}^2$  further 
reduce identification power, since  the effect of inversion of 
the mass hierarchy can be partly mimicked by changes of these parameters;

(v) the sample of $\nu_\mu$ events is contaminated by contributions 
from $\nu_\tau$ and $\nu_e$ charged current (CC)  
interactions and neutral current (NC) interactions 
of all neutrino species. 
In particular,  $\nu_\tau$'s generated via oscillations produce tau leptons,  
which decay in $18\% $ cases into muons, thus appearing as  $\nu_\mu$ events.  
Also $\nu_e$ and NC interactions   
can mimic $\nu_\mu$ events due to muon - pion misidentification.  
These events 
produce an additional effective smearing of the oscillatory pattern.   

All this renders the quest of the neutrino mass hierarchy difficult. 
Some (probably modest) developments of technology are required.
This includes the selection of certain geometry of the detector,
the upgrade of the optical modules, further developments of 
the time analysis of events, {\it  etc.}
On the other hand,   some particular ways to analyse the information 
obtained can also improve the sensitivity. 

In this connection, we explore 
improvements of the sensitivity to the neutrino MH  due to the measurement of the inelasticity,  
$y \approx 1 - E_\mu/E_\nu$,  of the  charged current $\nu_\mu$~events. 
As we will show,  this new ingredient in the analysis enables 
us to alleviate some of the problems mentioned above. 
In particular, it allows to effectively 
separate the $\nu_\mu$ and $\bar\nu_\mu$ signals,  
and thus to reduce  the partial cancellation of their contributions to the
MH asymmetry.  The idea was mentioned in \cite{Ribordy:2012gk}. 
Using the inelasticity will also allow to reduce kinematical 
smearing effect and degeneracy of parameters. It  will lead to a better flavor 
identification of the $\nu_\mu$ events. 

The paper is organized as follows. In Sec.~\ref{sct:2}, we describe the $\nu_\mu$ events, 
their detection characteristics, relevant kinematics and cross-sections. 
In Sec.~\ref{sct:3}, 
possible  improvements of the sensitivity to the mass hierarchy and other 
neutrino parameters due to the inelasticity measurements are discussed. 
In Sec.~\ref{sct:4}, we compute the 
three dimensional distributions of events in the reconstructed neutrino energy,  $E_\nu$, 
the  muon zenith angle, $\theta_\mu$,  and $y$ variables, which take into account the 
kinematical smearing. We explore the properties of 
these 3D distributions and find the corresponding hierarchy asymmetry plots.  
In Sec.~\ref{sct:5}, we present results of  smearing of the distributions 
over the finite experimental resolutions of observables: the muon and hadron cascade energies and 
the muon angle. We then compute the total significance of identification of the mass 
hierarchy, and its dependence on  possible accuracy of measurements 
of  the energies and angles (experimental smearing). 
We estimate how measurements of inelasticity reduce the effect of degeneracy of the mass hierarchy and  
mass splitting $\Delta m^2_{32}$.    
Sec.~\ref{sct:6} contains discussion of the results and outlook.

\section{$\nu_\mu$ events and inelasticity} 
\label{sct:2}

\subsection{$\nu_\mu$ events}

In this paper we concentrate on the $\nu_\mu$ events induced 
by the charged current  weak interactions:  
\be
\nu_\mu + N \rightarrow \mu + h,  
\label{eq:react}
\ee
where $h$ refers to the hadron system in the final state.  
Observables associated to the reaction~(\ref{eq:react}) 
are the  energy of the muon $E_\mu$, its direction characterized 
by the zenith and azimuth angles $\theta_\mu$ and $\phi_\mu$, 
and the cascade energy  (the total energy in hadrons), $E_h$. 

At the energies we consider,  from a few GeV up to $\approx$30 GeV, the cascade direction
is not meaningful on an event basis. Indeed, the cascade energy 
is shared in a variable mixture between light mesons and heavier hadrons, 
which leads to a highly random and anisotropic Cherenkov photons emission.
Contrary to muons, the energy release from cascades
is approximately ``point-like", given the sparsely instrumented 
detector arrays under investigation. 

The reconstruction of the $\nu_\mu$ event consisting of
recorded photons (hits) from the combined emission from a vertex shower
and a muon track in this case, can be
performed well at low energy using prescriptions in~\cite{Ribordy:2006qd}. 
The visible cascade energy, the muon track length 
({\it i.e.} its energy) and incoming muon direction can be extracted. 
Moreover, the impact of the short scattering length  
of Cherenkov photons for a dense detector
in ice is expected to be rather mild as it will be argued later.
As most hits are undelayed,  a good reconstruction accuracy 
of the muon incoming direction as well as a  clear separation 
of the shower and muon signatures are expected.

Thus, the set of observables $\{E_\mu, \theta_\mu, \phi_\mu, E_h\}$, 
constitutes a rather exhaustive description of the CC $\nu_\mu$ interaction.
The original neutrino energy is determined through
\be
E_\nu  = E_h + E_\mu - m_N,    
\label{eq:Econs}
\ee
where $m_N$ is the nucleon mass. 

\subsection{Inelasticity and kinematics of the process}

The inelasticity $y$ is defined as 
\be
y \equiv 
\frac{E_\nu - E_\mu}{E_\nu}, 
\label{def-y}
\ee

Let us consider the angle between the neutrino and the produced muon, $\beta$. 
The square of the transfer momentum, $q^2$, equals 
\be
q^2 \equiv (p_\nu - p_\mu)^2 = - 2 E_\nu (E_\mu - |{\bf p}_\mu| \cos \beta ) + m_\mu^2, 
\nonumber
\ee
so that  $Q^2 \equiv - q^2$ equals 
\be
Q^2 =  2 E_\nu E_\mu \left(1  - \frac{|{\bf p}_\mu|}{E_\mu} \cos \beta \right) - m_\mu^2 . 
\nonumber
\ee
This gives 
\be
\cos \beta = \frac{E_\mu}{|{\bf p}_\mu|} \left(1 - \frac{Q^2 + m_\mu^2}{2 E_\nu E_\mu} \right). 
\label{cosb}
\ee

In terms of the Bjorken variable  
\be
x \equiv \frac{Q^2}{2 (p_N q)} = \frac{Q^2}{2 m_N (E_\nu - E_\mu)} 
\nonumber
\ee 
we have 
\be 
Q^2 = 2 x y  m_N E_\nu,  
\nonumber
\ee
where we used (\ref{def-y}). Insertion of  this expression into 
(\ref{cosb}) gives
\be
\cos \beta = \frac{E_\mu}{|{\bf p}_\mu|} 
\left[1 - \frac{2 x y  m_N E_\nu + m_\mu^2}{2 E_\nu E_\mu} 
\right].
\label{cosb1}
\ee
Notice that with decrease of  ${\bf p}_\mu$, 
$\cos \beta \rightarrow \pm 1$ when $x\rightarrow 0,1$. 
We can rewrite (\ref{cosb1}) as 
\be
\cos \beta =   1 - 2 x y \zeta(E_\nu,x,y),  
\nonumber
\ee
where
\be
\zeta(E_\nu,x,y) =\frac{m_\mu^2 + 2 m_N E_\nu x y - 2 E_\nu(E_\mu  - |{\bf p}_\mu|)}{4 E_\nu x y |{\bf p}_\mu|}. 
\nonumber
\ee
Here $E_\mu = E_\mu(E_\nu,y)$ and  ${\bf p}_\mu = {\bf p}_\mu(E_\nu,y)$. 
If ${\bf p}_\mu  \approx E_\mu \gg m_\mu$,  
we obtain neglecting  $m_\mu$   
\be
\zeta(E_\nu,y)  \approx \frac{m_N}{2 E_\mu}.
\nonumber
\ee

Let us find the limits in which $c_\beta \equiv \cos \beta$ changes. 
Varying $x$, we obtain for $x = 0$ that  
$c_\beta \approx E_\mu/|{\bf p}_\mu| > 1$, so that 
$c_\beta^{\rm max} = 1$. For not very small 
$|{\bf p}_\mu|$ (and we will consider $|{\bf p}_\mu| > m_N$) 
the minimal value of $c_\beta$ corresponds to 
$x = 1$: 
\begin{eqnarray}
c_\beta^{\rm min} &=& \frac{E_\mu}{|{\bf p}_\mu|} 
\left[1 - \frac{2 m_N (E_\nu - E_\mu) + m_\mu^2}{2 E_\nu E_\mu} \right] \nonumber\\
&\approx& 
\frac{E_\mu}{|{\bf p}_\mu|} 
\left[1 - \frac{2 m_N}{E_\mu} 
\left(1 - \frac{E_\mu}{E_\nu}\right) \right], 
\label{cbeta-min}
\end{eqnarray}
or $\sin \beta^{\rm min}/2 \sim  \sqrt{y m_N / E_\mu}$. 

For a given muon direction,  the neutrino 
direction is determined by the angle $\beta$ and the azimuthal angle 
$\phi$    
with respect to the plane formed by the muon momentum and axis $x$. 
It is straightforward (see Appendix~\ref{appendix:A}) to find the relation between 
the neutrino zenith angle $\theta_\nu$   
and the muon zenith angle $\theta_\mu$: 
\be
c_\nu =  c_\beta c_\mu + s_\beta s_\mu c_\phi,   
\label{eq:anglrel}
\ee 
where $c_\nu \equiv \cos \theta_\nu$, 
$c_\mu \equiv \cos \theta_\mu$ and $c_\phi \equiv \cos \phi$. 

According to  (\ref{eq:anglrel}) for fixed $\beta$ the maximal 
and minimal values of $c_\nu$ correspond to 
$c_\phi = \pm 1$ and equal  
\be
c_\nu^{\rm max}  =  \cos(\theta_\mu - \beta),  ~~~~~
c_\nu^{\rm min}  =  \cos(\theta_\mu + \beta).  
\label{eq:c-limit}
\ee

\subsection{Cross-sections}
 
In our calculations of Secs.~\ref{sct:4} and~\ref{sct:5}
we will use the deep inelastic scattering  (DIS) cross-section only.  
We neglect the contributions of the single pion production  and  
quasi-elastic scattering processes. This will lead to conservative 
estimations of sensitivities, 
as the reduced momentum transfer translates into smaller angle 
between the muon and the neutrino directions. 
In any case the relative importance of
these processes  becomes negligible above $\approx 5$~GeV,   
and below $\approx 5$~GeV, detector resolutions and effective volumes are strongly limited.

The differential CC cross-sections  of the $\nu$ and $\bar\nu$ DIS   
on an isonucleon N=$\frac{1}{2}$(n+p) equal  
\begin{eqnarray}
&&\frac{{\rm{d}}^2\sigma_\nu^{\rm CC}}{{\rm{d}}x{\rm{d}}y}(E_\nu,x,y) = 
\frac{G_F^2 m_{\rm{N}} x E_\nu}{\pi} 
\nonumber \\ 
&&\,\,\,\,\,\,\,\,\,\,\,\,\times [ (q+s-c) + (1-y)^2 (\bar{q} -\bar{s}+\bar{c})] ,  
\label{eq:crossnu}
\\
&&\frac{{\rm{d}}^2\sigma_{\bar\nu}^{\rm CC}}{{\rm{d}}x{\rm{d}}y}(E_\nu,x,y) = 
\frac{G_F^2 m_{\rm{N}} x E_\nu}{\pi} 
\nonumber \\
&&\,\,\,\,\,\,\,\,\,\,\,\,\times [ (\bar{q}-\bar{s}+\bar{c}) + (1-y)^2 (q+s-c)],  
\label{eq:crossantinu}
\end{eqnarray}
where $q \equiv u+d+s+c$,  
$\bar{q} \equiv \bar{u} + \bar{d} + \bar{s} + \bar{c}$ 
and the quark densities $u =  u(x, Q^2)$, {\it etc.},   
are described by the  CTEQ5 parton distribution functions 
in the standard $\overline{\rm MS}$ scheme~\cite{cteq5}, 
valid down to $Q^2\approx1$~GeV$^2$. 

The limits of $x-$integration of the cross sections are 
in the interval $\{ x_{{\rm min}}, x_{\rm max} \} = \{ x(c_\beta=1) , 1\}$,  
and $y$--integration runs from 0 to $y_{\rm max}=1-m_\mu/E_\nu$. 

Integrating the cross-sections (\ref{eq:crossnu}) and  
(\ref{eq:crossantinu}) over $x$ 
we obtain  
\begin{eqnarray}
\frac{{\rm d}\sigma_\nu^{\rm CC}}{{\rm d}y} &=& [-a_0 - a_1 (1 - y)^2] \, 10^{-38}{\rm cm}^2
\frac{E_\nu}{\rm 1~GeV}, \nonumber \\
\frac{{\rm d}\sigma_{\bar\nu^{\rm CC}}}{{\rm d}y} &=& [-b_0 - b_1 (1 - y)^2)]  \,10^{-38}{\rm cm}^2
\frac{E_\nu}{\rm 1~GeV}, \nonumber
\end{eqnarray}
where $a_0=0.72$, $a_1=0.06$, $b_0=0.09$ and $b_1=0.69$.  
Then the normalized inelasticity distributions equal 
\begin{eqnarray}
p_\nu & \equiv &-\frac{1}{\sigma_\nu }\frac{{\rm{d}}\sigma_\nu}{{\rm{d}}y} 
\approx\frac{a_0+a_1 (y-1)^2}{a_0+a_1/3},\label{eq:pnuY} \\
p_{\bar\nu} & \equiv & -\frac{1}{\sigma_{\bar\nu}}\frac{{\rm{d}}\sigma_{\bar\nu}}{{\rm{d}}y} 
\approx\frac{b_0+b_1 (y-1)^2}{b_0+b_1/3}.  
\label{eq:pnubarY}
\end{eqnarray}
Here,
we have dropped the very 
weak $E_\nu-$dependence in the range of interest and considered the limit $m_\mu\rightarrow0$.

\subsection{Number of events}

The number of neutrino and antineutrino events in the case of NH,  
$N_{\nu}^{NH}$ and $N_{\bar{\nu}}^{NH}$ in a given 
$ij-$ bin of the size $\Delta_i \cos \theta_\nu$,  $\Delta_j E_\nu$ 
equals
\be
N_{\nu}^{NH} =   
\int_{\Delta_i \cos \theta_\nu}  {\rm d}\cos\theta_\nu 
\int_{\Delta_j E_\nu}  {\rm d} E_\nu ~
\rho_{\nu}^{NH} (E_\nu, \cos \theta_\nu),
\nonumber
\ee
and for $N_{\bar{\nu}}^{NH}$  one needs to substitute 
$\rho_{\nu}^{NH} \rightarrow \rho_{\bar{\nu}}^{NH}$.   
Here 
\ba
\rho_{\nu}^{NH} \equiv  
2\pi N_A n_{\rm ice} V_{\rm eff} T 
\sigma^{\rm CC} \Phi_\mu^0 \left[P_{\mu \mu}^{NH} + 
\frac{1}{r} P_{{\rm e} \mu}^{NH}  \right],
\label{eq:numb-emu} \\
\rho_{\bar{\nu}}^{NH} \equiv 
2\pi N_A n_{\rm ice} V_{\rm eff} T 
\bar{\sigma}^{\rm CC} \bar{\Phi}_\mu^0   \left[\bar{P}_{\mu \mu}^{NH} +
\frac{1}{\bar{r}} \bar{P}_{{\rm e} \mu}^{NH}  \right].  
\label{eq:anumb-emu}
\ea
In (\ref{eq:numb-emu}, \ref{eq:anumb-emu}),
\be
r \equiv \frac{\Phi_\mu^0}{\Phi_{\rm e}^0}, ~~~ 
\bar{r} \equiv \frac{\bar{\Phi}_\mu^0}{\bar{\Phi}_{\rm e}^0}  
\nonumber
\ee
are the flavor ratios, where $\Phi^0_\alpha = \Phi^0_\alpha(E_\nu, \theta_\nu)$ 
are the neutrino fluxes at production; 
$P_{\alpha\beta}^{NH}$ and $\bar{P}_{\alpha\beta}^{NH}$ are the 
$\nu_{\alpha} \rightarrow \nu_{\beta}$ oscillation probabilities 
for neutrinos and antineutrinos.  
$V_{\rm eff}(E_\nu)$ is the effective volume of a detector, 
$\rho_{\rm ice}$ is the ice density, $N_A$ is the Avogadro number, and  
$T$ is the exposure time.  

For the effective mass of the detector we take~\cite{ARS}
\be
\rho_{\rm ice} V_{\rm eff}(E_\nu) = 14.6\times \left[\log(E_\nu/{\rm GeV})\right]^{1.8} \,{\rm Mt}.
\nonumber
\ee
We keep the same effective volume as in~\cite{ARS}
in spite of several recent re-evaluations for two reasons: 
(i) for easier comparison of results with those in \cite{ARS} 
and some other publications,  and (ii) because the 
final configuration of the detectors are not yet determined. 
If the  effective volume 
(which also depends on the criteria of selection of events) 
is  reduced by factor 3 - 4, 
the significance for the same exposure period will be reduced by factor 1.7 - 2.0.  

Expressions for the inverted mass hierarchy is obtained with substitution  NH $\rightarrow$ IH.  
Recall that in the $2 \nu$ approximation, 
when effects of  1-2 mixing and mass splitting are neglected, 
there are relations between the probabilities for normal and inverted hierarchies 
\be
P_{\alpha\beta}^{NH} = \bar{P}_{\alpha\beta}^{IH}, ~~~~ 
P_{\alpha\beta}^{IH} = \bar{P}_{\alpha\beta}^{NH}. 
\label{eq:relp}
\ee 
That is, an inversion of the mass hierarchy is equivalent to 
switching neutrinos and antineutrinos. 
In the three neutrino mixing context, the relations (\ref{eq:relp})
are not exact (see, e.g., Fig. 2 in \cite{ARS}), 
especially for the  core crossing trajectories.

\section{Impact of inelasticity determination. Qualitative picture}
\label{sct:3}

\subsection{Inelasticity and separation of neutrino and antineutrino
signals}  

The hierarchy asymmetries in the  neutrino and antineutrino 
channels have opposite signs. 
In fact, in expressions for the difference of numbers 
of events for NH and IH  (see \cite{ARS}) all the terms are proportional 
to the factors $(1 - \kappa_\mu)$ and $(1 - \kappa_{\rm e})$, 
where 
\be
\kappa_\mu \equiv
\frac{{\bar \sigma}^{\rm CC} \bar{ \Phi}_\mu^0}
{\sigma^{\rm CC} \Phi_\mu^0}
\nonumber
\ee
is the ratio of cross-sections and fluxes of the muon  antineutrinos and neutrinos at the production, 
and $\kappa_{\rm e}$ is defined similarly.  
The ratios $\kappa_\mu$ and $\kappa_{\rm e}$ depend on neutrino energy and direction and equal approximately
$0.4-0.6$.   
The $\nu$ and $\bar{\nu}$  contributions to the number of  events  partially cancel 
each other. 
So,  in this case the MH determination  relies 
on the non-equal $\nu$ and $\bar\nu$  fluxes and cross-sections.
The separation of the neutrino and antineutrino signals allows to further reduce  
the cancellation and therefore to enhance the significance. 
As follows from Eqs (\ref{eq:pnuY}-\ref{eq:pnubarY}), the average value 
of $y$ is 50\% larger for $\nu$ than for $\bar\nu$.  
Therefore we can use the inelasticity to separate the $\nu$ 
and $\bar\nu$ signals.  

One possible procedure is to determine for each bin (a large number 
of events will allow to do this) the fraction of neutrino 
and antineutrino events by fitting its $y-$distribution:  

1. Select small enough bins in neutrino energy - zenith angle plane, 
   so that the oscillatory structures due to certain mass hierarchy are not averaged out
   (the bin size should be eventually optimized). The number of neutrino and antineutrino events in each bin equals 
\be
N_{\bar{\nu}} = N  \alpha ~~~~N_\nu = N  (1 - \alpha), 
\label{eq:alpha}
\ee
where $\alpha$ is the fraction of antineutrino events and we have omitted the bin 
indices. 

2. Measure the $y-$distribution of these events.

3. Fit the measured distribution with 
\be
p_{\nu \bar{\nu}}(y, \alpha) = (1 - \alpha) p_\nu (y) + 
\alpha p_{\bar{\nu}}(y),  
\label{eq:pnunubar}
\ee
where $p_\nu (y)$ and  $p_{\bar{\nu}}(y)$  are given in 
(\ref{eq:pnuY})  and  (\ref{eq:pnubarY}), 
thus, determining the fraction $\alpha$. \\ 

A possible enhancement of the sensitivity to the hierarchy due 
to the separation of the $\nu$ and $\bar{\nu}$  
signals can be estimated in the following way.  
In the described procedure there are two independent observables: 
the total number of events, $N = N_\nu + N_{\bar \nu}$,
and $\alpha$ extracted from the $y$-distribution
with the accuracy $\delta \alpha$. In general,  
\be
\delta \alpha  \approx \frac{\gamma}{\sqrt {N}}, 
\label{eq:gamma}
\ee
where $\gamma =  \gamma(\alpha ,  N)$. 
The error $\delta \alpha$ can be estimated using the method of moments. 
As we have two parameters, $\alpha$ and $\delta\alpha$, to extract, 
it is sufficient to calculate the first and second $y$-moments of  
$p_{\nu \bar{\nu }}(y,\alpha )$ given in  Eq.~(\ref{eq:pnunubar}).  Using  
expressions (\ref{eq:pnuY}--\ref{eq:pnubarY}), we obtain the 
average inelasticity $\langle y \rangle$: 
\be
\langle y (\alpha ) \rangle =  \int  y p_{\nu \bar{\nu }}(y,\alpha ) {\rm d}y 
\approx  0.494 - 0.174 \alpha . 
\label{eq:m1} 
\ee
The mean deviation $\sigma_{\bar y}^2\equiv\langle (\bar y - \langle y \rangle)^2\rangle$ 
of the average $\bar y$ value after $N$ measurements from the true value $\langle y \rangle$   is 
\begin{eqnarray}
\sigma_{\bar y}(\alpha, N)^2&=&\frac{\sigma_y(\alpha)^2}{N}
 =  \frac{1}{N}\left[\int y^2 p_{\nu \bar{\nu}}(y,\alpha) {\rm d}y - 
\langle y (\alpha ) \rangle^2 \right] 
\nonumber \\
&\approx& \frac{1}{N} \left(0.084 + 0.010\alpha - 0.030 \alpha^2 \right).  
\label{eq:m2}
\end{eqnarray}
From Eq.~(\ref{eq:m1}) and following a measurement  of $\bar y$, 
we obtain a measured  value $\tilde\alpha$. 
The average value $\delta\alpha$ is given by 
\be 
\delta\alpha = \sqrt{\langle  (\tilde\alpha - \alpha)^2 \rangle} 
= \frac{\sqrt{\langle  (\bar y - \langle y \rangle)^2\rangle}}{ 0.174} = 5.75\sqrt{\sigma_{\bar y}^2}. 
\ee
Substituting  $\sigma_{\bar y}^2$ from
Eq.~(\ref{eq:m2}), we obtain
\be
\delta \alpha(\alpha,N) \approx \frac{\gamma}{\sqrt{N}} 
\sqrt{1+0.115\alpha -0.362\alpha^2},
\label{eq:alphaDepDA}
\ee
where $\gamma = 1.66$. 

If $\alpha$  is not close to 1 (for all practical purpose $\alpha\lesssim0.5$), 
$\delta \alpha$ weakly depends on $\alpha$ and we can use Eq. (\ref{eq:gamma}).   

A detailed investigation of $\delta\alpha(\alpha, N)$ 
by means of the  maximum likelihood method confirms the estimate 
Eq.~(\ref{eq:alphaDepDA}) for large $N$. 
For $N\lesssim100$, this method shows a slight improvement 
with increasing $\alpha$ with respect to the method of moments. 
It is worthwhile to further explore this approach.

Let us find the errors $\sigma_\nu$ and $\sigma_{\bar{\nu}}$ 
in the determination of $N_\nu$ and $N_{\bar{\nu}}$. 
According to Eq. (\ref{eq:alpha}), variations of  $N_\nu$ can be written as 
\be
\delta N_\nu = (1 - \alpha) \delta N - N \delta \alpha  =
(1 - \alpha) \sqrt{N} - N \delta \alpha .
\nonumber
\ee
The variations $\delta N $ and $\delta \alpha $
are independent and therefore they sum up squared:
\be
\sigma_\nu^2 = (1 - \alpha)^2 N  + (\delta \alpha)^2 N^2, ~~~
\sigma_{\bar{\nu}}^2 = \alpha^2 N  + (\delta \alpha)^2 N^2. 
\label{eq:sigsq}
\ee
Assuming that the measured quantities $\tilde N$ and $\tilde\alpha$ are respectively 
distributed according to Poisson with mean $N$ and 
Gaussian with mean $\alpha$ and standard deviation $\delta\alpha$, 
the exact variance calculation of $N_{\nu, \bar\nu}(\tilde\alpha, \tilde N)$ 
leads to the same result as in~Eq.~(\ref{eq:sigsq}), provided that $N\gg\gamma^2$.

Using (\ref{eq:gamma}), we have for NH
\be
\frac{\sigma_{\nu}}{\sqrt{N^{NH}}} = \sqrt{(1 - \alpha)^2 + \gamma^2}, 
~~~~\frac{\sigma_{\bar \nu}}{\sqrt{N^{NH}}} = 
\sqrt{\alpha^2 + \gamma^2}.
\label{eq:sigmaext}
\ee

The hierarchy asymmetries in the neutrino and antineutrino channels
can then be written  as
\be
S_{\nu} = \frac{
N_{\nu}^{IH} - N_{\nu}^{NH}}{\sigma_{\nu}}, ~~~~~
S_{\bar \nu} =
\frac{
N_{\bar \nu }^{IH} - N_{\bar \nu }^{NH}}{\sigma_{\bar \nu}}.
\label{eq:ssbar}
\ee
Here we assume that NH is the true hierarchy and therefore 
corresponding number of events is what is measured.

If $S_{\bar{\nu}}$ and $S_{\nu}$ are independent,
the total significance equals
\be
S_{\rm tot}^{\rm sep} = \sqrt{S_\nu^2 + S_{\bar\nu}^2}~.  
\nonumber
\ee
It can be rewritten using Eqs. (\ref{eq:ssbar})  and (\ref{eq:sigmaext}) as 
\be
S_{\rm tot}^{\rm sep} = \frac{1}{\sqrt{N^{NH}}}
\sqrt{\frac{(N_{\nu}^{IH} - N_{\nu}^{NH})^2}{(1 - \alpha)^2 + \gamma^2}
+ \frac{(N_{\bar{\nu}}^{IH} - N_{\bar{\nu}}^{NH})^2}{\alpha^2 + \gamma^2}}.
\nonumber
\ee
For the significance without $\nu - \bar{\nu}$ separation we 
would have  
\be
|S_{\rm tot}| = \left|\frac{N_{\nu}^{IH} + N_{\bar{\nu}}^{IH}  -
N_{\nu}^{NH} - N_{\bar{\nu}}^{NH}}{\sqrt{N^{NH}}} \right| . 
\nonumber
\ee
Therefore the enhancement factor $R \equiv S_{\rm tot}^{\rm sep}/|S_{\rm tot}|$ due to separation 
of the neutrino and antineutrino signals equals
\be
R = 
\frac{1}{1 - \kappa_\mu f_P} 
\frac{1}{\sqrt{\alpha^2 + \gamma^2}}~ 
\sqrt{\frac{\alpha^2 + \gamma^2}{(1 - \alpha)^2 + \gamma^2}
+ (\kappa_\mu f_P)^2}.  
\label{eq:enhancement3}
\ee
Here 
\be
\kappa_\mu f_P = -  \frac{N_{\bar \nu}^{IH} - N_{\bar \nu}^{NH}}{N_{\nu}^{IH} - N_{\nu}^{NH}}, 
\nonumber
\ee
and 
\be
f_P \equiv  
\frac{ \bar{P}_{\mu \mu}^{NH} - \bar{P}_{\mu \mu}^{IH}  
+ \frac{1}{\bar{r}} \left(\bar{P}_{\rm e \mu}^{NH} - 
\bar{P}_{\rm e \mu}^{IH}  \right)}{P_{\mu \mu}^{IH} - P_{\mu \mu}^{NH} 
+ \frac{1}{r} \left( P_{\rm e \mu}^{IH} - P_{\rm e \mu}^{NH} \right)}. 
\nonumber
\ee
If $\bar{r} = r$, in the $2\nu$ approximation we would 
have  $f_P = 1$. 

In Eq. (\ref{eq:enhancement3}) the minus sign in the denominator 
of the first factor reflects the partial cancellation of the hierarchy 
asymmetries from  the neutrino and antineutrino channels. 
The second factor describes the reduction of enhancement due to the
error in the separation of the neutrino and antineutrino signals.
The expression is valid if $\alpha$ is not very close to 0 or 1. 

Notice that the enhancement factor $R$  
does not depend explicitly on the number of events. The number of events 
is mainly encoded in $\kappa_\mu f_P$ and in  $\gamma$. 
The value of 
$\alpha$ changes from bin to bin. 
For $\alpha = 0.50 ~~(0.32)$,  $\kappa_\mu = 0.4$ and 
$f_P = 1$ we obtain $R = 1.05~~ (1.01)$.

The enhancement factor is very close to unity. However, a slight improvement 
on the determination of $\gamma$  leads to a substantial increase of $R$. 
A 10\% decrease of $\gamma$ leads to $R=1.15~~ (1.10)$ for $\alpha=0.50$ (0.32). 

Notice that according to (\ref{eq:ssbar}), the ratio 
\be
\frac{S_{\bar \nu}}{S_{\nu}} = - f_p \kappa_\mu 
\frac{\sigma_{\nu}}{\sigma_{\bar\nu}}
\nonumber
\ee
is negative and 
$\nu$ and $\bar{\nu}$ asymmetries
have opposite signs. 

For ideal separation, $\gamma = 0$, we would have  
\be
R =   
\frac{1}{1 - \kappa_\mu f_P} 
\frac{1}{\alpha}~ 
\sqrt{\frac{\alpha^2}{(1 - \alpha)^2}
+ (\kappa_\mu f_P)^2}~ . 
\nonumber
\ee
It gives $R = 3.6$ (3.2) for  $\alpha = 0.50$ (0.32).
This number can be considered as the maximal possible 
enhancement.

Notice that the estimations presented above differ from the  estimations 
in the case in which  the
numbers of $\nu_\mu$ and $\bar{\nu}_\mu$ events are 
measured independently (in our previous consideration these numbers correlate). 
In the latter,  $\sigma_{\nu} = \sqrt{N_{\nu}}$, 
$\sigma_{\bar{\nu}} = \sqrt{N_{\bar{\nu}}}$
and the enhancement factor equals
\be
R =
\frac{1}{1 - \kappa_\mu f_P}
\frac{1}{\sqrt{\alpha}}~
\sqrt{\frac{\alpha}{1 - \alpha}
+ (\kappa_\mu f_P)^2}~ .
\nonumber
\ee
If $f_P = 1$ and $\kappa_\mu \approx 0.5$,  we obtain 
$R = 2.4$ for $\alpha = 0.32$ . 

The above estimations have been done for a single bin and 
one should average the enhancement factor over all the  bins. 
Since $R$ depend weakly on $N$,  the estimation for  
$\alpha \sim 0.5$ give good idea about the overall enhancement.  

Notice that the weak  enhancement factor  we obtain is due to
the  error of the 
separation parameter, $\delta \alpha$. This is confirmed by 
exact computations in Sec.~\ref{sct:5}. 

\subsection{Inelasticity and reconstruction of neutrino direction}

The dominant source of sensitivity loss for the determination
of the neutrino mass hierarchy follows from the angular smearing
of the oscillograms \cite{ARS}, and in particular, the kinematical 
smearing due to the angle between the neutrino and muon directions.   
Indeed, according to (\ref{cosb1}),
\be
\sin^2{\frac{\beta}{2}} \approx \frac{Q^2}{4E_\nu E_\mu} \approx 
\frac{m_N xy}{2 E_\mu}.  
\label{eq:beta2}
\ee 
From this relation 
with $\langle x \rangle\approx0.3$ we find that 
the average angle which characterizes the kinematical smearing  
is
\be
\langle \beta  \rangle \approx \frac{0.75}{\sqrt{E_\nu/{\rm{GeV}}}} 
\sqrt{\frac{y}{1-y}}. 
\label{eq:kinsmear}
\ee
Then for the average values $y_\nu\approx0.5$  
and $y_{\bar\nu}\approx0.3$ it equals 
$\langle \beta_\nu \rangle \approx {0.75}/{\sqrt{E_\nu/{\rm{GeV}}}}$ 
and 
$\langle \beta_{\bar\nu} \rangle  \approx 
{0.5}/{\sqrt{E_\nu/{\rm{GeV}}}}$.  
Using these estimations we find that $\Delta c_\nu$ is larger 
than the region of the same sign hierarchy asymmetry for $E_\nu < 6$ GeV. \\

According to (\ref{eq:beta2}), interactions with small $y$ 
correspond to small scattering angles. Thus, the selection of events 
with small $y$ reduces the interval of possible values of $\beta$. 
For instance,  for a sample with $y<0.3$,
the average inelasticity is about $\langle y\rangle\approx0.14$. 
Then according to (\ref{eq:kinsmear}) the  average angle between the muon 
and  neutrino incoming directions equals  
$\langle \beta \rangle \approx {0.13}/{\sqrt{E_\nu/{\rm{GeV}}}}$. 
The sample however retains about 30\% of neutrino 
and 55\% of antineutrino events, thus having lower statistics. 

At small $y$, the angular reconstruction error 
of the muon itself is small, as the muon carries most of 
the neutrino energy and there are less hits from the cascade, which otherwise 
worsen the reconstruction of the muon direction.

However, at small $y$, the difference of cross-sections 
of neutrinos and antineutrinos 
becomes smaller (they are equal at $y = 0$).
Therefore the separation of the neutrino and antineutrino signal 
becomes difficult, and the cancellation of neutrino 
and antineutrino signals in the hierarchy asymmetry becomes stronger. \\

For large $y$, on the other hand, the contribution 
of $\bar{\nu}$ is strongly suppressed, which eliminates  
the $\nu - \bar{\nu}$ cancellation.  
But for events with large $y$ the reconstruction of the neutrino direction 
is very poor.  Furthermore, identification of the 
$\nu_\mu$ events becomes difficult (see below).

\subsection{Inelasticity, systematic errors and degeneracy of parameters} 

In~\cite{ARS}, the method is mainly based on the differential measurement
of the neutrino-induced muon flux from different incoming directions
and at various energies, avoiding some sources of systematic 
uncertainties (especially the correlated ones).  
The approach adopted in this paper goes
a step beyond with the additional sensitivity to 
the $\nu_\mu$ -- $\bar\nu_\mu$ admixture or $y$  providing
the method an even stronger immunity to sources 
of systematic uncertainties.

The degeneracy of the neutrino  parameters reduces significantly 
the sensitivity to the mass hierarchy \cite{ARS}. The
problem may  be alleviated, but not avoided, in the future by more precise measurement 
of neutrino parameters in  MINOS, T2K, NOvA and in reactor experiments. 
The use of inelasticity in analyses  will reduce the  impact
of degeneracies.  Indeed, effects of uncertainties, 
e.g. in $\Delta m^2_{32}$ and $\theta_{23}$, are nearly the same for $\nu$ and $\bar\nu$, while 
the $y$-distributions for $\nu$ and $\bar\nu$ are different.  
Therefore  measurements of inelasticity will allow to somehow
separate effects. A quantitative study of the corresponding improvements will be given in Sec. \ref{sct:5}. 

\subsection{Inelasticity and identification of $\nu_\mu$ events}

As discussed in \cite{ARS},  tau neutrinos contaminate the $\nu_\mu$ sample by about 5\%  contribution.  
Oscillation effect on this contribution differs from the one 
on the true $\nu_\mu$ events. This leads to a kind of 
additional smearing, which cannot be neglected.  
The inelasticity observable enables us to further suppress the 
number of  $\nu_\tau \rightarrow \mu$
events in a sample because of the specific vertex kinematics  
of tau neutrino interactions: Rather large 
showers are produced and the angle between the muon 
and tau neutrino is large, as the muon is sharing energy 
with two other neutrinos. Therefore  this class of events has in average rather 
large effective $y$, and restricting an analysis to small $y$ will 
allow to disentangle at least partly the $\nu_\tau$ contribution.
Quantitative analysis of this suppression is beyond the scope this paper.  

For large $y$, due to the low energy of muon, 
the probability of  misidentification of the $\nu_\mu$ events with the 
CC $\nu_{{\rm e}, \tau}$ events as well as the  NC events of all neutrino species   
becomes large. 
Indeed, there can be  confusion 
between the charged pion and muons as they both have a long decay length 
($\lambda_{\pi^\pm}\approx56\,\rm m$ at 1 GeV) and propagate with low energy loss rate 
(the dominant ionization energy loss  limits  their range 
to $\lesssim5\,\rm m$ per GeV).  
However, the energy distributions of muon and pions strongly differ: 
the simulation with GENIE~\cite{genie} of 10 GeV $\nu_\mu$ interactions 
shows that the most probable $E_{\pi^\pm}$ is of the order 
of a few 100's MeV, so that  $E_{\pi^\pm}\approx1\,{\rm GeV}$ 
is already unlikely high. The reaction favors events with higher $\pi$ multiplicity 
rather than events with  higher $\pi$ energies. 

\section{3D - distributions and oscillograms}\label{sct:4}

As we saw in the previous section  
the separation of the neutrino and antineutrino signals requires
measurements of the $y-$distribution in a wide range of $y$,  
and especially for large $y$,  where the difference of the 
neutrino and antineutrino cross-section is maximal. 
On the other hand, good reconstruction of the neutrino directions
requires  selection of events with small $y$. 
In a sense, improvements of the sensitivity due to 
$\nu -\bar{\nu}$ separation and narrowing 
the angular distribution  are incompatible. 
Small $y$ are preferred also for the identification of 
the $\nu_\mu$ events and the disentanglement of  the $\nu_\mu$ from $\nu_{e, \tau}$ 
events. Therefore, one expects that the best sensitivity to the 
neutrino mass hierarchy is for the intermediate range of $y$.  
Here the  interplay of different effects occurs, which requires 
a combined description using the differential characteristics in 
$y$ and also in $x$ since the angle depends on $x$ too.

\subsection{Densities of events. Oscillograms for different $y$}

The density of the $\nu_\mu$ events as function of 
$E_\nu$, $c_\mu$, $y$ equals  
\be 
n_\nu^{NH}(E_\nu, c_\mu, y) =  
\frac{1}{2\pi}\int_{c_\beta^{\rm min}}^{1} {\rm d}c_\beta \int_0^{2\pi} 
{\rm d}\phi ~\frac{{\rm d}^2 \sigma_\nu^{\rm CC}}{{\rm d}c_\beta {\rm d}y} 
\frac{\rho_\nu^{NH} (E_\nu, c_\nu )}{\sigma_\nu^{\rm CC} (E_\nu)}, 
\nonumber
\ee
where $\rho_\nu^{NH}$ is defined in (\ref{eq:numb-emu}); 
$c_\nu$ is given in (\ref{eq:anglrel}), and 
the lower limit of integration,  
$c_\beta^{\rm min}$, is defined in (\ref{cbeta-min}). 
Similar expressions can be written for  antineutrinos  
and for the IH case.

Using the relation Eq.~(\ref{eq:anglrel}),
we change the integration variables,
$d\phi \rightarrow  dc_\nu$:   
\begin{eqnarray}
\label{final}
n_\nu^{NH} (E_\nu, c_\mu, y) =  
\frac{1}{\pi} \int_{c_\beta^{\rm min}}^{1} {\rm d}c_\beta 
\frac{{\rm d}^2 \sigma_\nu^{\rm CC}}{{\rm d} c_\beta {\rm d}y} (x(c_\beta), y) 
\nonumber\\
\times  \int_{c_\nu^{\rm min}}^{c_\nu^{\rm max}} {\rm d}c_\nu 
\frac{1}{\sqrt{h(c_\mu, c_\beta, c_\nu)}}
 ~\frac{\rho^{NH}_\nu (E_\nu, c_\nu)}{\sigma_\nu^{\rm CC}(E_\nu)}. 
\label{eq:dens2}
\end{eqnarray}
Here 
\be
h \equiv (s_\mu s_\beta)^2 - (c_\nu - c_\mu c_\beta)^2 ,  
\nonumber
\label{def-h}
\ee
and $1/\sqrt{h}$, is essentially the Jacobian of transition 
to new variables according to (\ref{eq:anglrel}). 
The limits of integration $c_\nu^{\rm max}$ and $c_\nu^{\rm min}$ 
will be specified later.  
Notice that appearance of an additional 
factor 2 in the expression (\ref{final}) is due to twofold ambiguity at the transition 
from $\phi$ to $c_\nu$. 

Let us make another change of the integration variable:  
$c_\beta \rightarrow x$. Using the equality 
\be 
\frac{{\rm d}^2 \sigma_\nu^{\rm CC}}{{\rm d} c_\beta {\rm d} y}{\rm d}c_\beta = 
\frac{{\rm d}^2 \sigma_\nu^{\rm CC}}{{\rm d} x {\rm d} y} {\rm d} x,
\nonumber
\ee 
we obtain from (\ref{eq:dens2}) 
\begin{eqnarray}
n_\nu^{NH} (E_\nu, c_\mu, y) &=& \frac{1}{\pi} \int_{x_{\rm min}}^{x_{\rm max}} {\rm d}x 
\int_{c_\nu^{\rm min}}^{c_\nu^{\rm max}} {\rm d}c_\nu 
\frac{d^2 \sigma_\nu^{\rm CC}}{d x d y}  
\nonumber\\ 
\times&& \frac{1}{\sqrt{h(c_\mu, c_\beta, c_\nu)}} 
\frac{\rho^{NH}_\nu (E_\nu, c_\nu)}{\sigma^{\rm CC}_\nu (E_\nu)}.
\nonumber
\end{eqnarray}
Here $x_{{\rm min}, {\rm max}}$  
correspond  to the values  $c_\beta(E_\nu,x,y)=\pm 1$. 
In turn,  the limits of integration over $c_\nu$ correspond to 
$h = 0$, {\it i.e.} to the borders of the interval of the positivity condition: $h \geq 0$. 
Indeed, the expression for $h$ can be rewritten as 
\be
h = - [c_\nu - \cos(\theta_\mu - \beta)] \times [c_\nu - \cos(\theta_\mu + \beta)],  
\nonumber
\label{hnew}
\ee 
where $\beta = \beta(x, y, E_\nu)$ is determined in Eq.~(\ref{cosb}). 
Then the limits $c_\nu^{\rm max} = \cos(\theta_\mu - \beta)$ and 
$c_\nu^{\rm min} = \cos(\theta_\mu + \beta)$ follow immediately. 

Changing the order of integrations over $x$ and $c_\beta$,
we obtain
\begin{eqnarray}
n_{\nu}^{\rm{NH}}(E_\nu, c_\mu, y) &=&   
\frac{1}{\pi} \int_{|\theta_\mu-\theta_\nu|\le\beta_0}
{\rm d}c_\nu \rho_{\nu}^{\rm{NH}}(E_\nu, c_\nu)
\nonumber \\
 \times && 
 g_{\nu}(E_\nu,y,c_\nu,c_\mu)~, 
\label{eq:pho-f}
\end{eqnarray}
where 
\begin{eqnarray}
g_{\nu}(E_\nu,y,c_\nu,c_\mu) &\equiv& 
\frac{1}{\sigma_{\nu}^{\rm CC}(E_\nu)} 
\int_{x^{-}}^{x^{+}} {\rm d}x\,
\frac{{\rm{d}}^2\sigma_{\nu}^{\rm CC}(E_\nu,x,y)}{{\rm{d}}x{\rm{d}}y}
\nonumber \\
 \times &&
\frac{
1}{\sqrt{s_\beta^2 s_\mu^2 - (c_\nu - c_\beta c_\mu)^2}}.  
\label{eq:g-f}
\end{eqnarray}
Here $s_\beta$ and $c_\beta$ are functions of $E_\nu,\, x$ and $y$.
The function $g_{\nu}$ does  not depend on the mass hierarchy and
essentially play the role of the kinematic smearing function.  

Writing similar expressions for IH and $\bar{\nu}$,
we obtain the densities of the events for NH and IH:  
\begin{eqnarray}
n^{\rm{NH, IH}}(E_\nu, c_\mu, y) =   
n_{\nu}^{\rm{NH, IH}}(E_\nu, c_\mu, y) +  
n_{\bar\nu}^{\rm{NH, IH}}(E_\nu, c_\mu, y) 
\nonumber\\
= \frac{1}{\pi} \int_{|\theta_\mu-\theta_\nu|\le\beta_0} {\rm d}c_\nu
\left[
\rho_{\nu}^{\rm{NH, IH}}(E_\nu, c_\nu)
g_{\nu}(E_\nu,y,c_\nu,c_\mu)~~~~\right.
\nonumber\\
\left. + \rho_{\bar{\nu}}^{\rm{NH, IH}}(E_\nu, c_\nu)
g_{\bar\nu}(E_\nu,y,c_\nu,c_\mu)
\right]. ~~~~ 
\label{eq:nfinal}
\end{eqnarray}
Introducing $\rho^{\rm{NH}} \equiv \rho_{\nu}^{\rm{NH}} + 
\rho_{\bar{\nu}}^{\rm{NH}}$, we can rewrite the expression in 
(\ref{eq:nfinal}) as 
\begin{eqnarray}
n^{\rm{NH}}(E_\nu, c_\mu, y) = ~~~~~~~~~~~~~~~~~~~~~~~~~~~~~~~~~~~~~~~~~~~~~   
\nonumber\\
\frac{1}{\pi} \int_{|\theta_\mu-\theta_\nu|\le\beta_0} {\rm d}c_\nu
\rho^{\rm{NH}}(E_\nu, c_\nu)~ G (E_\nu,y,c_\nu,c_\mu), 
\nonumber
\label{eq:nfinal1}
\end{eqnarray}
where 
\be
G (E_\nu,y,c_\nu,c_\mu) \equiv 
g_{\nu} \frac{\rho_{\nu}^{\rm{NH}}}{\rho^{\rm{NH}}} 
+ g_{\bar\nu} \frac{\rho_{\bar{\nu}}^{\rm{NH}}}{\rho^{\rm{NH}}}.  
\nonumber
\ee
The function $G$  can be immediately compared 
with the Gaussian smearing function, 
which was used in 
\cite{ARS} embedding both kinematic and experimental 
resolution effects.

Let us find the limits of integration over $x$ 
in  Eq.~(\ref{eq:g-f}). According to Eq. (\ref{cosb1})
\be 
x(c_\beta)= \frac{2E_\nu(E_\mu-|{\bf p}_\mu| c_\beta)
- m_\mu^2}{2m_N E_\nu y},
\nonumber
\ee
which  imposes the lower and upper bounds to $x$:
\be 
x^\pm = x(\cos{(\theta_\nu\pm \theta_\mu)}). 
\nonumber
\ee
For a given $\theta_\nu$ and $\theta_\mu$, the minimal angle between 
the muon and the neutrino is $\beta=|\theta_\mu-\theta_\nu|$. 
The maximal angle $\beta$ is given  by $c_{\beta, {\rm max}}=\cos{(\theta_\nu + \theta_\mu)}$.

The integration over $c_\nu$  in Eq.~(\ref{eq:pho-f}) runs 
from $\cos(\theta_\mu + \beta_0)$ to $\cos(\theta_\mu - \beta_0)$,  
where $c_{\beta_0}=c_\beta(E_\nu, x=1,y)$ and  
$\theta_\mu \pm\beta_0$ is restricted by the interval $0 - \pi$.

\subsection{Kinematical smearing function}

\begin{figure}[h]
\centering
 \includegraphics*[width=0.35\textwidth]{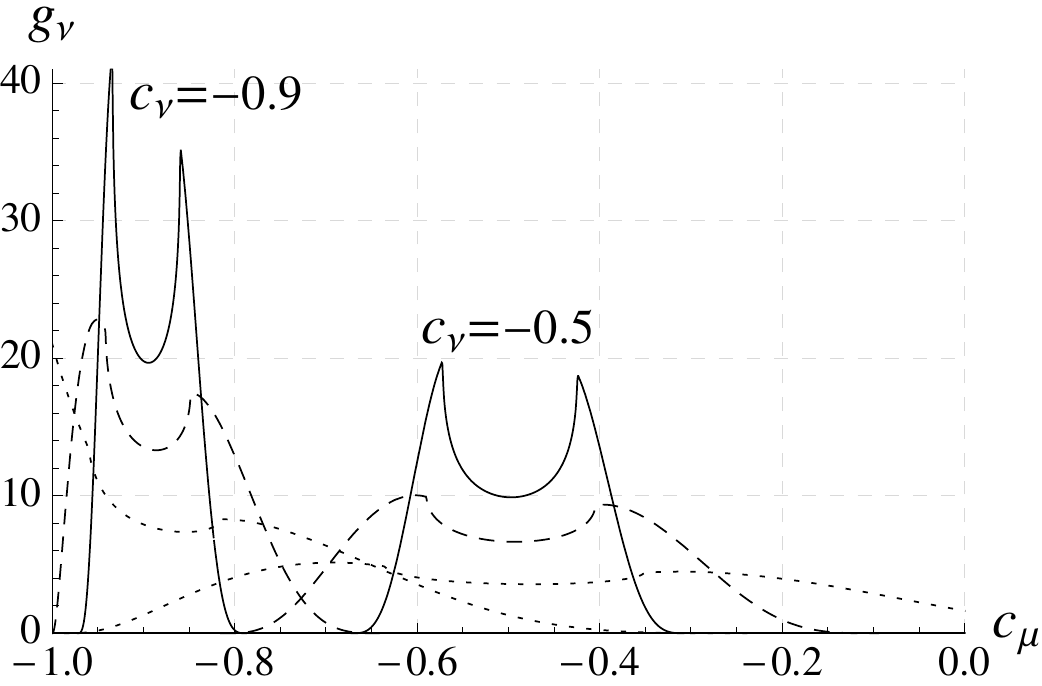}
 \includegraphics*[width=0.35\textwidth]{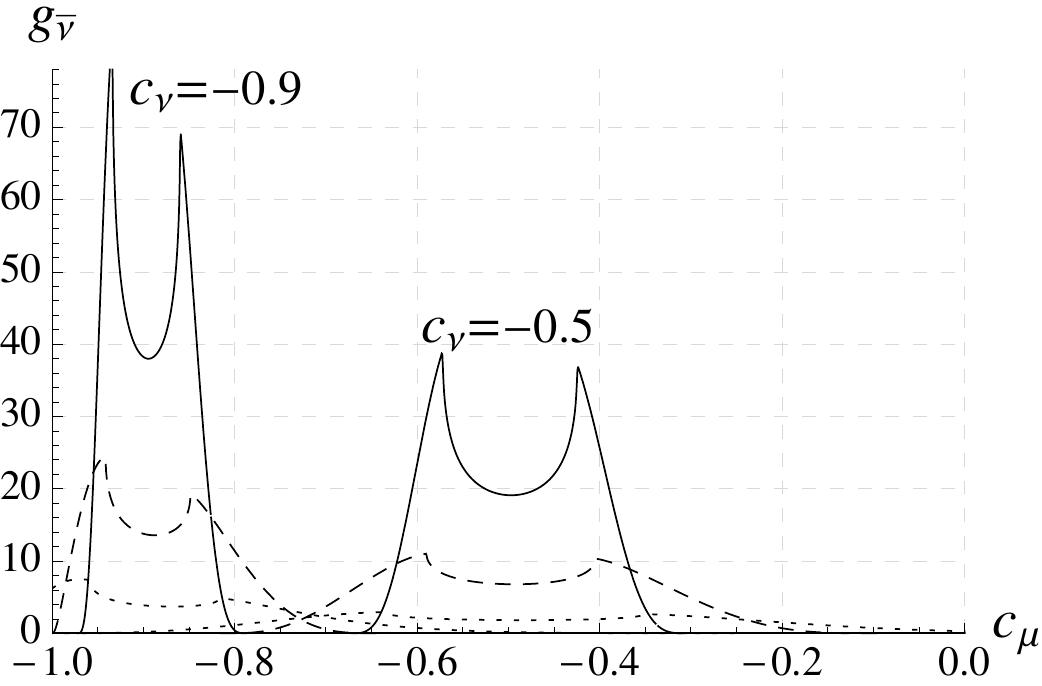}
\caption{The angular smearing functions for neutrinos (upper panel)
and antineutrinos (bottom panel) 
for  $E_\nu=10\,\rm GeV$ and  different values of $c_{\nu}$ (numbers at the curves). 
Solid, dashed and dotted curves are respectively for $y=0.2$, $y=0.5$ and $y=0.8$.
\label{gnuanu}}
\end{figure}

According to Eq.~(\ref{eq:pho-f}), the functions $g_{\nu,{\bar\nu}}(E_\nu,y,c_\nu,c_\mu)$ in 
(\ref{eq:g-f}) can be considered as  the smearing functions 
over the neutrino angle. 
Fig.~\ref{gnuanu} shows dependence of $g_\nu$ and $g_{\bar\nu}$ 
on  $c_\mu$ for several values of $c_\nu$ and $y$.

The smearing functions differ from the Gaussian function 
assumed in \cite{ARS}. 
They have two peaks with a local minimum in between; 
there are no exponential tails;  
the central parts are at $c_\mu \approx c_\nu$. 
The asymmetry of the peaks becomes stronger with $c_\mu$ approaching $\pm 1$;  
the width of the functions increases with $y$.  
The functions $g$ are similar for neutrinos and antineutrinos.  
As expected, for antineutrinos the overall normalization decreases with 
increase of $y$, whereas for neutrinos normalization changes weakly.  

The properties of $g_{\nu,\bar\nu}$ can be readily understood from  the expression 
for $h$. Indeed, $g_{\nu,{\bar\nu}}$ have inverted (and also smoothed) 
shapes with respect to that of $\sqrt{h}$. In particular, peaks of  $g_{\nu,{\bar\nu}}$  
correspond to zeros of $h$, the minima of $g_{\nu,{\bar\nu}}$ correspond to the maxima of $h$, {\it etc.}. 

The function $h$ can be rewritten as 
\be
h = s_\beta^2 - (c_\nu^2 + c_\mu^2) + 2 c_\nu c_\mu c_\beta, 
\nonumber
\ee
which is obviously symmetric with respect to the interchange   
\be 
c_\nu \leftrightarrow c_\mu . 
\nonumber
\ee
As a consequence,  $g_{\nu,{\bar\nu}}$ also obeys this symmetry. 

Introducing
\be
r \equiv 2 \zeta x y = 2 \sin^2\frac{\beta}{2},
\nonumber
\ee
we can present $h$ as 
\be
h = 2r (1 - c_\nu c_\mu) - r^2 - (c_\nu - c_\mu)^2. 
\nonumber
\ee
Then defining the difference 
$\Delta \equiv c_\nu - c_\mu$, 
we have
\be
h (\Delta) = s_\mu^2(2r - r^2) - (\Delta + r c_\mu)^2. 
\nonumber
\ee
So, $h$,  as function of $\Delta$, is an inverted parabola with  its
maximum shifted to $\Delta = - r c_\mu$.   
In agreement with Fig.~\ref{gnuanu}, $h (\Delta)$ is not symmetric 
with respect to $\Delta = 0$ or $c_\nu = c_\mu$,  
and  the minimum of $g_{\nu,{\bar\nu}}$ is shifted with respect to 
$c_\mu = c_\nu$.  
This also leads to difference  of heights of peaks. 

Zeroes of $h$ are at 
\be
\Delta = - r c_\mu \pm \sqrt{s_\mu^2(2r - r^2)}. 
\nonumber
\ee
According to (\ref{eq:c-limit}) in terms of angles the zeros of $h$ are given by 
$ \cos \theta_\nu = \cos(\theta_\mu \pm \beta)$.
So, the width of the smearing function increases with $\beta$. 
In turn, according to (\ref{eq:kinsmear}),
$\beta \propto \sqrt{y}$, and consequently, the width 
increases with $y$, as we mentioned before.  

It is easy to understand the appearance of 
peaks in $g_{\nu,{\bar\nu}}(c_\nu)$ at the borders of allowed interval  
using the following graphical representation. 
The neutrino vector is on the surface of the  
cone with angle $\beta$ and axis along the muon momentum. 
With change of $\phi$, the neutrino vector moves on the surface 
of the cone.  
The maximum and minimum of $c_\nu$ given by (\ref{eq:c-limit}) 
correspond to the neutrino vector situated   
in the plane formed by the muon vector and the axis $z$ and the neutrino vector is 
moving perpendicularly to this plane. 
Therefore around these positions the $z$-projection 
of the neutrino vector does not change appreciably, 
and so the integration over $\phi$ leads to bigger contribution. 


\subsection{Oscillograms for different values of $y$}

We will use the general formulas obtained 
in the previous sections to 
compute the oscillograms and asymmetry distributions for different values of $y$. 
The functions $\rho_{\nu, \bar\nu}^{\rm{NH, IH}}(E_\nu, c_\nu)$ 
are taken from~\cite{ARS}.

\begin{figure*}[ht!]
\centering
 \includegraphics*[width=0.4\textwidth]{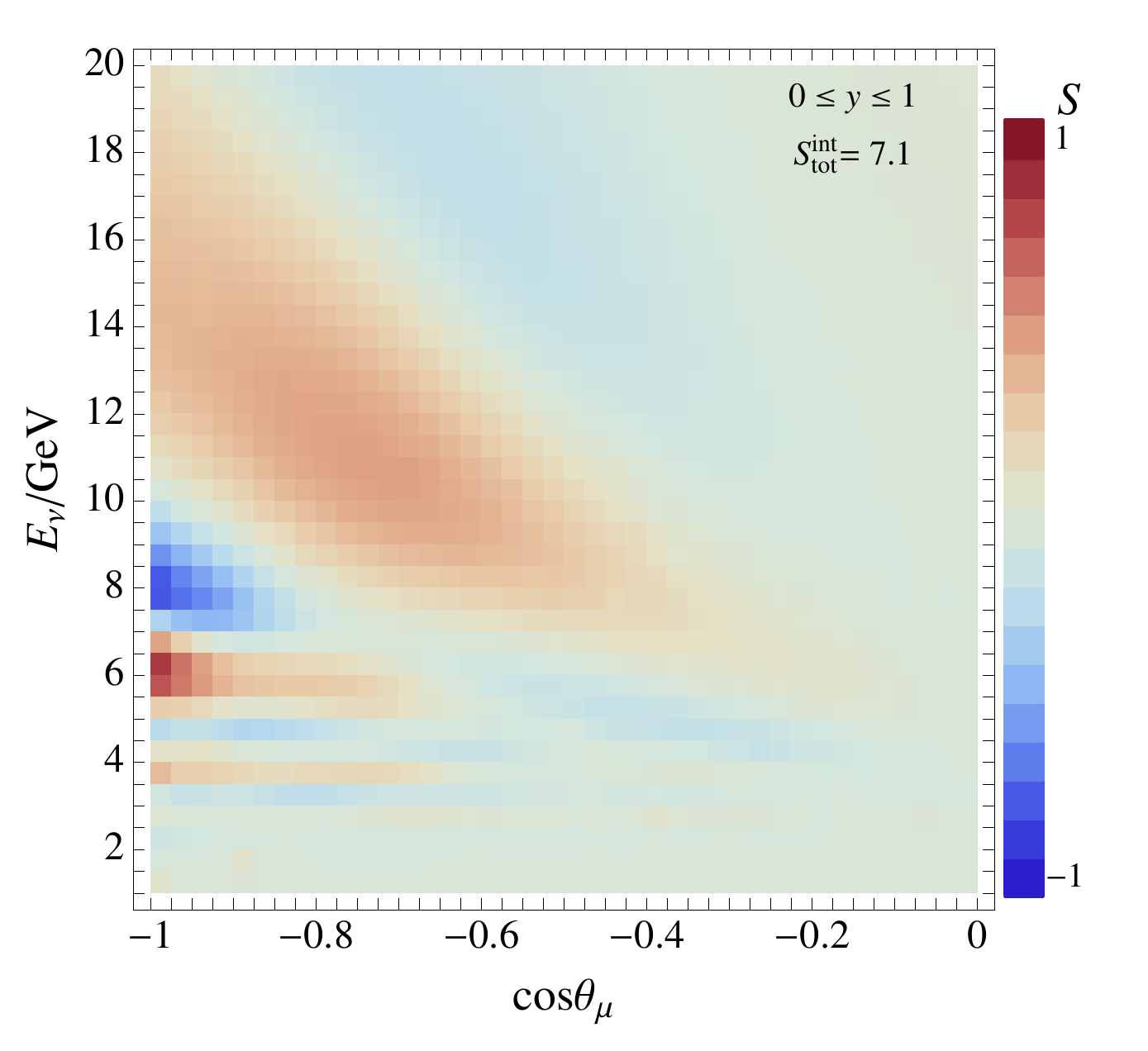}~~~~~~
 \includegraphics*[width=0.4\textwidth,type=pdf,ext=.pdf,read=.pdf]{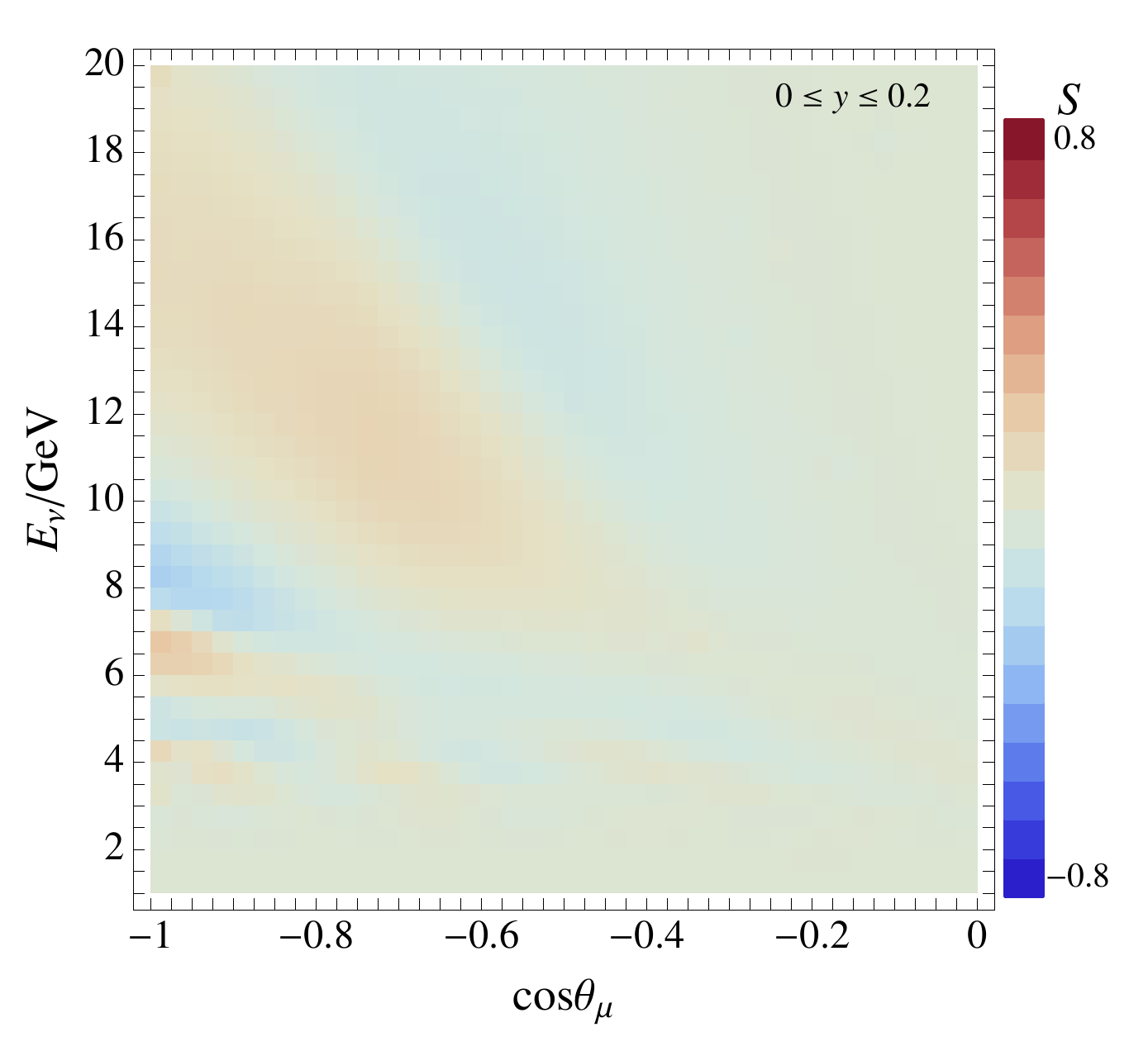}
 \includegraphics*[width=0.4\textwidth,type=pdf,ext=.pdf,read=.pdf]{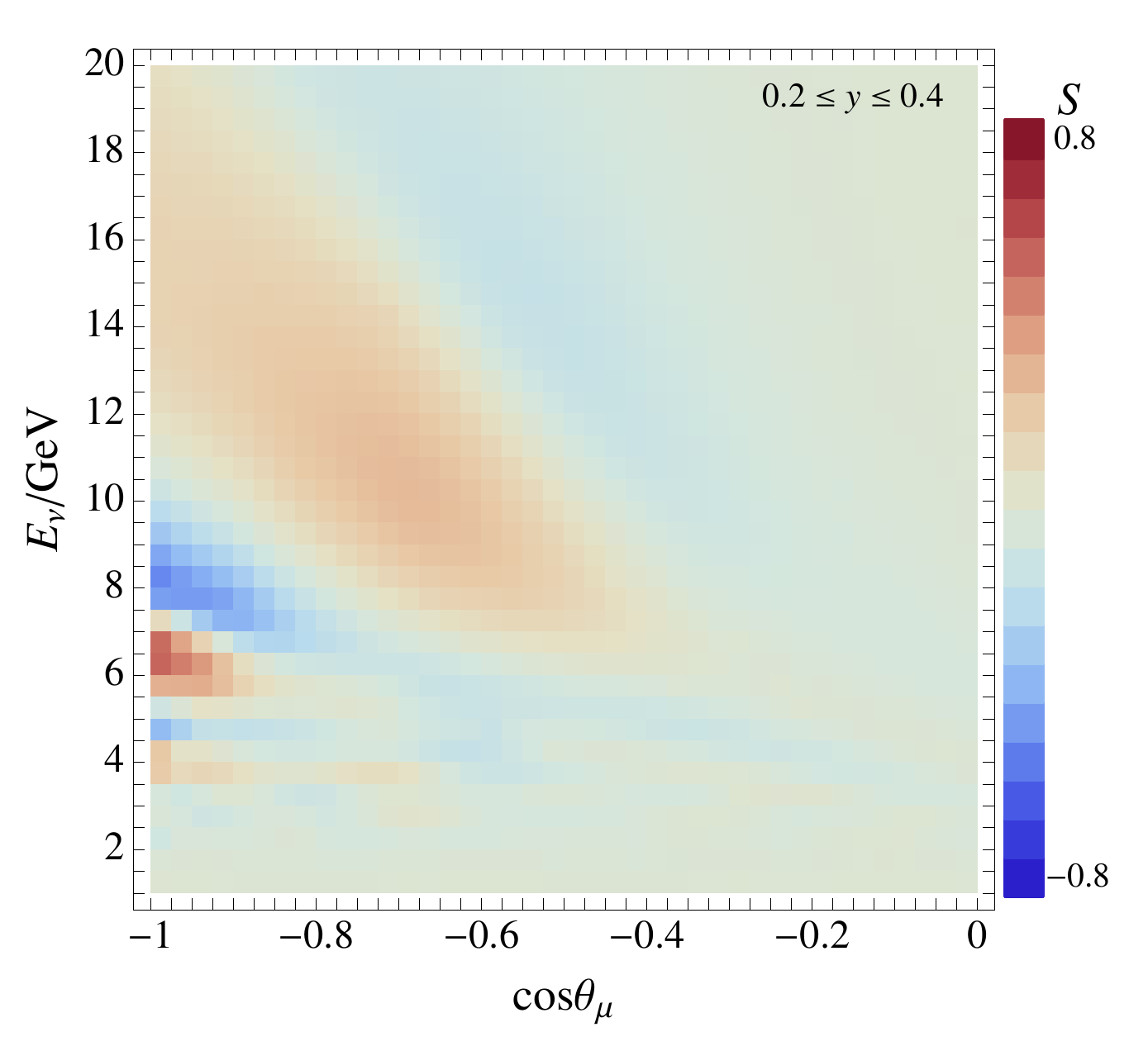}~~~~~~
 \includegraphics*[width=0.4\textwidth,type=pdf,ext=.pdf,read=.pdf]{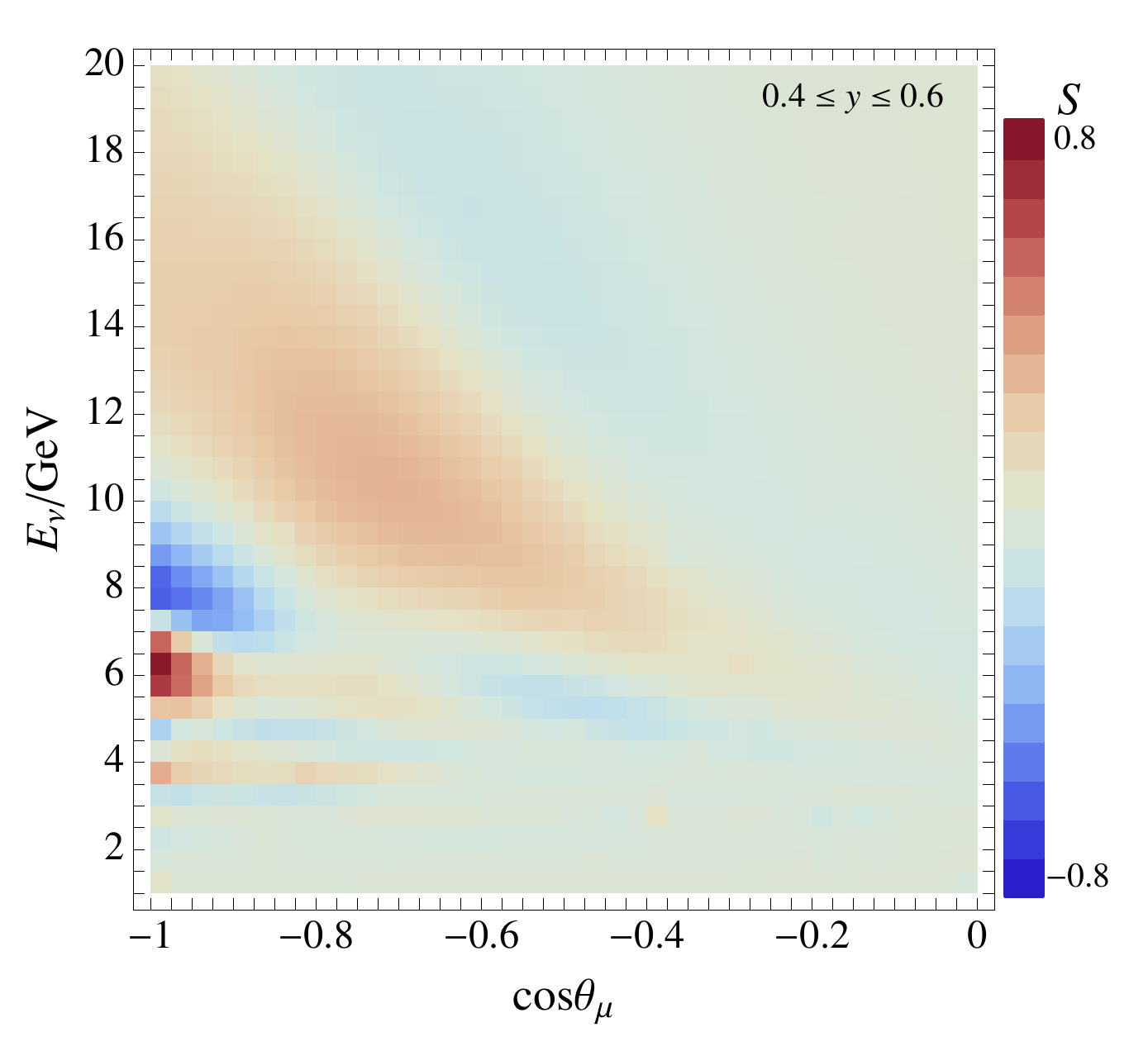}
 \includegraphics*[width=0.4\textwidth,type=pdf,ext=.pdf,read=.pdf]{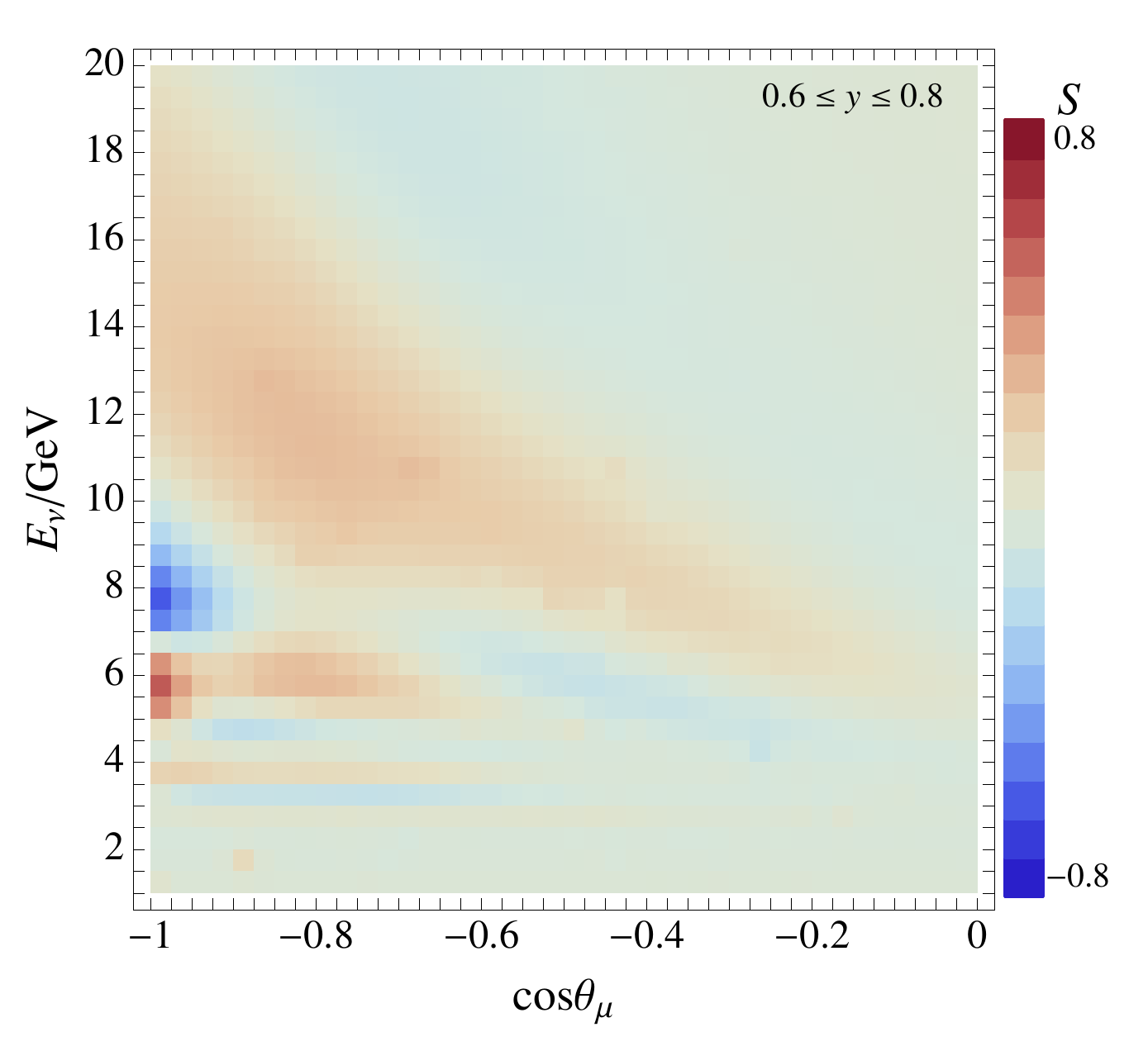}~~~~~~
 \includegraphics*[width=0.4\textwidth,type=pdf,ext=.pdf,read=.pdf]{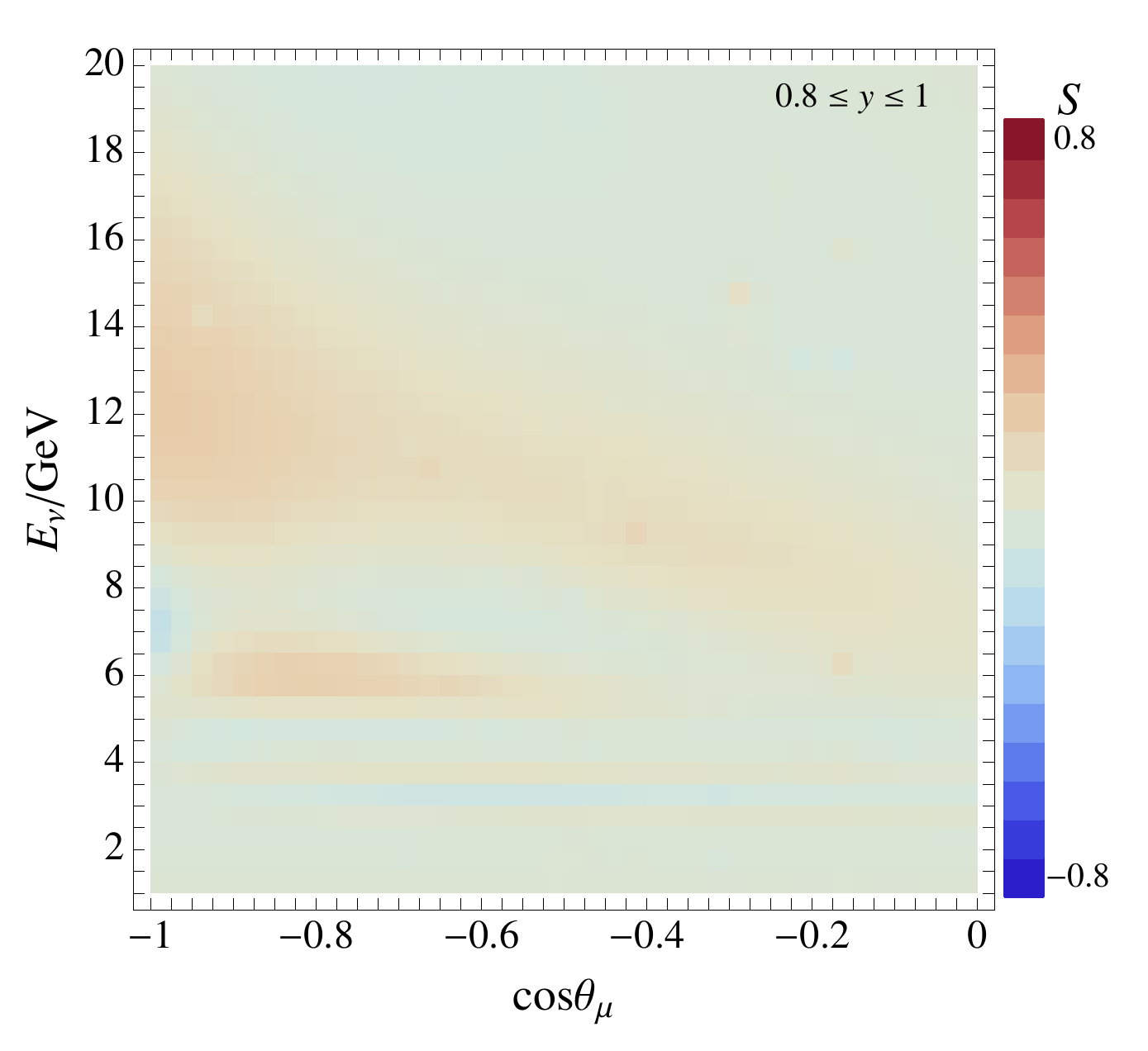}
\caption{The hierarchy asymmetry distributions 
after the kinematical smearing for various inelasticity ranges and for 1 year of exposure.
\label{fig:raw-muOsc-yrange}}
\end{figure*}

Fig.~\ref{fig:raw-muOsc-yrange} shows the 
$E_\nu - \cos \theta_\mu$  binned distribution of the 
hierarchy asymmetry with the inelasticity ($y-$dependence) 
and kinematical smearing taken into account.  
Different panels in this figure correspond to different 
$y-$intervals  $\{y_{\rm min}, y_{\rm max}\}$.  
The asymmetry in these intervals has been computed  
in the following way.
We first used very small $y-$bins $\Delta y \ll (y_{\rm max} - y_{\rm min})$. 
We computed the asymmetry  in each of these small bins 
$S_k = S(y_k, E_\nu, \cos \theta_\mu)$ and  then  
the total asymmetry in the interval  $\{y_{\rm min}, y_{\rm max}\}$ 
as $S_{\rm tot} = \sqrt{\sum S_k^2}$ (the sum runs over all 
small $y-$bins in the interval $\{y_{\rm min}, y_{\rm max}\}$), and the sign  
is the same as for dominant contribution.  
In practice the  
summation over small bins was substituted by integration: 
\be
S(y_{\rm max}, y_{\rm min}, E_\nu, \cos \theta_\mu) = 
\left[\int_{y_{\rm min}}^{y_{\rm max}} {\rm d}y 
\frac{(n^{\rm IH} - n^{\rm NH})^2}{n^{\rm NH}}
\right]^{1/2}. 
\nonumber
\ee
Here $n^{\rm NH, IH}$ is the number of events in the bin 
$\Delta E_\nu \,\Delta \cos \theta_\mu $ given in (\ref{eq:nfinal}).

The first panel of Fig.~\ref{fig:raw-muOsc-yrange} 
corresponds to  $\{y_{\rm min}, y_{\rm max}\} = \{0, 1\}$,
the others - to various intervals with $y_{\rm max} - y_{\rm min} = 0.2$. 
The first panel is the sum of contributions described in other panels. 
As we see the biggest contribution comes from 
the intermediate region $y\in\{0.3,0.7\}$.
 Indeed, at small $y$ the hierarchy asymmetry
is suppressed due to strong cancellation
of the nearly equal contributions from neutrinos and
antineutrinos (recall that at $y \sim 0$ the
$\nu$ and $\bar{\nu}$ cross-sections become equal).
At large $y$, the asymmetry is suppressed due to
strong smearing over the angle between
muon and neutrino.  
With the increase of $y$, the region of strong asymmetry
first shifts smaller $E_\nu$ and larger $\cos \theta_\mu$, and
then move to larger $E_\nu$ and  $\cos \theta_\mu = - 1$.
The region expands in horizontal ($\cos \theta_\mu$) direction
for small $y$.

The total significance (given by integration over the 
first panel with $0\le y\le1$) equals   
\be 
|S_{\rm tot}|= 
\left[
\int{\rm d} c_\mu \int{\rm d}E_\nu \int_0^1 {\rm d}y\, 
\frac{(n^{\rm IH}-n^{\rm NH})^2}{n^{\rm NH}}
\right]^{1/2}. 
\label{eq:Sy} 
\ee
For exposure $T = 1$ year, this leads to  $|S_{\rm tot}| = 8.43$. 

If the $y-$dependence is not used, the densities of events should 
be integrated over $y$ before computing $S$. This gives  
\be
|S^{\rm int}_{\rm tot}| = 
\left[
\int{\rm d}c_\mu\int{\rm d}E_\nu 
\frac{(\int_0^1 {\rm d}y\, 
(n^{\rm IH}-n^{\rm NH}))^2}{{\int_0^1 {\rm d}y \,n^{\rm NH}}}
\right]^{1/2}. 
\label{eq:Syint}
\ee 
For 1 year exposure we obtain from (\ref{eq:Syint}) 
$|S^{\rm int}_{\rm tot}|  = 7.11$,  which is about $15\%$ 
smaller than  in the  case when $y$-distribution is used according to Eq. (\ref{eq:Sy}). 

The following comments are in order:

(i) The kinematical smearing strongly 
reduces the total significance:  
for the ideal reconstruction of the neutrino energy 
and direction we would obtain  
$|S^{\nu+\bar\nu}_{\rm tot}|= 23.7$ during 1 year 
even without $y-$information. 
This number can be considered as maximal achievable significance.  
It should be compared with $|S^{\rm int}_{\rm tot}| = 7.11$ 
obtained from~(\ref{eq:Syint}).   
Note that $|S_{\rm tot}| = 8.43$ can be obtained with an ideal detector 
having  perfect resolutions (see Sec.~\ref{sct:5}).

(ii) The increase of significance by about 15\% 
with  $y$-distribution 
is better than the one predicted from our qualitative 
discussion in Sec. \ref{sct:3}, using $\gamma$ derived 
from the method of moments.  This is probably related 
to the fact that the characteristics of the $y-$distribution 
are more fully exploited. 

Note also that in our treatment 
the cross-sections have been restricted to the DIS 
approximation, thus exhibit cutoffs at small $y$, 
a region of good angular resolution. Therefore a more 
complete description of the cross sections will recover 
the events in the small $y-$region, further enhancing the significance.

\section{Significance of determination of mass hierarchy
with inelasticity}\label{sct:5}

In the previous section we have taken into account 
the kinematical smearing - the integration over the angle 
between the neutrino and muon, $\beta$. 
Besides this, one should perform  
the experimental smearing over the observables:  
the energy of muon and  cascade as well as the direction of muon   
due to finite experimental  energy and angular 
resolutions.  

\subsection{Experimental resolution functions}

We present here the significance of the identification 
of the  mass hierarchy, considering various scenarios for the widths   
$\sigma_{\mu, h}(E_\nu, y)$ and $\sigma_\psi(E_\nu,y)$ 
of the energy and angular resolution functions. 
We use the notation $\tilde{x}$  for the reconstructed value  
of the observable $x$.

\subsubsection{Energy resolution}

We assume the Gaussian energy resolution functions of the cascade and muon  
with widths $\sigma_{h}$ and $\sigma_{\mu}$ correspondingly. 
Then the neutrino energy resolution is itself the Gaussian function 
(sum of two normal distributions):  
\be
g_{E_\nu}(\tilde{E}_{\nu}, E_\nu)=  
\frac{1}{\sqrt{2\pi}\sigma_{E_\nu}}
\exp{\left[-\frac{1}{2}\frac{(\tilde{E}_{\nu}-E_\nu)^2}{\sigma_\nu^2}\right]}  
\label{eq:genu}
\ee
with width 
\[
\sigma_\nu (y)=\sqrt{\sigma_{\mu}^2 + \sigma_{h}^2}~,
\]
which depends on $y$. 

We consider two cases for the energy resolution of muons and cascades:

(i) $\sigma_{\mu, h} = b E_{\mu, h}$, which gives 
$\sigma_{\mu} = b E_\mu 
\approx b (1 - y)E_\nu$  and 
$\sigma_h = b E_h \approx b  y  E_\nu$,  so that  
\be 
\sigma_\nu(y)= b E_\nu \sqrt{1-2y+2y^2}~; 
\nonumber
\ee

(ii) $\sigma_{\mu, h} = \sqrt{b E_{\mu, h}}$,   then 
\be
\sigma_\nu =\sqrt{b E_\nu}~,
\nonumber
\ee
which has the same form as $\sigma_{\mu, h}$.
We make here a reasonable simplification 
that $b$ is the same for cascade and muon.  
In fact, this is true only 
if we assume that most of the energy of the cascade is visible like for a muon. 
More likely, it has a bit smaller Cherenkov photon yield per GeV 
and subject to greater event by event fluctuations.

In the case $(i)$, we use $b = 0.3$ in order to compare with the results from~\cite{ARS}.
In the case $(ii)$, we take  $b$ in the range 
$0.35\le  b \le 0.7$, which is derived from an 
estimated number of detected photons 
$n_{\rm hit}/{\rm GeV}\approx1 - 3$~\cite{Demiroers:2011am}.
Eq.~(26) in~\cite{Demiroers:2011am} predicts 
$n_{\rm hit}/{\rm GeV}\approx1.5$ for a mean distance 
of about 10~m between the Cherenkov light emitter 
and an optical module of IceCube type. The range 
is extended in the mentioned limits, because the 
precise topology and technology (for instance, the 
photo-detection efficiency and area of the optical 
modules) of a dense array are not yet precisely known.

We then obtain the energy resolution given by the 
statistical uncertainty of the number of expected hits:
\begin{eqnarray}
\sigma_{\mu, h} &=& \frac{\delta\left(E_{\mu, h} \, n_{\rm hit}/{\rm GeV}\right)}{n_{\rm hit}/{\rm GeV}}  \\
&\approx& \sqrt{\{0.35, 1\}E_{\mu, h}}\approx \{0.6,1\}\surd{E_{\mu, h}}. 
\nonumber
\end{eqnarray}

\subsubsection{Inelasticity resolution function}

The inelasticity distribution $g_y(E_\nu, y)$ can be derived  straightly 
from  $E_{\mu, h}$ distributions, 
$g_{\mu, h}(\tilde{E}_{\mu, h},E_{\mu, h})$,  described above. 
We show in the Appendix~\ref{appendix:B}
that it is nearly Gaussian in most cases 
of interests. It deviates from Gaussian, showing  
enhanced tails, when $E_\mu$ and $E_h$ are both small. 
In our computations we use the Gaussian function with width  
\be
\sigma_y  =  \frac{1}{\surd2}
\left(
\frac{E_h + \sigma_h}{E_h + \sigma_h + E_\mu - 
\sigma_\mu} - y
\right).
\label{eq:resy}
\ee

Notice that we could perform smearing using immediately $E_\mu$ 
and $E_h$ without introducing $y$, and if needed,  
introduce $\tilde{y}$ after smearing.

\subsubsection{Angular resolution}

The angle $\psi$ between the true and the reconstructed muon
directions is described by the normalized distribution
\be
g_\psi =\frac{2\psi}{\sigma_\psi^2}
\exp{\left(-\frac{\psi^2}{\sigma_\psi^2}\right)},
\label{eq:gpsi}
\ee
which is derived from the 2D Gaussian distribution.
The interval $\psi\le\sigma_\psi$ encloses 63\% , and  
$g_\psi$ peaks at $\sigma_\psi/\surd{2}$. The width 
$\sigma_\psi$ is generically a function of $E_\mu$, 
which has the form~\cite{Ribordy:2006qd}
\[
\sigma_\psi = \psi_0 \sqrt{\frac{m_N}{E_\mu}}. 
\]
Here $\psi_0$ depends on the detector medium  (ice, water) 
and its topology.
The IceCube detector is sparsely instrumenting 
a medium of relatively short scattering length.  
Therefore a large number of photons 
will not travel on a straight path between the 
Cherenkov light emission point and the detection 
location. On the contrary in ANTARES (and similarly 
in ORCA), the photons are detected un-delayed.
This is the main reason why IceCube has worser 
angular resolution than the ANTARES detector. 
Therefore, we consider $\psi_0$ values in a range 
reflecting common angular resolutions achieved 
in water and by a sparse array in ice. Note, however, 
that one reasonably expects a substantially improved 
angular resolution in ice with a smaller and denser array (PINGU), 
{\it i.e.} a global reduction of scale: in this case  
the short scattering length will be of relative importance 
and many Cherenkov photons will reach  the optical modules closest from their 
emission point rather undelayed.

The angular resolution of 
an event with 60 hits is about $5^\circ$ in IceCube 
and better than $2^\circ$ ($n_{\rm hit}/{\rm GeV} \approx1.5$) in ANTARES. 
Therefore $15^\circ \lesssim \psi_0 \lesssim 30^\circ$.
The range $1\le n_{\rm hit}/{\rm GeV}\le3$ leads 
to $8.5^\circ \lesssim \psi_0 \lesssim 40^\circ$.

The smearing function 
for the zenith angle of muon,  $g(c_\mu, \tilde{c}_\mu)$,  
can be computed using the smearing function for 
$\psi$ (\ref{eq:gpsi}) as  
\be
g(c_\mu, \tilde{c}_\mu) = \frac{1}{\pi} \int _{|\theta_\mu -\tilde{\theta}_\mu|}^\pi \,
\frac{g_{\psi}(\psi)\,
{\rm d}\psi}{\sqrt{\tilde{s}_\mu^2 s_\psi^2 - (\tilde{c}_\mu c_\psi-c_\mu)^2}}. 
\label{eq:gcmu}
\ee
The denominator here appears similarly  
to that at  the variable change 
$\phi\rightarrow c_\nu$ performed in Eq.~(\ref{final}). 
The function $g (c_\mu, \tilde{c}_\mu)$ 
is normalized, which follows from  normalization of  
$g_{\psi}$.

\begin{figure}[h]
\centering
 \includegraphics*[width=0.35\textwidth]{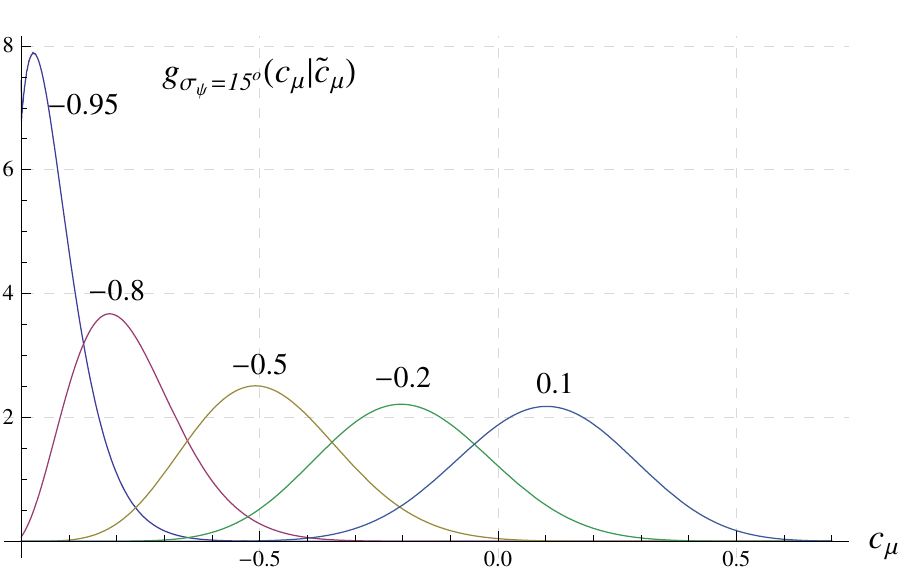}
\caption{Angular smearing function in 
$\cos{\theta_\mu}$  
for $\sigma_\psi=15^\circ$ and several values of $\tilde{c}_\mu$.
\label{fig:thetaDistr}}
\end{figure}

Fig.~\ref{fig:thetaDistr} shows the angular smearing
function $g(c_\mu, \tilde{c}_\mu)$ 
for $\psi_0 =15^\circ$ and several values of $\tilde{c}_\mu$. 
Notice that $g_{\psi}(c_\mu, \tilde{c}_\mu) \tilde{s}_\mu$
is not symmetric and it increasingly deviates from the normal distribution,
when approaching $\theta_\mu = 180^{\circ}$.

\subsection{Distributions with experimental smearing}

We calculate the distribution of events smeared over 
the experimental resolution functions.  
We convert the 3-D distributions in the parameter space  
$E_\nu - y - c_\mu$ into observed 
parameters space $\tilde{\theta}_\mu, \tilde{E}_\nu, \tilde{y}$
convoluting $n$ with the resolution distributions for $E_\nu$, $y$ and $\psi$: 
\be
\hat{n}^{\text{IH},\text{NH}}_{\nu,\bar{\nu }}(\tilde{\theta}_\mu, \tilde{E}_\nu, \tilde{y})
= n^{\text{IH},\text{NH}}_{\nu, \bar{\nu}}(\theta_\mu, E_\nu, y) *(g_{\psi}~g_y~g_{\nu}).
\nonumber
\ee
The convolution is performed sequentially in order indicated in the 
last brackets. The smearing functions are taken according to 
Eqs. (\ref{eq:genu}), (\ref{eq:gcmu}) and (\ref{eq:resy}) 
(the width of Gaussian function for $g_y$).  
Values of  $n^{\text{IH},\text{NH}}_{\nu, \bar{\nu}}$ 
outside region $1<E_\nu/{\rm GeV}<20$ 
and for $\theta_\mu<90^\circ$ are set to the values taken at the boundaries.

Integrating over the bins, we obtain the binned oscillogram: 
\be
N_{ijk}(\tilde{c}_{\mu i}, \tilde{E}_{\nu j},\tilde{y}_k)= 
\int_{\rm{bin(ijk)}} {\rm d}\tilde{c}_\mu {\rm d}\tilde{E}_{\nu} {\rm d}\tilde{y}
\,\hat{n}^{\text{IH},\text{NH}}_{\nu,\bar{\nu }}(\tilde{\theta}_\mu, \tilde{E}_\nu, \tilde{y}).
\nonumber
\ee

\begin{figure*}[ht!]
\centering
\includegraphics*[width=0.4\textwidth,type=pdf,ext=.pdf,read=.pdf]{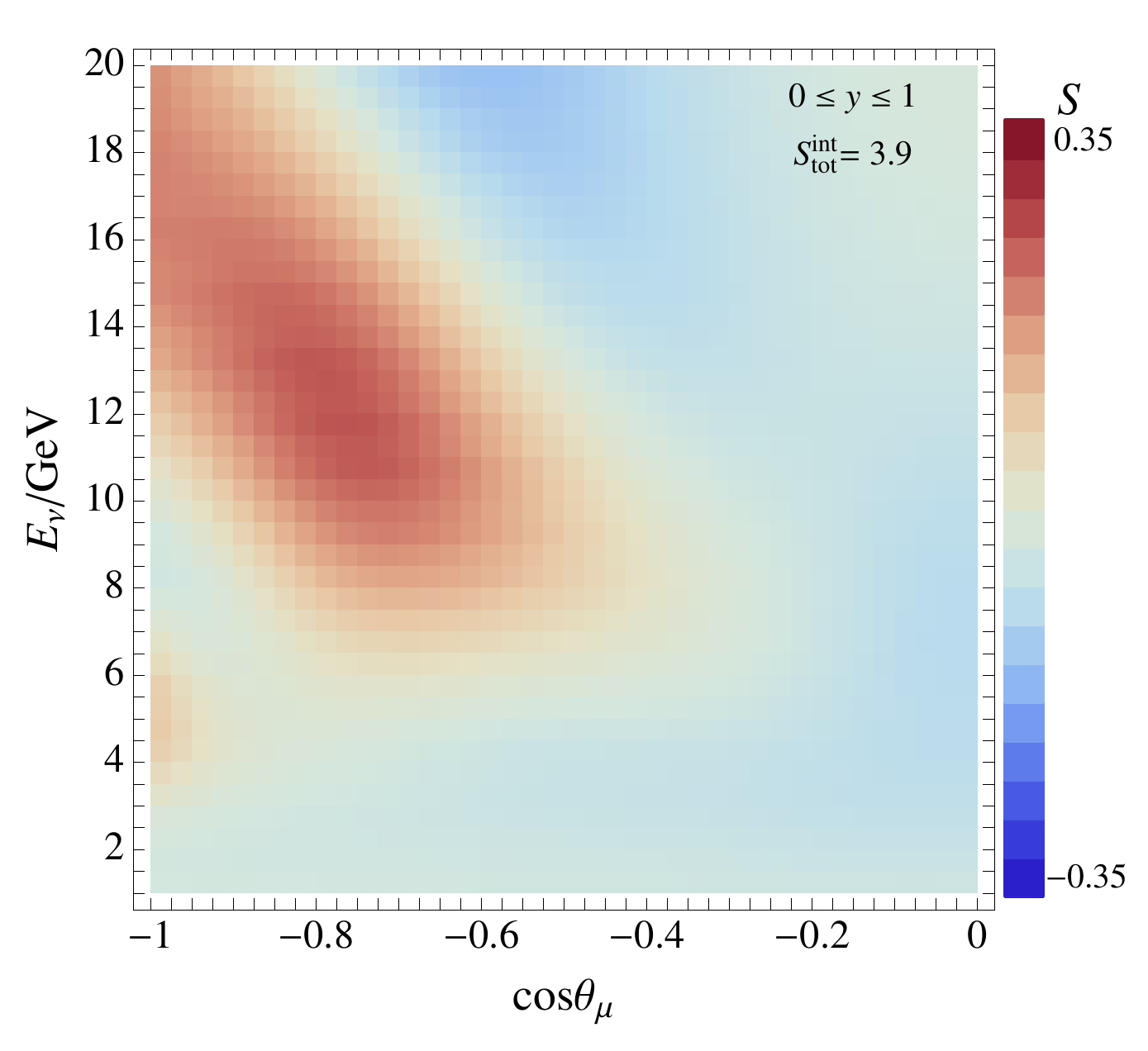}~~~~~~
 \includegraphics*[width=0.4\textwidth,type=pdf,ext=.pdf,read=.pdf]{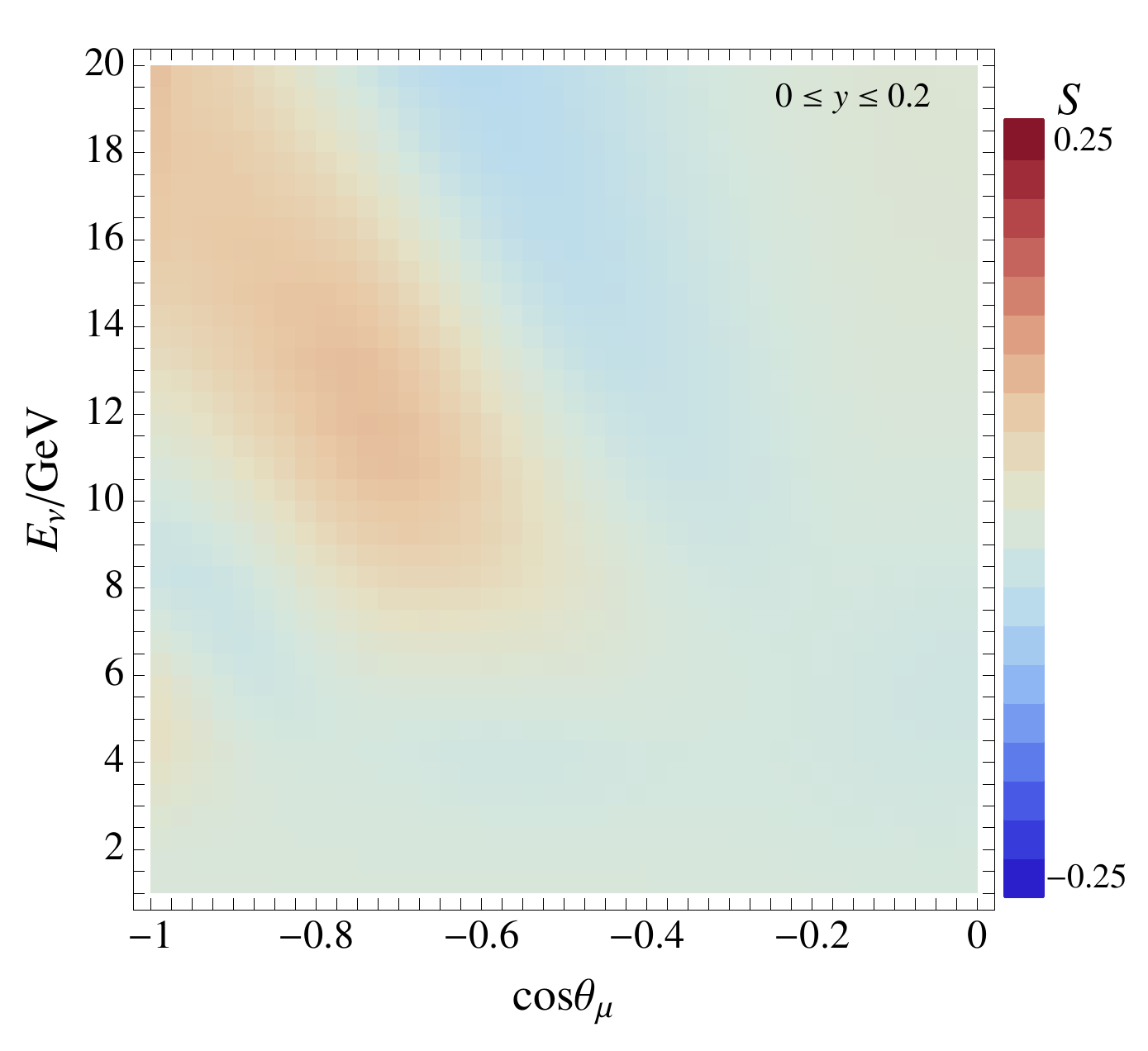}
 \includegraphics*[width=0.4\textwidth,type=pdf,ext=.pdf,read=.pdf]{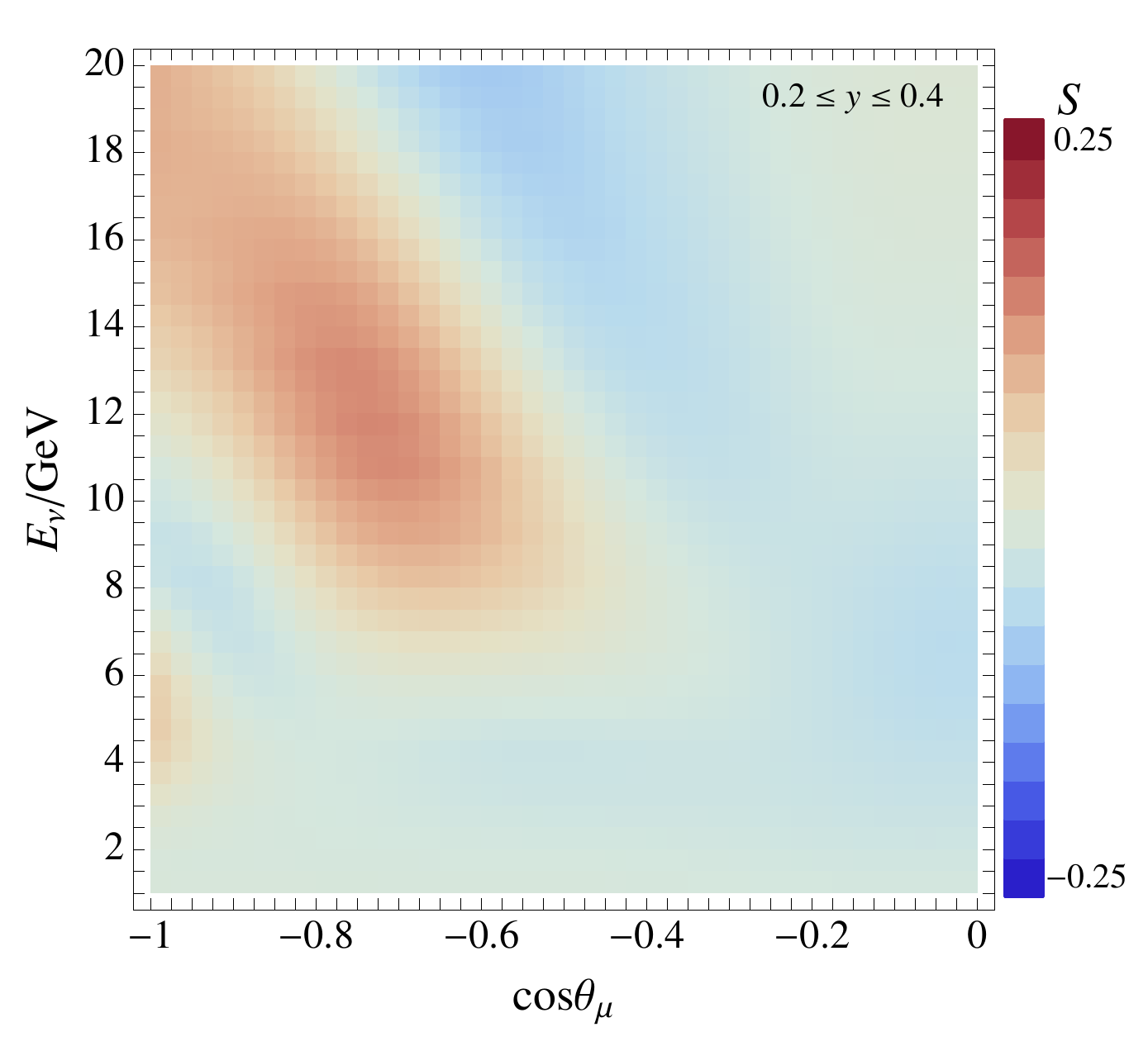}~~~~~~
 \includegraphics*[width=0.4\textwidth,type=pdf,ext=.pdf,read=.pdf]{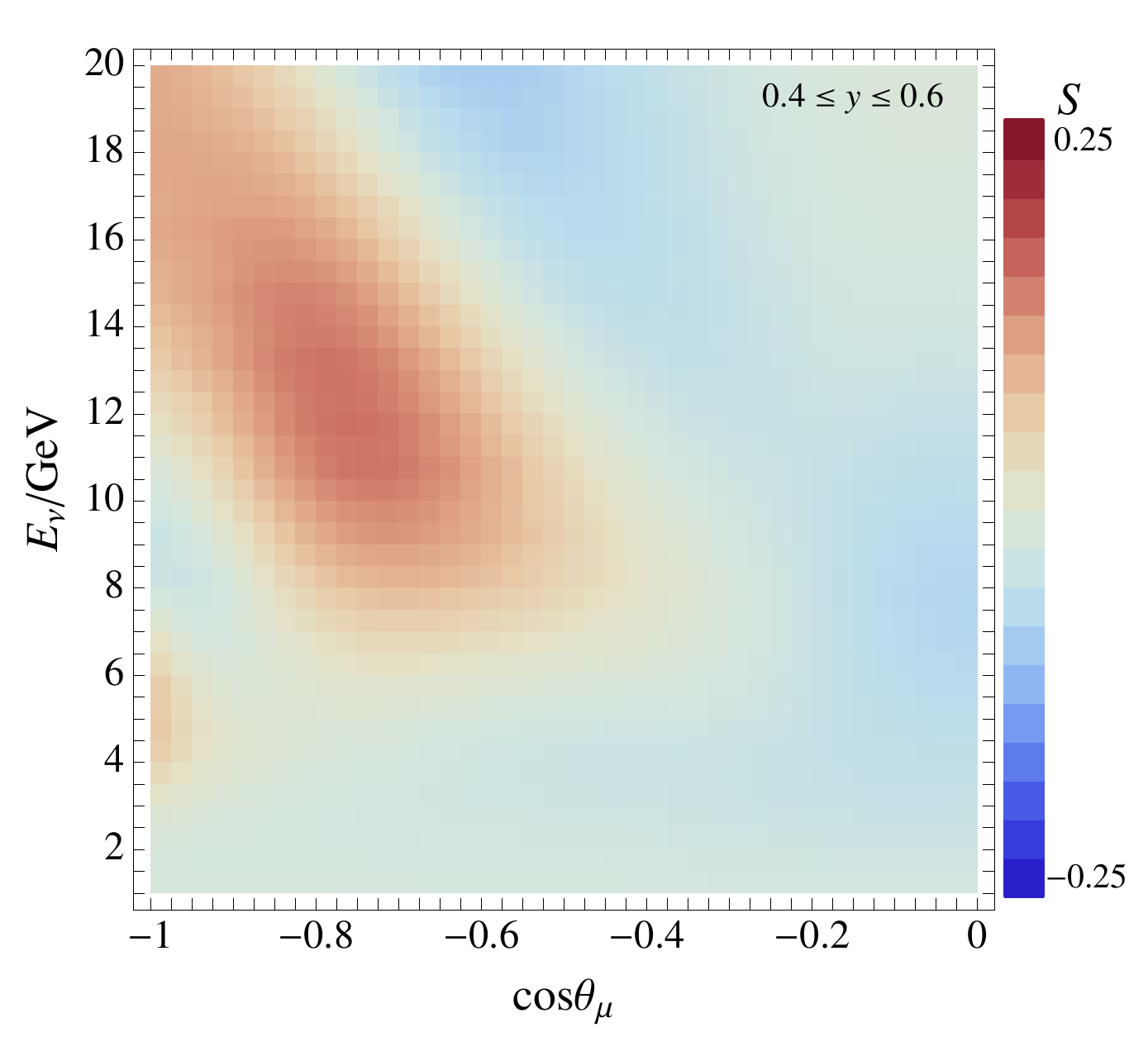}
 \includegraphics*[width=0.4\textwidth,type=pdf,ext=.pdf,read=.pdf]{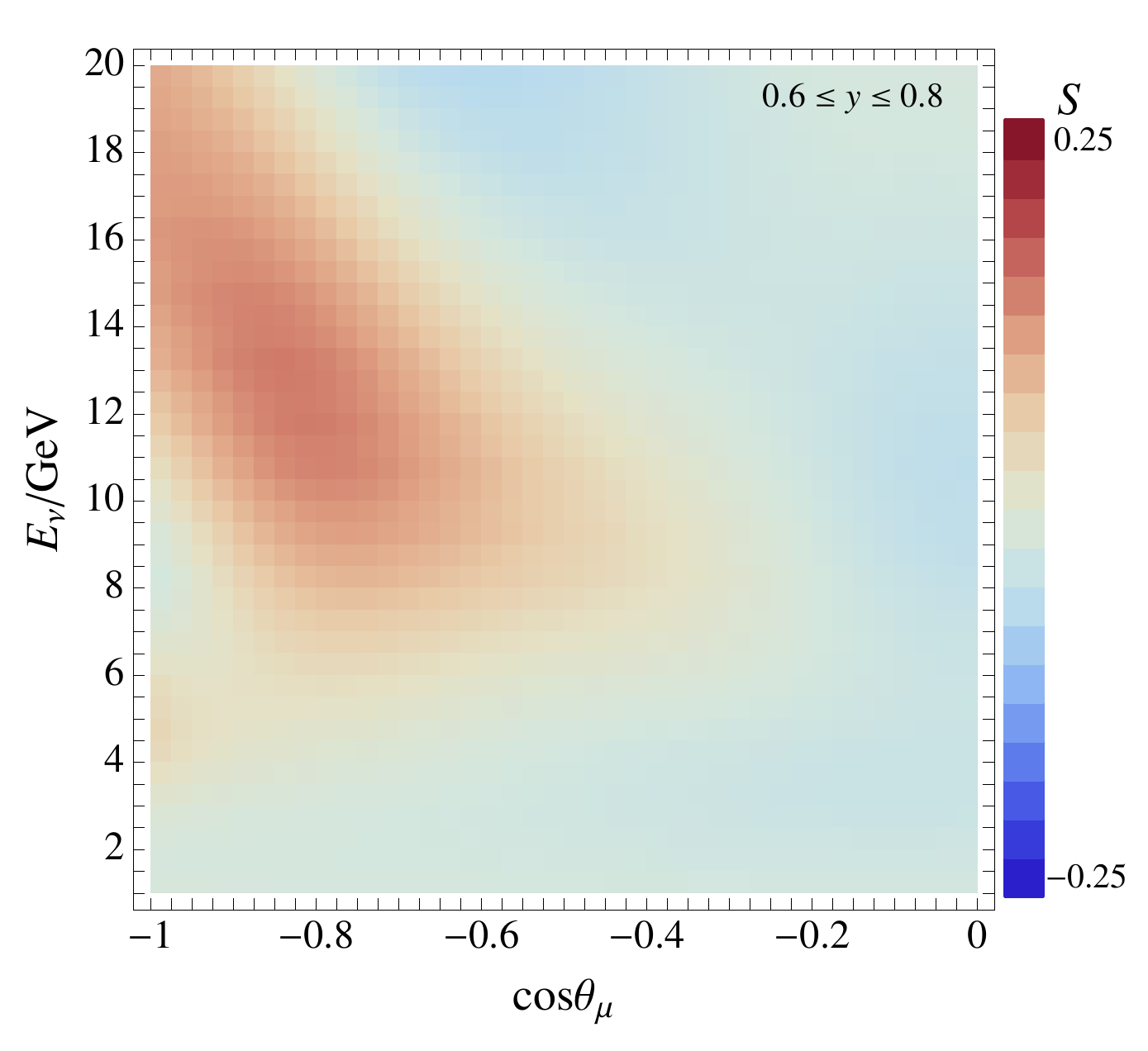}~~~~~~
 \includegraphics*[width=0.4\textwidth,type=pdf,ext=.pdf,read=.pdf]{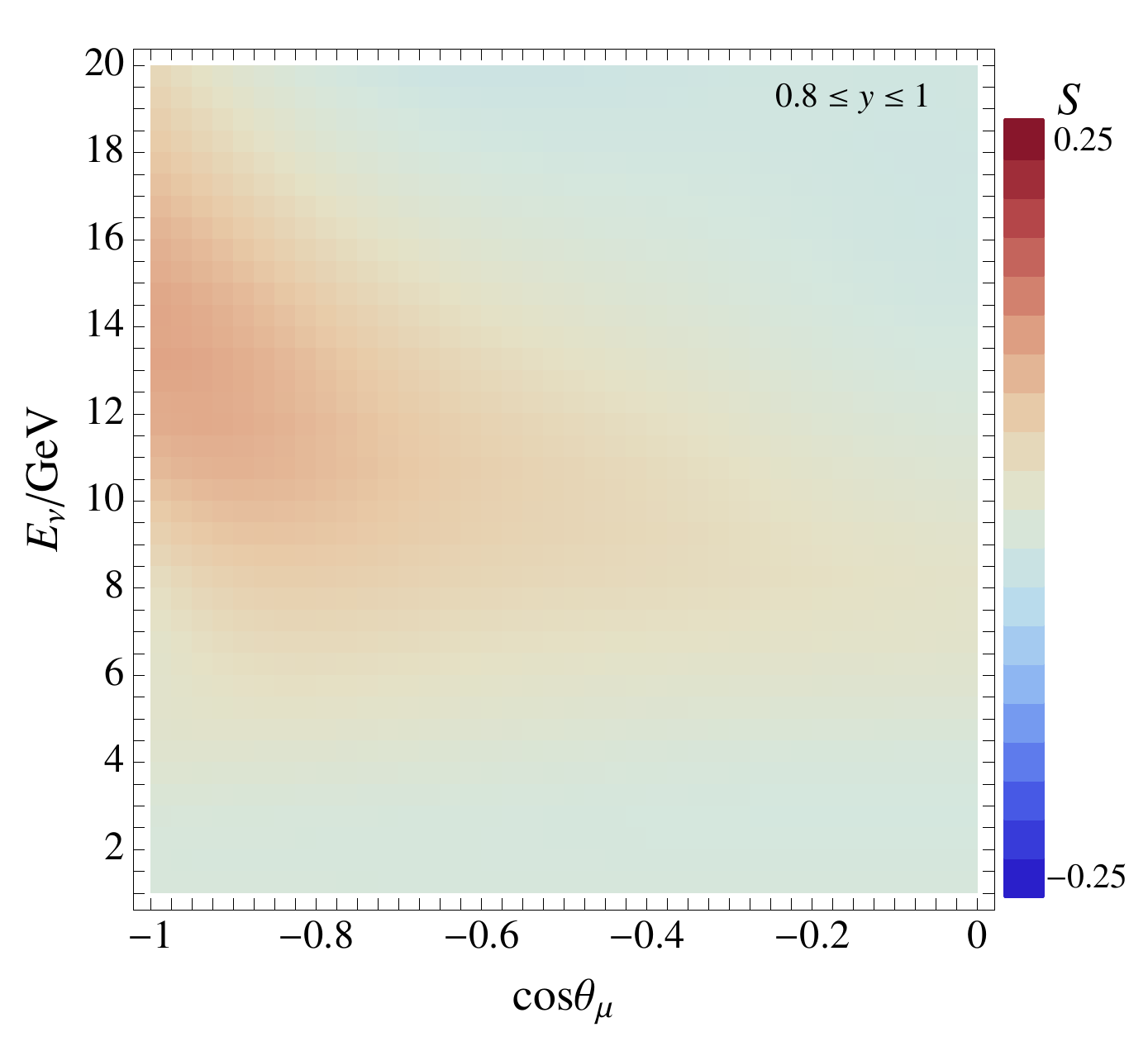}
\caption{The hierarchy asymmetry distribution
in the ($E_\nu - \cos \theta_\mu$) plane for different intervals of the inelasticity and for 1 year of exposure. The experimental smearing of the distributions was performed with the energy and 
angular resolutions $\sigma_E=\sqrt{0.7E},\, \psi_0 =20^\circ$.
\label{fig:sch1-muOsc-yrange}}
\end{figure*}

\begin{figure*}[ht!]
\centering
 \includegraphics*[width=0.4\textwidth,type=pdf,ext=.pdf,read=.pdf]{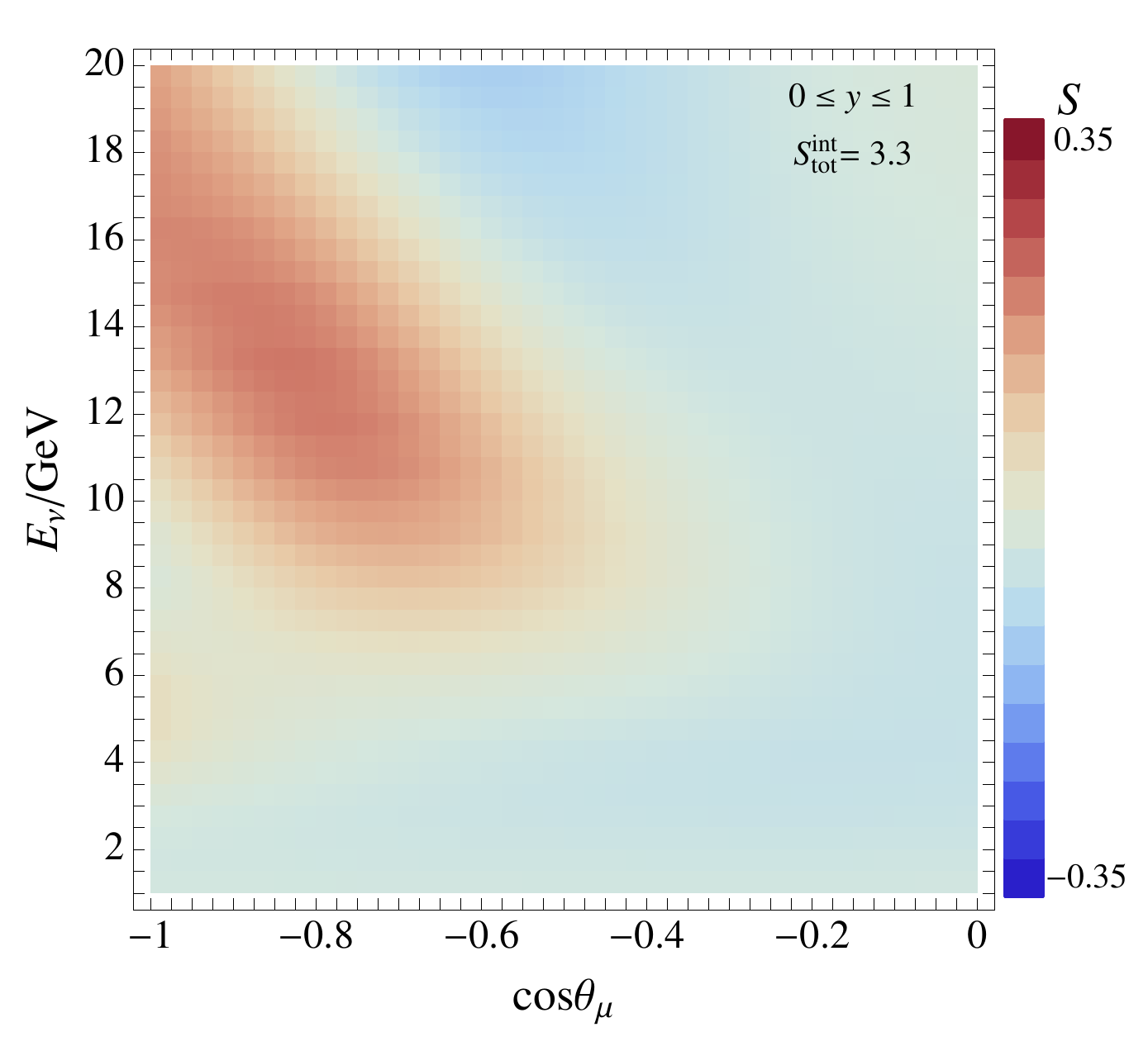}~~~~~~
 \includegraphics*[width=0.4\textwidth,type=pdf,ext=.pdf,read=.pdf]{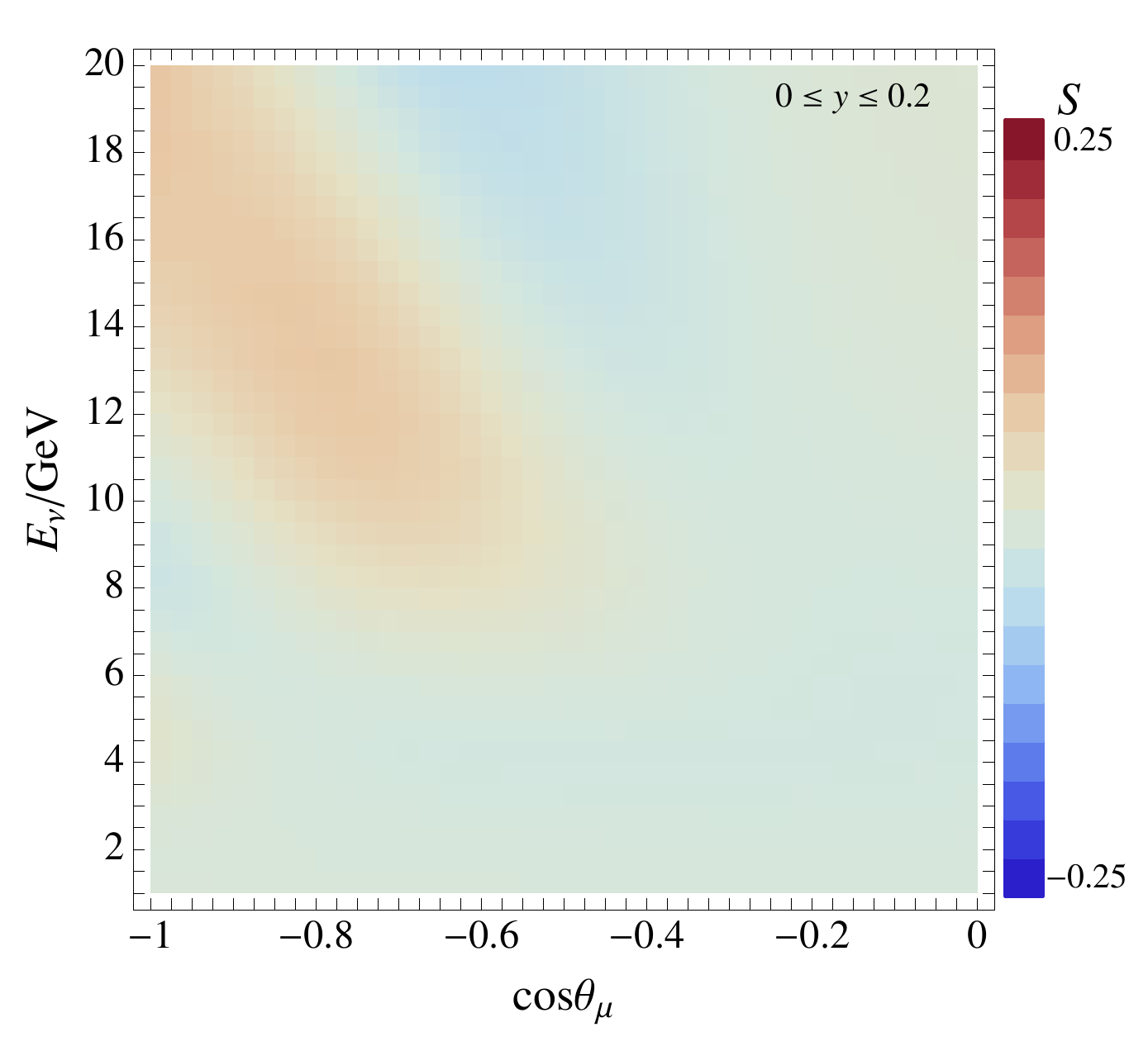}
 \includegraphics*[width=0.4\textwidth,type=pdf,ext=.pdf,read=.pdf]{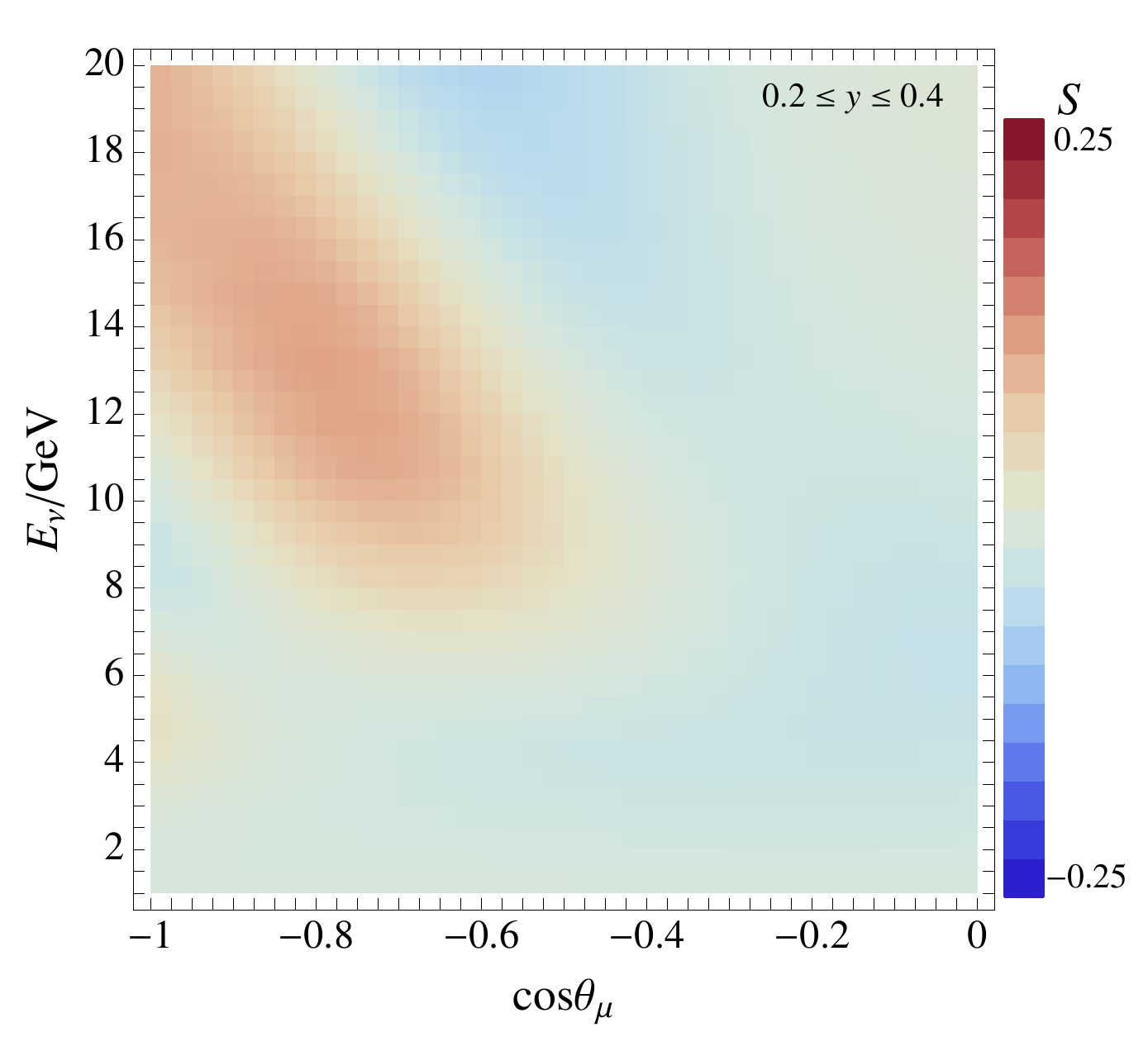}~~~~~~
 \includegraphics*[width=0.4\textwidth,type=pdf,ext=.pdf,read=.pdf]{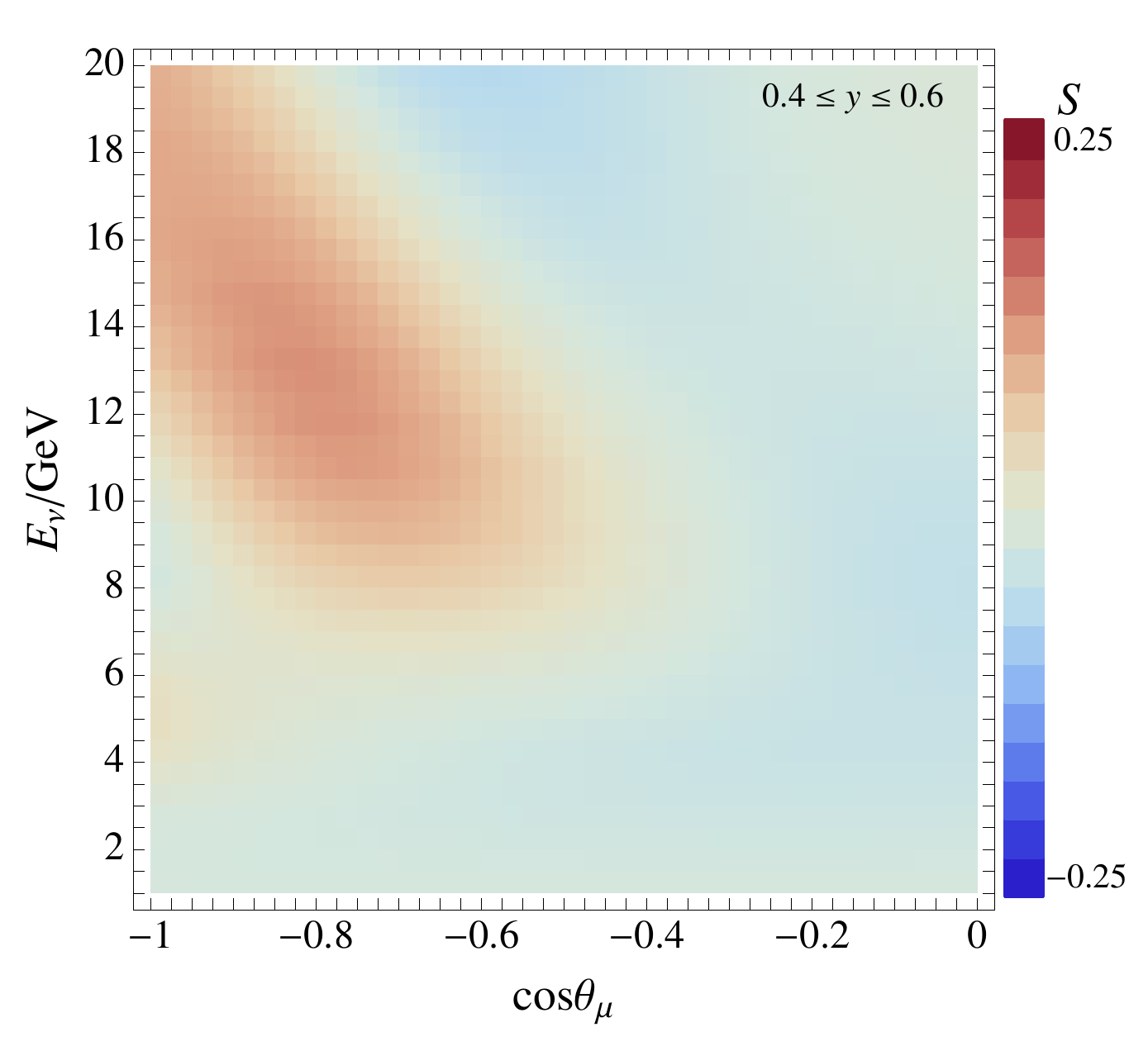}
 \includegraphics*[width=0.4\textwidth,type=pdf,ext=.pdf,read=.pdf]{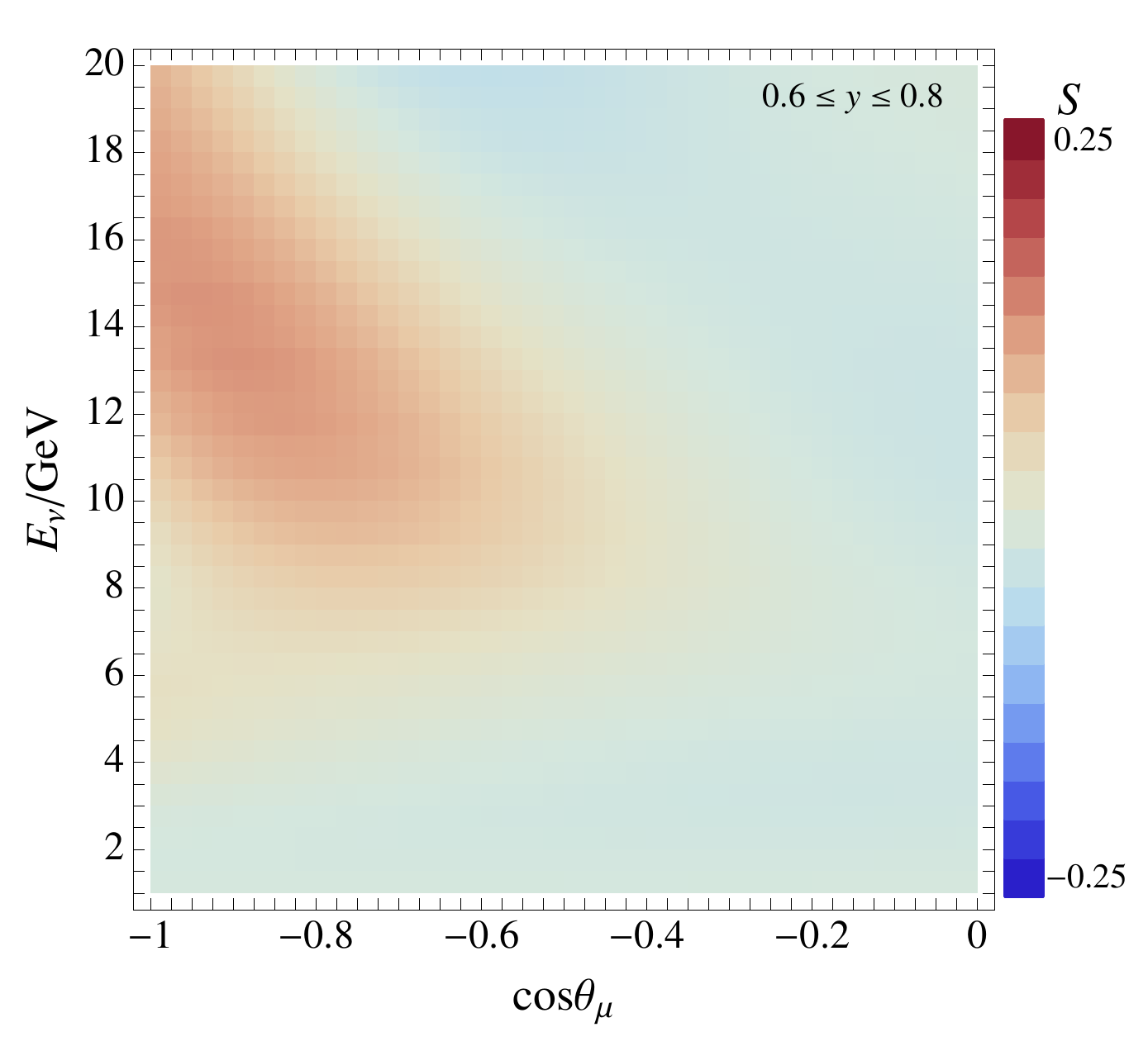}~~~~~~
 \includegraphics*[width=0.4\textwidth,type=pdf,ext=.pdf,read=.pdf]{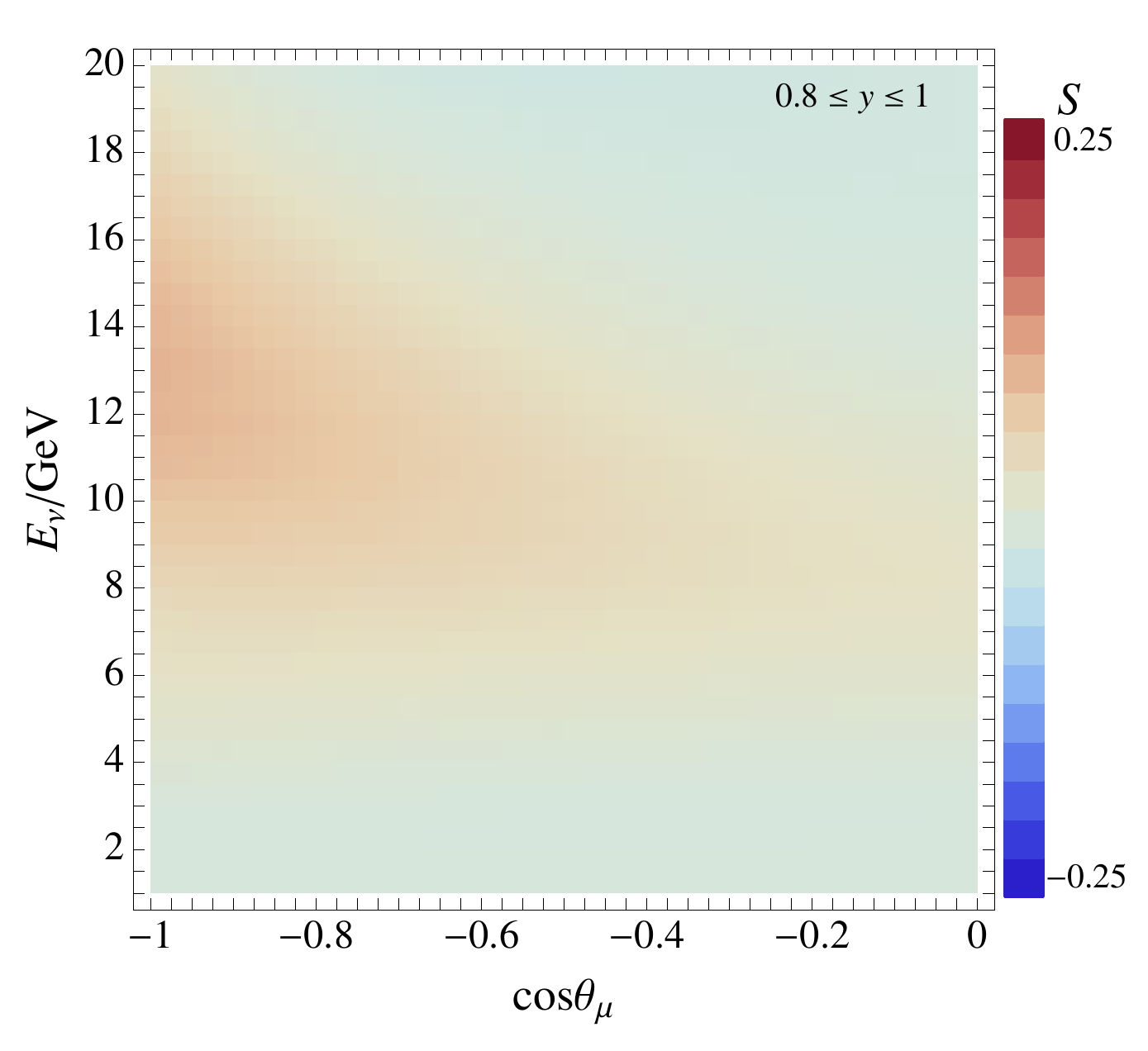}
\caption{The same as in Fig.~\ref{fig:sch1-muOsc-yrange} with  
$\sigma_E=\sqrt{0.7E}$, and $\psi_0 = 40^{\circ}$.
\label{fig:sch2-muOsc-yrange}}
\end{figure*}

\begin{figure*}[ht!]
\centering
 \includegraphics*[width=0.4\textwidth,type=pdf,ext=.pdf,read=.pdf]{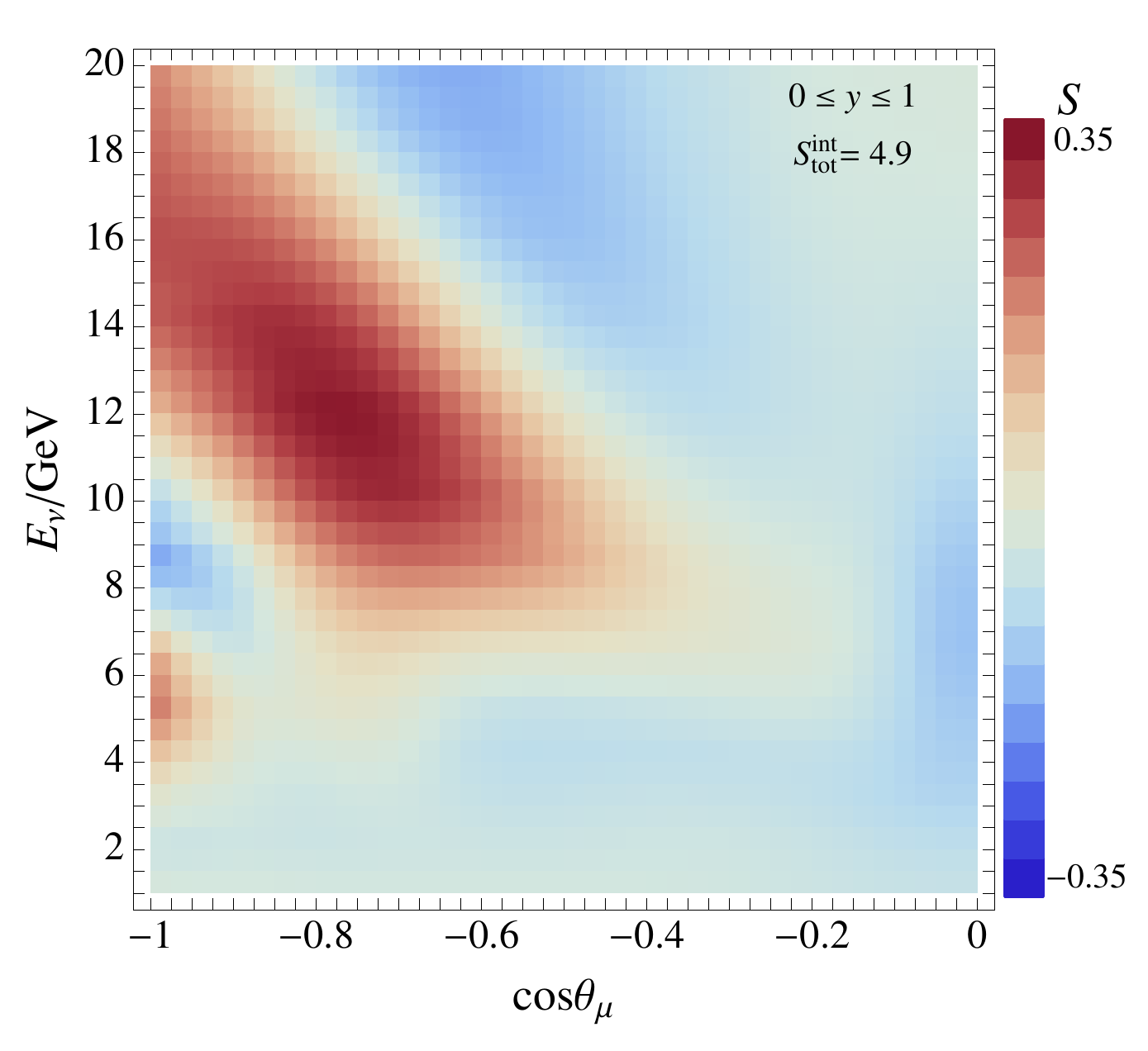}~~~~~~
 \includegraphics*[width=0.4\textwidth,type=pdf,ext=.pdf,read=.pdf]{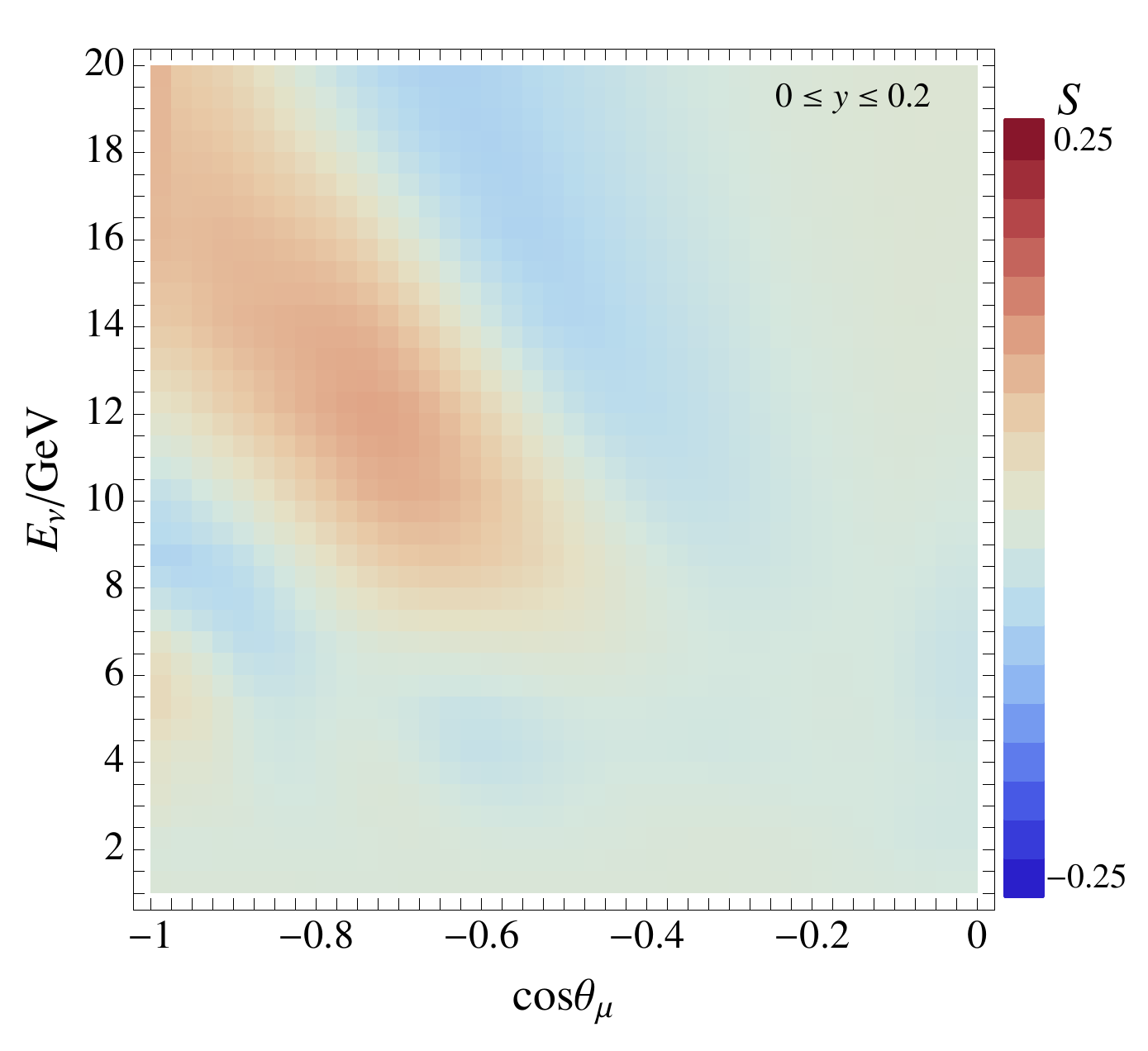}
 \includegraphics*[width=0.4\textwidth,type=pdf,ext=.pdf,read=.pdf]{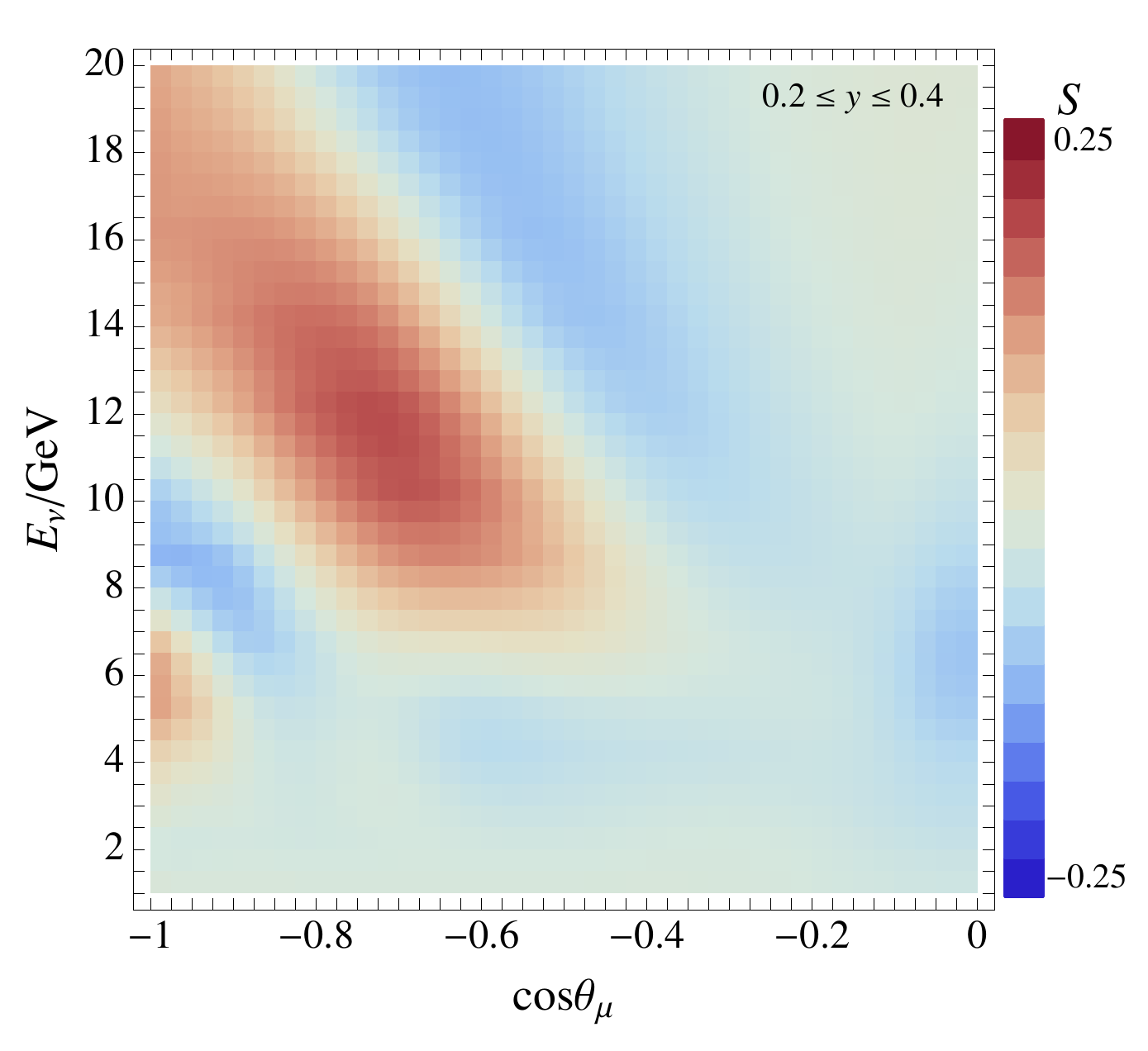}~~~~~~
 \includegraphics*[width=0.4\textwidth,type=pdf,ext=.pdf,read=.pdf]{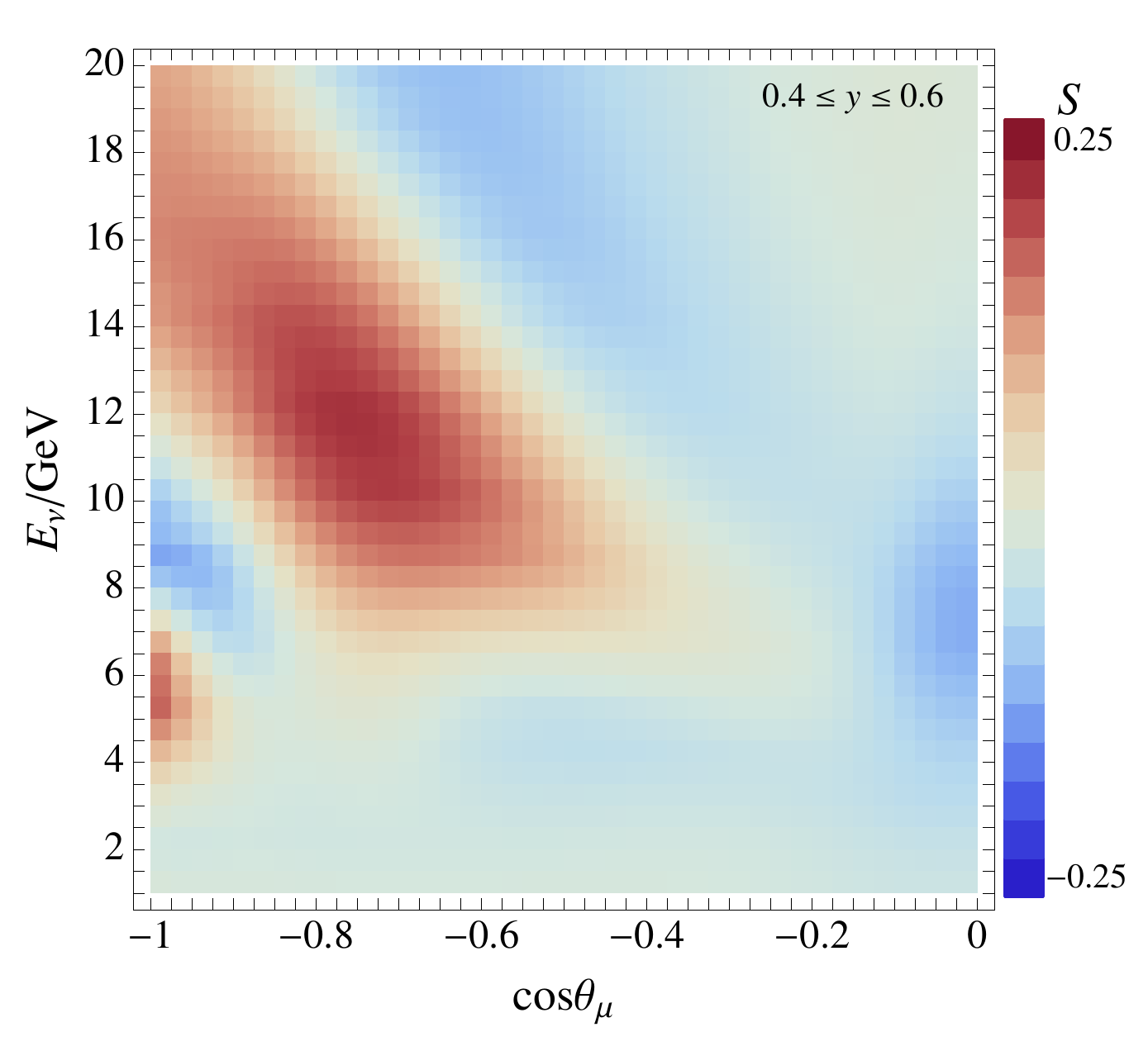}
 \includegraphics*[width=0.4\textwidth,type=pdf,ext=.pdf,read=.pdf]{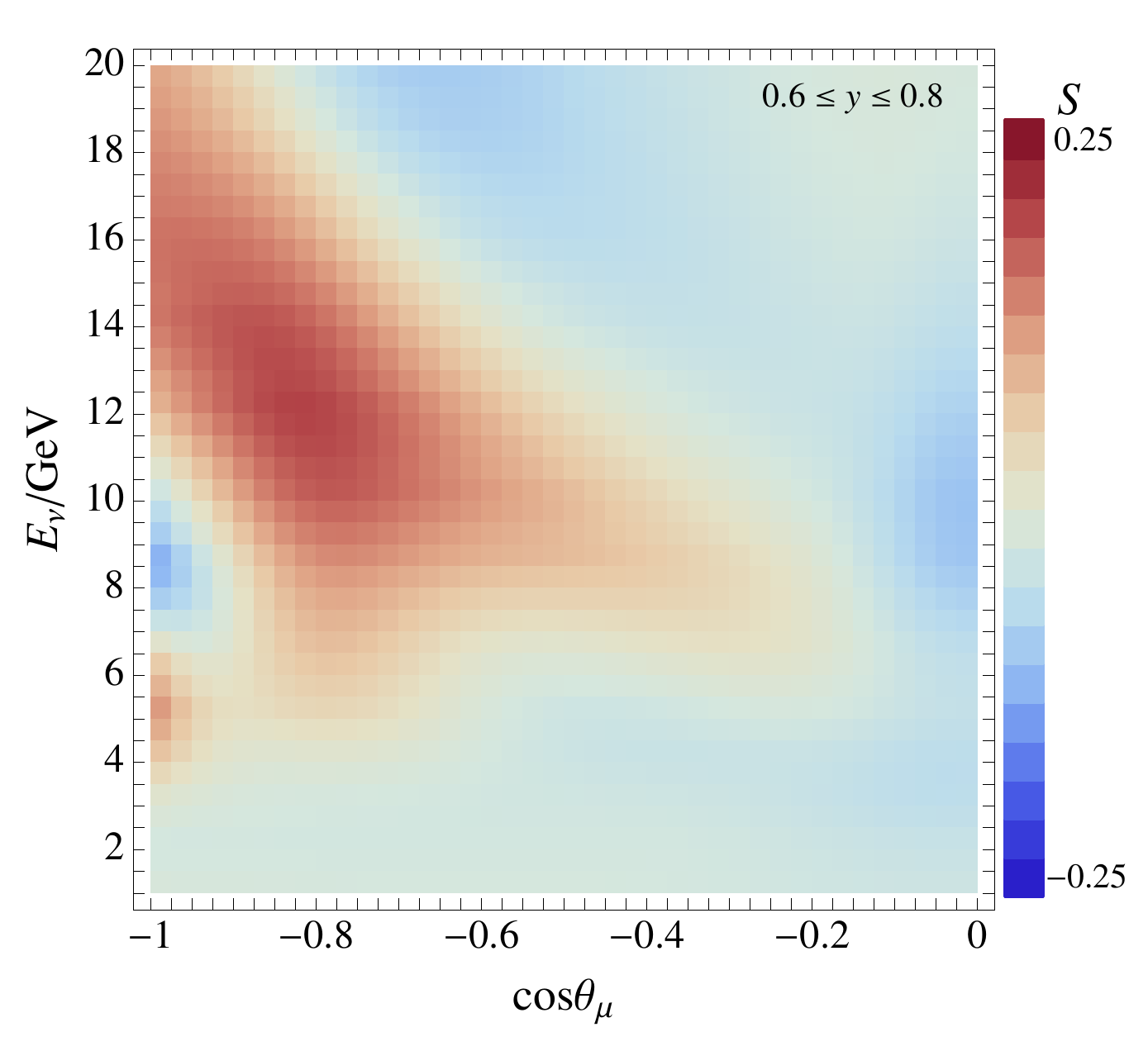}~~~~~~
 \includegraphics*[width=0.4\textwidth,type=pdf,ext=.pdf,read=.pdf]{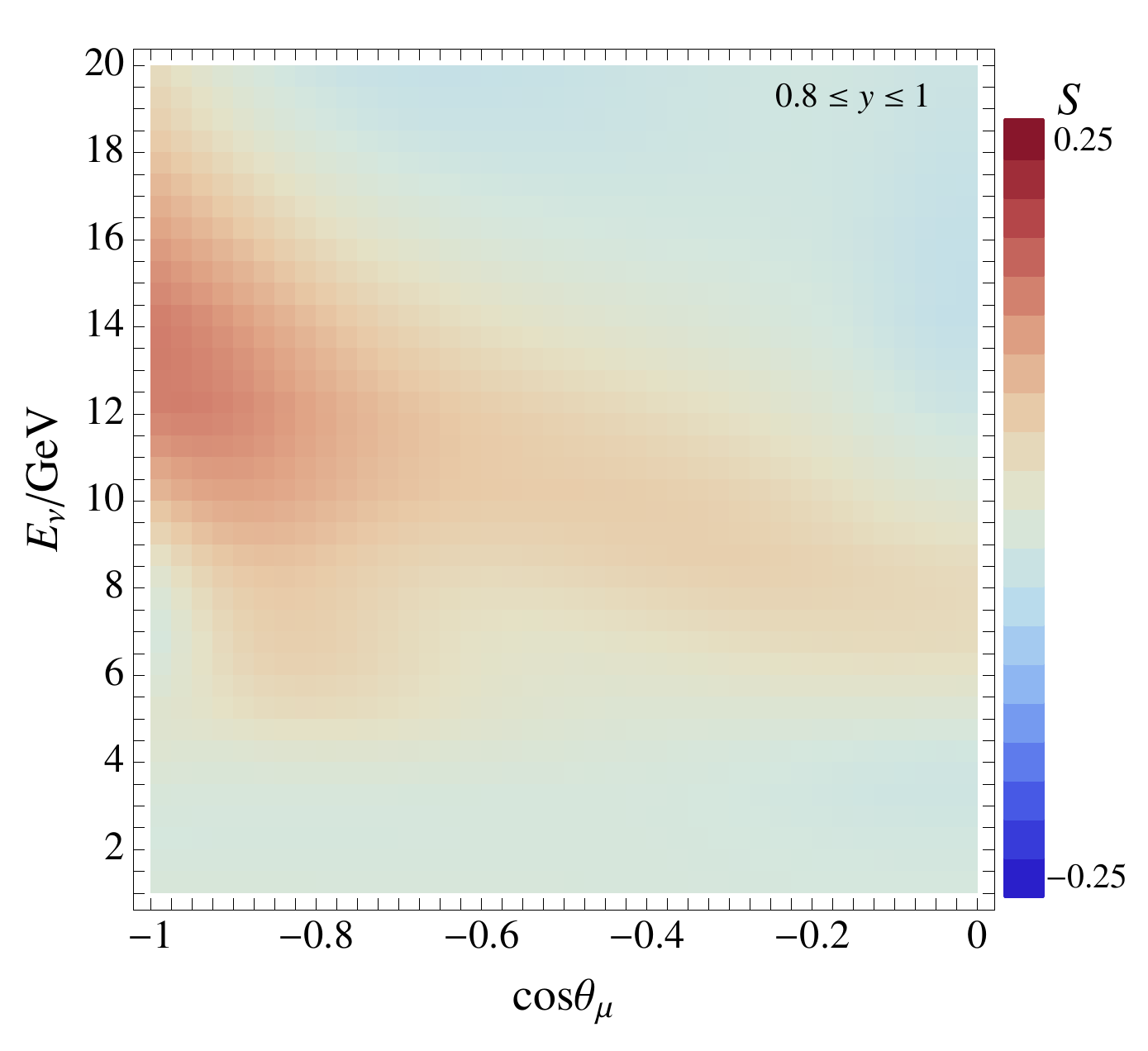}
\caption{The same as in Fig.~\ref{fig:sch1-muOsc-yrange} with $\sigma_E=\sqrt{0.35E}$ and $\psi_0 =10^{\circ}$.
\label{fig:sch3-muOsc-yrange}}
\end{figure*}

The smeared distributions in the plane $E_\nu - \cos \theta_\mu$ 
for different intervals of $y$ and different resolutions are shown in Figs. \ref{fig:sch1-muOsc-yrange}, 
\ref{fig:sch2-muOsc-yrange}, 
\ref{fig:sch3-muOsc-yrange}.

In comparison with Fig.~\ref{fig:raw-muOsc-yrange}, the overall scale
of asymmetries is reduced by factor of $\sim 2$,
which quantifies the effect of experimental smearing.
Position and shape of the regions of strong asymmetry
follow to a large extent those in Fig.~\ref{fig:raw-muOsc-yrange}.

~\\

\subsection{Estimations of the total significance}

Total significances with the experimental smearing  
are calculated using Eqs. ~(\ref{eq:Sy}-\ref{eq:Syint}) 
with replacement $n\rightarrow\hat{n}$. 
Table~\ref{tab:s2} presents the total 
significance after one year of 
exposure for several experimental resolution scenarii,  
including the one with  only kinematical smearing. 

For comparison we also compute the significances obtained 
immediately from the neutrino oscillograms, 
which corresponds to exact reconstruction of the neutrino 
energy and direction. If $\nu$ and $\bar{\nu}$ distributions are 
measured independently, we would have 
$|S_{\rm tot}^\nu| = 46.8$,  
$|S_{\rm tot}^{\bar\nu}| = 43.8$, and the total significance 
$|S_{\rm tot}| = \sqrt{|S_{\rm tot}^\nu|^2 + |S_{\rm tot}^{\bar\nu}|^2}
= 64.1$. The latter is about 3 times larger than the 
total significance in the case when $\nu$ and $\bar{\nu}$ signals 
are not separated  $|S_{\rm tot}^{\nu+\bar\nu}| = 23.7$, in agreement 
with our qualitative result in sect.~\ref{sct:3}. \\

\begin{table}[h]
\begin{tabular}{ccc|cccccc}
\hline
$\sigma_E$ &  & $\psi_0$ & & $|S_{\rm tot}|$ & & $|S_{\rm tot}^{\rm int}|$ & & $|S_{\rm tot}|/|S_{\rm tot}^{\rm int}|$ \\
\hline
0              & &  0           & &  8.43    & & 7.11    & & 1.19  \\
$\sqrt{0.35E}$ & & $10^{\circ}$ &  & 5.44  &  &  4.90    &  & 1.11   \\
$\sqrt{0.35E}$ & & $20^{\circ}$ & &  5.10  &  &  4.66    &  & 1.10 \\
0.3E          & & $20^{\circ}$ &  &  4.40  &  & 3.98     &  & 1.11 \\
$\sqrt{0.7E}$ & & $20^{\circ}$ &  &  4.19  &  &  3.87    &  & 1.08  \\
$\sqrt{0.7E}$ & & $40^{\circ}$ &  &  3.52  &  &  3.26    &  & 1.08 \\
\hline
\end{tabular}
\caption{The total significance of identification of the neutrino mass hierarchy 
for different experimental smearing scenarios and for 1 year of exposure. 
$S_{\rm tot}$ refers to analysis with inelasticity, whereas  $S_{\rm tot}^{\rm int}$ 
- for  $y-$integrated distributions analysis. The upper line with $\sigma_E = \psi_0 = 0$
corresponds to the kinematical smearing only. 
} 
\label{tab:s2}
\end{table}

In the realistic case of partial separation of the $\nu$ 
and $\bar{\nu}$  
signals, which takes place when $y-$information is included, 
and after the kinematical smearing 
the significance decreases strongly: 
down to 8.43 after one year (so that after 3 years of exposure we would have  
$|S_{\rm tot}^{\rm 3\,yr}|= 14.6$). 
 This number further reduces 
down  to $|S_{\rm tot}^{\rm 3\,yr}|\approx 6.1$ 
after the experimental smearing  in our worst case scenario 
($\sigma_{\mu,h} =\sqrt{0.7E_{\mu,h}}$, $\sigma_\psi=40^{\circ}$). 

Systematic uncertainties likely play a relatively mild role 
in degrading these results, as the measurements are differential 
from neighboring locations (bins in $E_\nu$ and $c_\mu$), 
and the systematic uncertainties between neighboring 
bins with different asymmetries are strongly correlated. 
In addition, the $y$-distribution must be a superposition 
of the $y$-distributions of neutrino and of antineutrino events, 
strongly constraining its shape and providing information 
related to the systematic effect in the $y-$dimension.
However, we have also found  negligible degradation 
of the significance (about 1\%) if introduced 
as in~\cite{ARS}, at level of 10\%. 
This is due to the small bin size of our oscillograms.

\begin{figure}[h]
\centering
 \includegraphics*[width=0.35\textwidth]{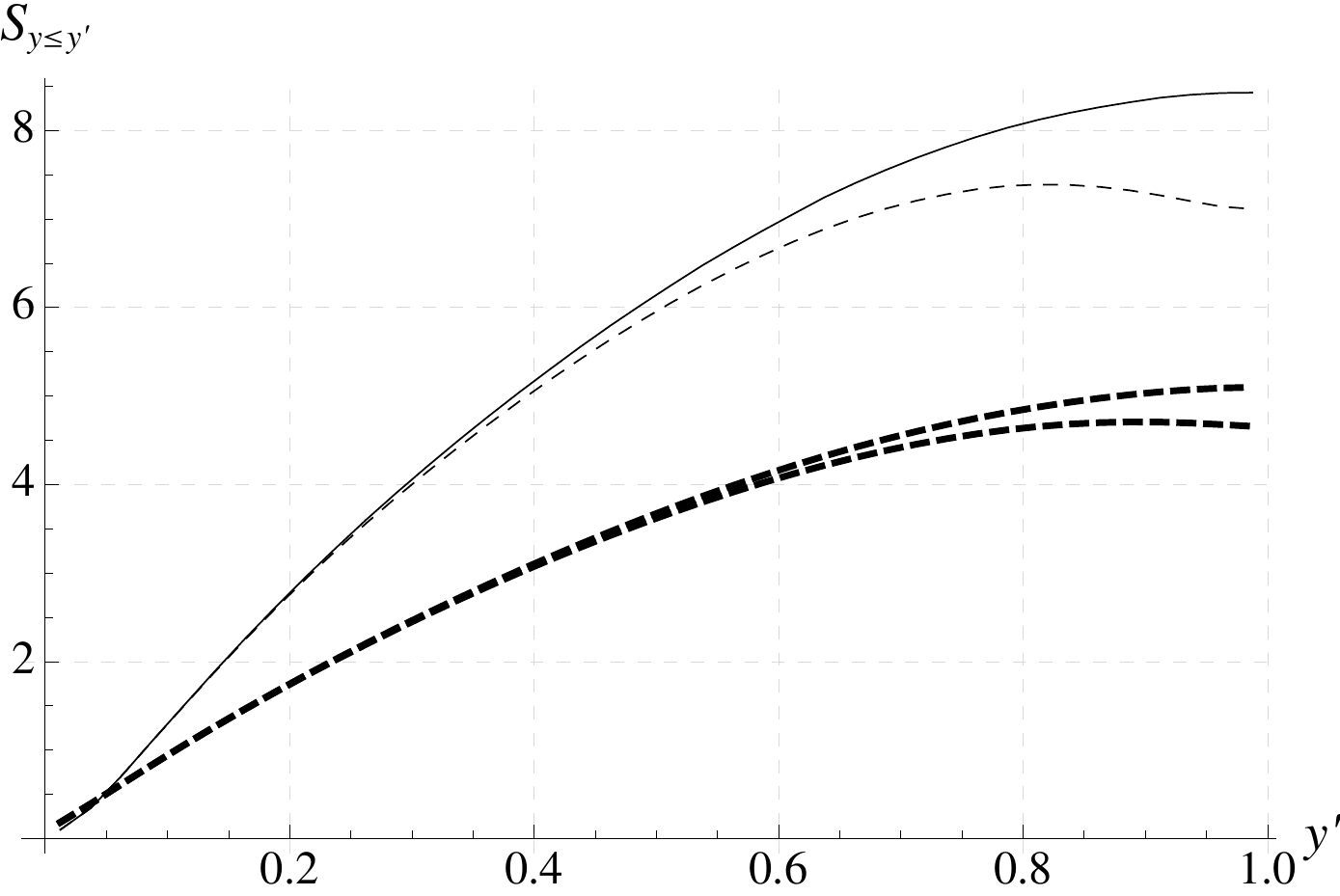}
\caption{Dependence of the significance on the upper limit of integration 
over $y$. 
Thin and thick curves respectively correspond to kinematical 
smearing only and  the experimental smearing added. 
Solid and dashed curves respectively show $S_{\rm tot}$ and $S_{\rm tot}^{\rm int}$.}
\label{fig:StotVSy}
\end{figure}

Fig.~\ref{fig:StotVSy} shows the dependence  of the significance 
on the upper limit of integration over $y$ for the case $\sigma_{\mu,h} =\sqrt{0.35E_{\mu,h}}$ and   
$\psi_0 = 20^{\circ}$ and for kinematical smearing only. 
The dashed curves are for $y-$integrated significances.

According to Fig.~\ref{fig:StotVSy}, the increase  of the significance is sustained 
up to higher $y$ for  curves including experimental smearing. 
This is due to the contribution to the asymmetry from larger $E_\nu$, 
whose relative importance in  smearing decreases. Also the difference 
between $y-$differential and $y-$integrated significances is relatively small after 
the experimental smearing. 

\subsection{$\Delta m_{32}^2$ degeneracy}

The effect of an inversion of the mass hierarchy (especially at large energies) 
is rather similar to a shift of the oscillation probabilities 
in the energy scale (see \cite{ARS}). 
This is equivalent to a change of $\Delta m_{32}^2$. 
Therefore the effect of hierarchy can be partly mimicked by a change of $\Delta m_{32}^2$. 
Indeed, the pattern of distribution  of the 
quantity  
\be
S^{\delta} \equiv \frac{N^{\rm NH} (\Delta m_{32}^2 + \delta) 
- N^{\rm NH}(\Delta m_{32}^2)}{\sqrt{N^{\rm NH} (\Delta m_{32}^2)}}
\label{eq:sdelta}
\ee
in $E_\nu - c_\mu$ plane is rather similar for certain values  of the shift parameter,  
$\delta$,  to the hierarchy asymmetry pattern: 
\be
S^{\rm MH} \equiv \frac{N^{\rm IH} (\Delta m_{32}^2)
- N^{\rm NH}(\Delta m_{32}^2)}{\sqrt{N^{\rm NH} (\Delta m_{32}^2)}}. 
\label{eq:smh}
\ee
Since $\Delta m_{32}^2$ is known with some error, 
this parameter degeneracy degrades the sensitivity to the mass hierarchy. 
To quantify  the effect 
the following significance has been computed in \cite{ARS}: 
\be
S^{{\rm MH} - \delta} \equiv \frac{N^{\rm IH} (\Delta m_{32}^2 + \delta)
- N^{\rm NH}(\Delta m_{32}^2)}{\sqrt{N^{\rm NH} (\Delta m_{32}^2)}}, 
\label{eq:sdiff}
\ee
where $\delta$ has been considered as a free parameter. 
This would correspond to NH as the true hierarchy 
and $\Delta m_{32}^2$ as the true value. 
The  true distribution $N^{\rm NH} (\Delta m_{32}^2)$ 
is then fitted by  IH distribution with arbitrary values of 
$\Delta m_{32}^2$. It has been found in \cite{ARS} that 
the minimum $S^{{\rm MH} - \delta}_{\rm min}$  is reached for 
$\delta \approx -  0.5 \sigma (\Delta m_{32}^2)$, 
where $\sigma (\Delta m_{32}^2)$ is the present $1\sigma$ 
accuracy of determination of $\Delta m_{32}^2$ from the global fit \cite{gfit}. 
The minimal value (for 1 year of exposure and no inelasticity information) 
$S^{{\rm MH} - \delta}_{\rm min} = 3.8$ should be 
compared with $S^{{\rm MH} - \delta} (\delta = 0) = 6.0$, 
thus showing reduction of the significance by a factor 1.6. 

Future measurements at accelerators 
will reduce the error by factor of 2,  
which means that no significant improvement is expected.

Let us show how information about inelasticity 
(or usage of 3D distributions) may help.
As we mentioned before, the effect of a $\Delta m_{32}^2$ change 
is nearly the same for  neutrinos and antineutrinos, 
whereas $y-$distributions are different.

We construct  the distribution $S^{\delta, {\rm int}}$ 
and the residual asymmetry plot $S^{\delta, {\rm int}} - S^{{\rm MH}, {\rm int}}$ 
which can be rewritten according to (\ref{eq:sdelta}) and  (\ref{eq:smh}) as  
\be 
S^{\delta, {\rm int}} - S^{\rm int} = 
- \frac{N^{\rm IH} (\Delta m_{32}^2)
- N^{\rm NH}(\Delta m_{32}^2 - \delta)}{\sqrt{N^{\rm NH} (\Delta m_{32}^2)}}. 
\nonumber
\ee
After substitution $(\Delta m_{32}^2-\delta) 
\rightarrow \Delta m_{32}^2$ this residual asymmetry 
essentially coincides with the quantity 
$S^{{\rm MH} - \delta}$ (\ref{eq:sdiff}) computed in \cite{ARS}.

\begin{figure*}[ht!]
\centering
 \includegraphics*[width=0.4\textwidth]{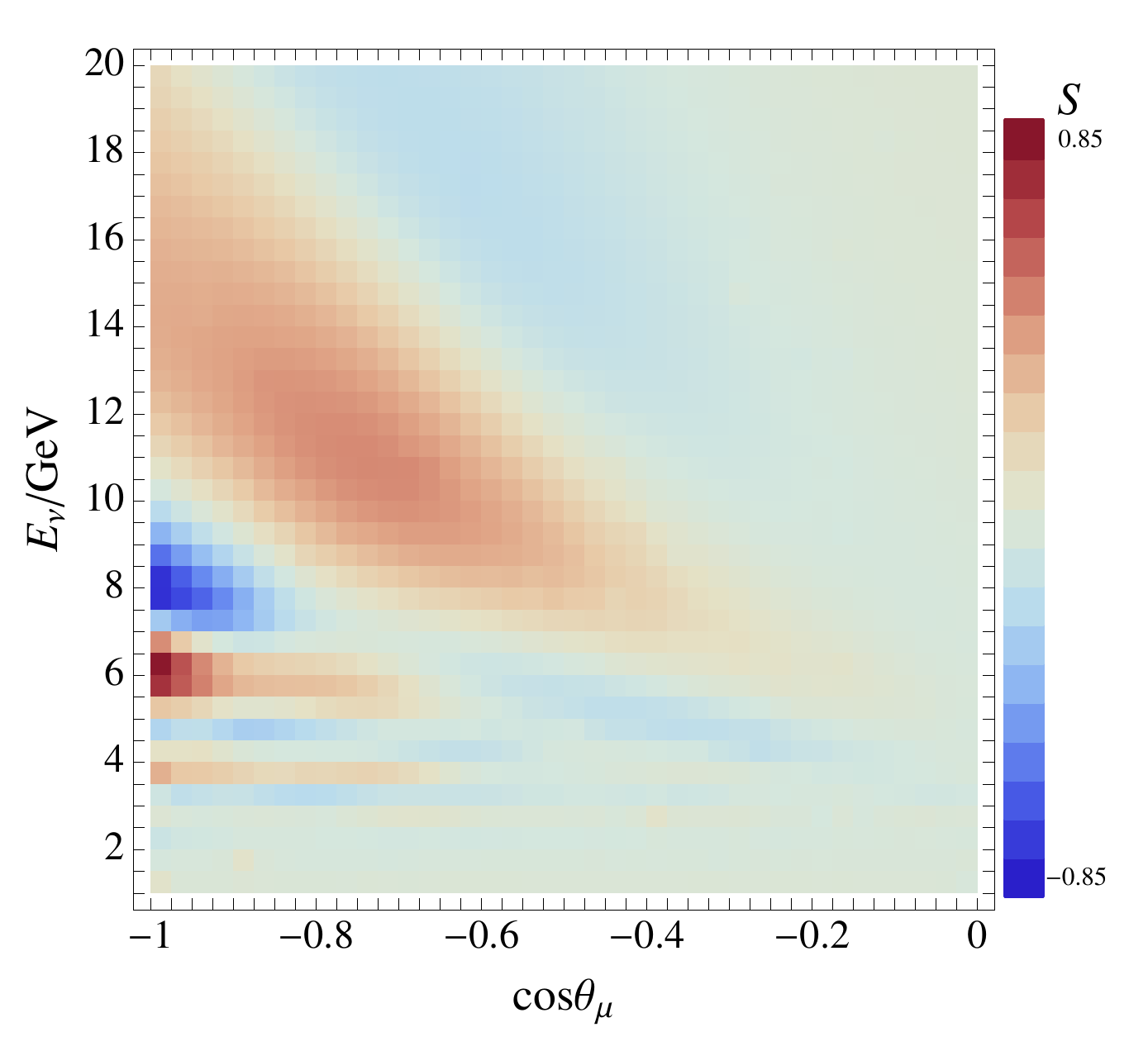}~~~~~~
 \includegraphics*[width=0.4\textwidth]{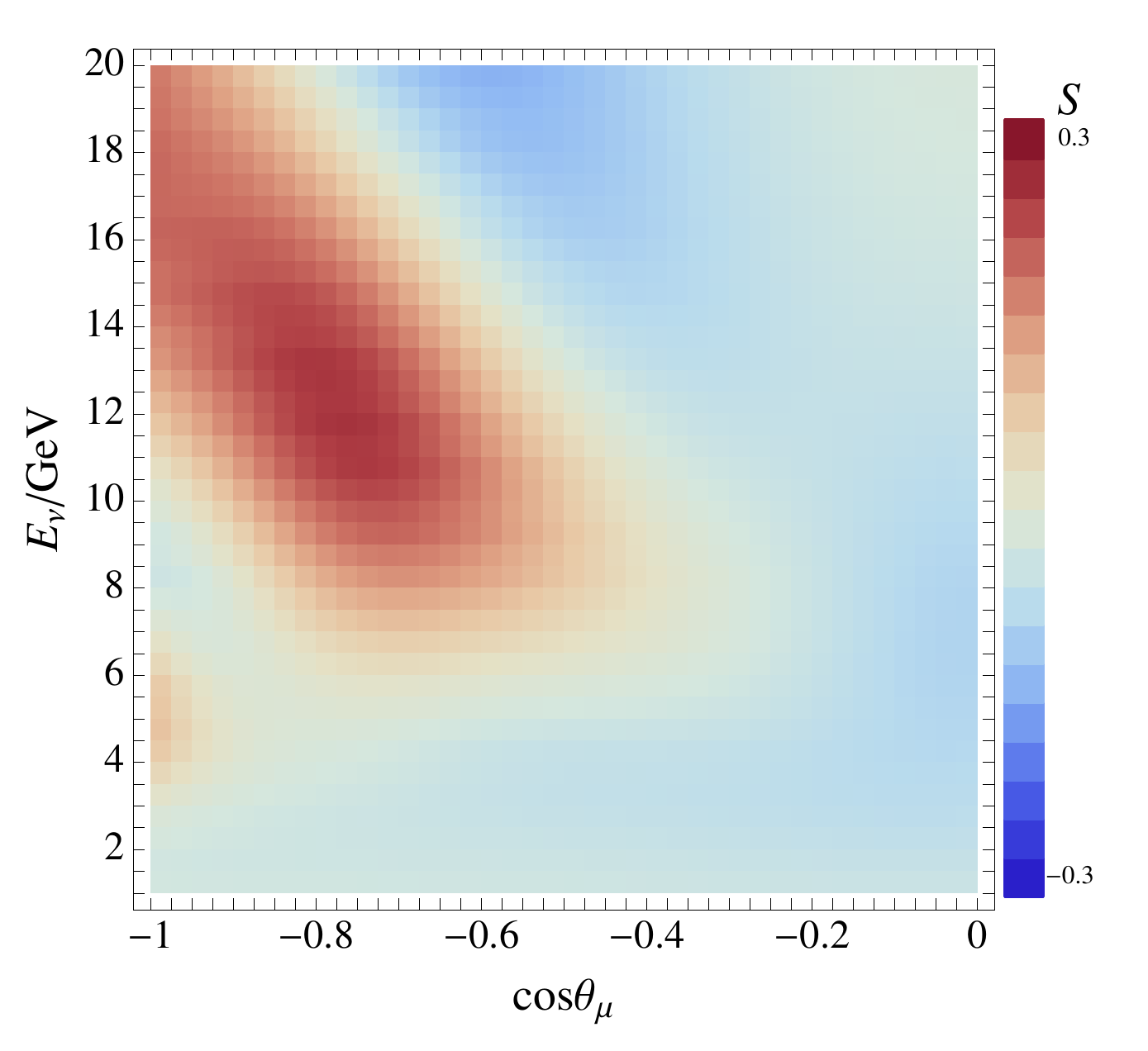}
\caption{Asymmetry plots $S^{\rm int}$ with the kinematical smearing 
only (left) 
and after application of the experimental smearing 
$\sigma_E=\sqrt{0.7E}, \sigma_\psi=20^o \sqrt{\frac{m_{\rm p}}{E_\mu}}$ (right), for 1 year of exposure.
\label{fig:S235}}
\end{figure*}

\begin{figure*}[ht!]
\centering
 \includegraphics*[width=0.4\textwidth]{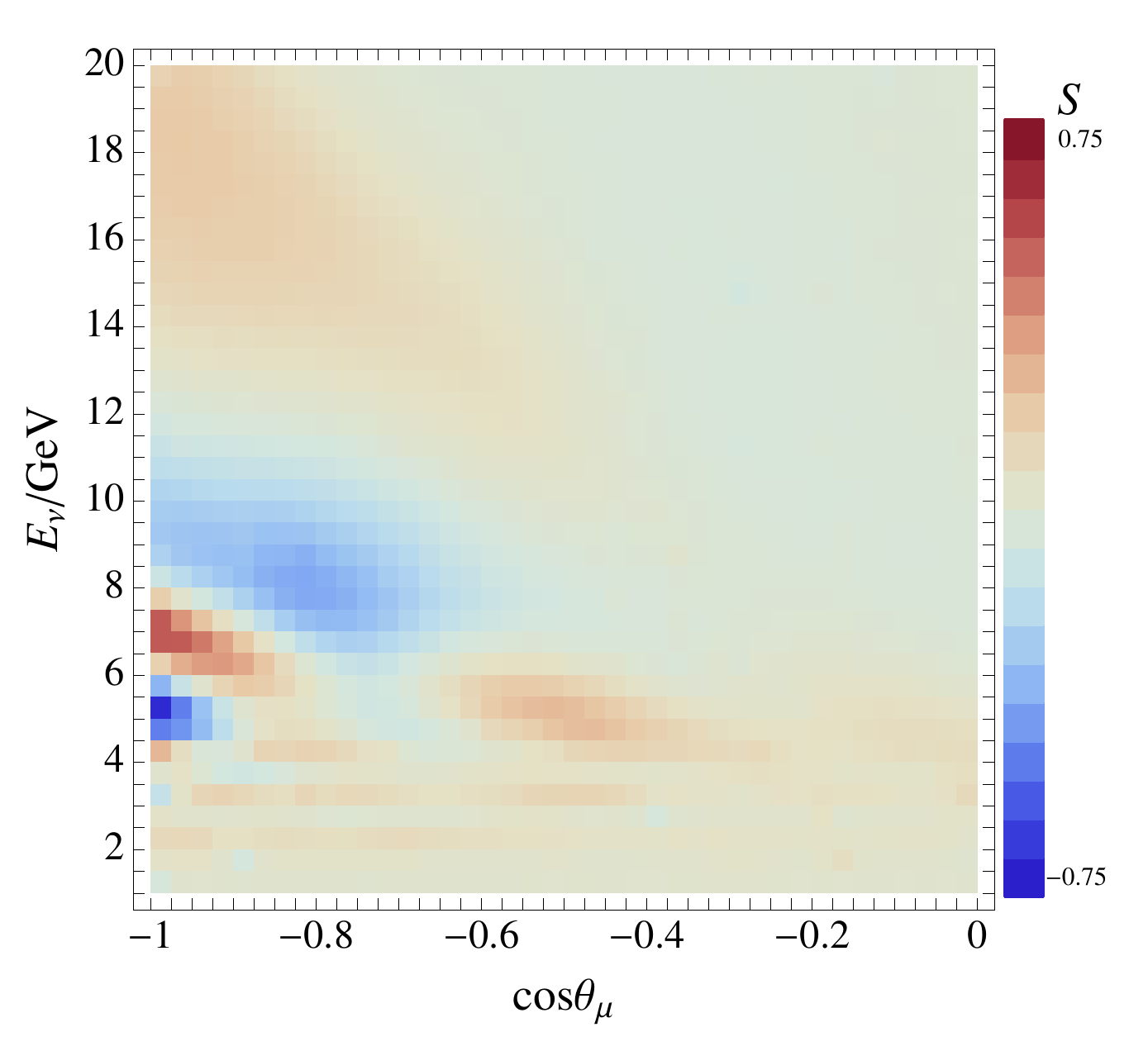}~~~~~~
 \includegraphics*[width=0.4\textwidth]{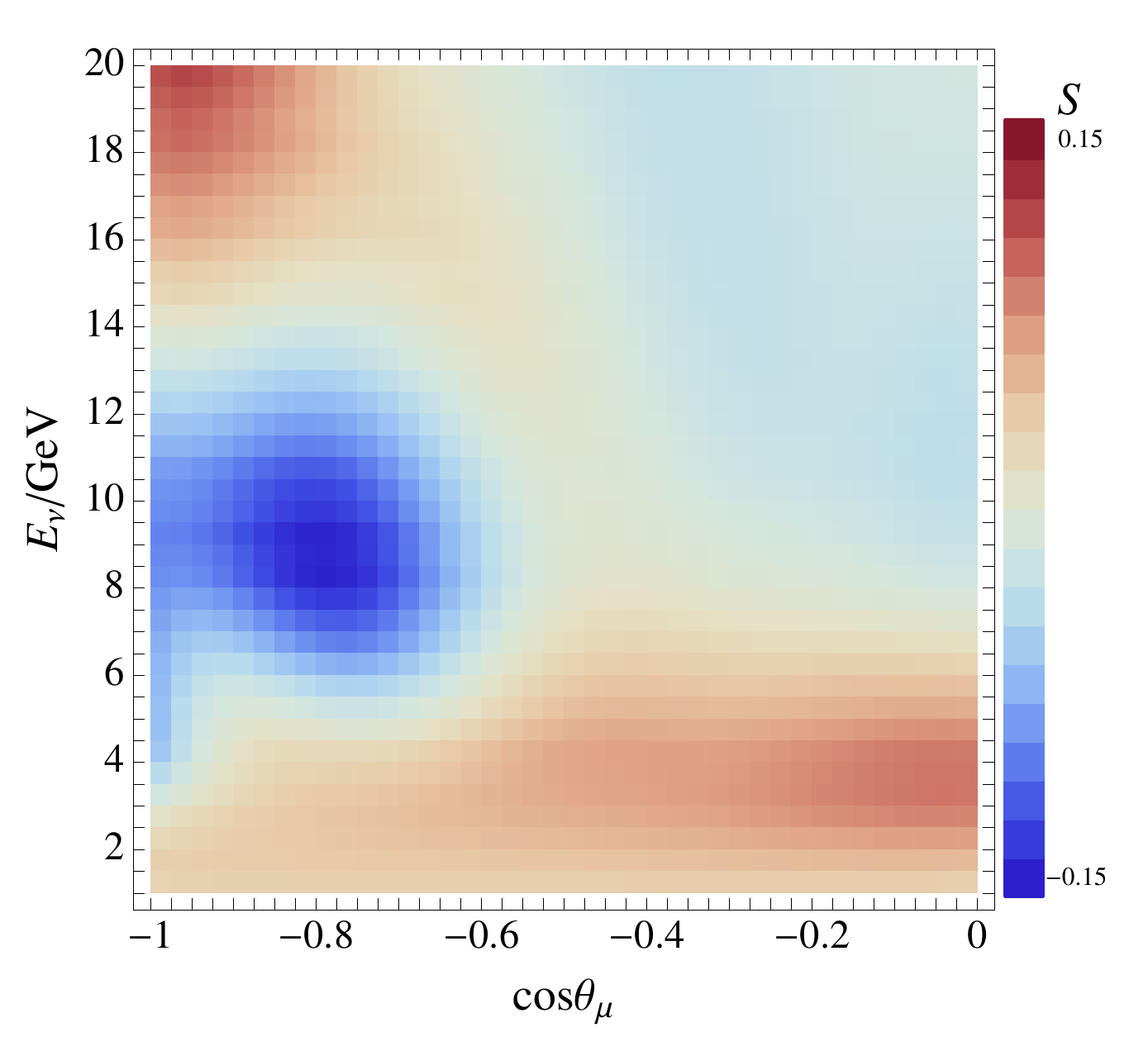}
\caption{Residual asymmetry plots $(S^{\delta, int} - S^{int})$  with the kinematical 
smearing only (left) and after application of the experimental smearing 
$\sigma_E=\sqrt{0.7E}, \sigma_\psi=20^o \sqrt{\frac{m_{\rm p}}{E_\mu}}$ (right), for 1 year of exposure.
\label{fig:SSprime}}
\end{figure*}

On the left panels of Figs.~\ref{fig:S235}  and \ref{fig:SSprime}, 
we respectively show   the plots for the asymmetry  $S^{\rm int}$ and for 
the residual asymmetry $S^{\delta, {\rm int}} - S^{\rm int}$ 
for an ideal detector after kinematical smearing only.
We take  $\delta \approx -  0.5 \sigma (|\Delta m_{32}^2|)= -6\cdot 10^{-5}$~eV$^2$,
which corresponds to the maximal degeneracy effect according to \cite{ARS}. 
On the right panels of Figs.~\ref{fig:S235}  and \ref{fig:SSprime}, 
we present respectively the asymmetry  
$S^{\rm int}$ and the residual asymmetry $S^{\delta, {\rm int}}-S^{\rm int}$ 
after application of the experimental smearing. 
We used $\sigma_E=\sqrt{0.7E}$, and  $\psi_0 = 20^{\circ}$ 
as a realistic experimental resolution.  

The total significance  of determination of the 
mass hierarchy can be computed using the residual asymmetry as   
\be 
S_{\rm tot}^{\rm int}({\rm MH}-\delta)  =  
\sqrt{\int {\rm d}E_\nu \int {\rm d}c_\mu (S^{\delta,  {\rm int}}-S^{\rm int})^2}
\nonumber
\ee
for the distributions without $y-$information,  and 
\be 
S_{\rm tot}({\rm MH}-\delta) =  
\sqrt{\int {\rm d}E_\nu \int {\rm d}c_\mu \int {\rm d}y (S^\delta-S)^2} 
\nonumber
\ee
with $y-$distribution.  
For $\delta \approx -  0.5 \sigma (|\Delta m_{32}^2|)$ and 
after 1 year exposure we obtain $S_{\rm tot}^{\rm int}({\rm MH}-\delta) = 4.23$. 
This should be compared with $S_{\rm tot}^{\rm int}  = 7.11$ 
without degeneracy effect (see Table~\ref{tab:s2}). 
So, the degeneracy effect found in such a way corresponds closely to the one found in \cite{ARS}. 
With the $y-$distribution, we obtain $S_{\rm tot}^{{\rm MH}-\delta} = 6.03$ ($S_{\rm tot}  = 8.43$). 
Thus,  the total significance is enhanced by $\sim43$\% using the inelasticity. 

After the experimental smearing described above we find 
\be
S_{\rm tot}^{\rm int}({\rm MH}-\delta) =  1.93,~~~~
S_{\rm tot}({\rm MH}-\delta)  = 2.42
\nonumber
\ee
(without degeneracy we would have  
$S_{\rm tot}^{\rm int}=3.87$ and $S_{\rm tot}=4.19$, see Table~\ref{tab:s2}). 
The significance enhancement is reduced to about 25\%.

These results mean that the necessary exposure to ascertain the mass hierarchy with an ideal detector is a factor 2 larger if $y$ is not exploited.
For the detector with the above mentioned experimental resolutions, this factor is not as large but still significant, about 1.55.
The additional relative power of the inelasticity,  
as we already noticed from numbers in Table~\ref{tab:s2},  is greater,
when detector resolutions are better.

\section{Discussion and conclusions}
\label{sct:6}

Multi-megaton scale under-ice and under-water detectors 
of atmospheric neutrinos with low (few GeV's) energy thresholds 
open up new possibilities for the  determination of  neutrino properties. 
This includes the neutrino mass hierarchy, the deviation of the 2-3 mixing 
from maximal and high accuracy measurement of    
$\Delta m_{32}^2$. 

With a dense array of optical modules, it will be possible 
to identify different atmospheric neutrino events, 
and in particular, the $\nu_\mu$ CC events 
and determine their characteristics.  
For the $\nu_\mu$ events, it will be possible 
to measure not only the energy and the direction of the muon, but also 
the energy of accompanying hadron cascade. The latter then  
determines the  inelasticity. 
With $y$, one can construct the three-dimensional distributions
of events in  $(E_\nu, \cos \theta_\mu, y)$.

In this paper, we have explored various improvements
of sensitivity to the mass hierarchy, which will be possible with the inclusion of
the inelasticity in the analysis. The results can be summarized in the following way.

1. Inelasticity measurements provide a certain sensitivity 
to separate signals from neutrinos and antineutrinos.   
%
This, in turn,  reduces the cancellation of
the neutrino and antineutrino contributions to the
hierarchy as well as to CP-violation effects.

We find that, in the ideal case of complete  separation 
or independent measurement of $\nu$ and $\bar{\nu}$
signals, the significance of the hierarchy determination
increases by factor $\sim 2.2 - 3$. However,  finite accuracy of
the separation (extraction of the  parameter $\alpha$) 
reduces the effect down to (20 - 30) \%.
The best separation is in the range of large $y$ 
where, however,  the angular smearing becomes strong and
effect of mass hierarchy is averaged out. 

2. The selection of events with small $y$ allows one to reduce
the angle between the neutrino and muon directions and 
therefore reduce the kinematical smearing,
which is very strong  at low energies in the resonance region where
effect of mass hierarchy at the probability
level is the biggest one.  However, for small $y$, the effective
separation of the $\nu$ and $\bar{\nu}$ contributions worsen, 
and moreover, the  statistics decreases with cut in $y$. So, for fixed exposure,
the overall gain is rather modest.

Separation  of the $\nu$ and $\bar{\nu}$ improves with increase of $y$, 
while the neutrino angle reconstruction improves with decrease of $y$. 
Therefore  the analysis of these improvements
should be done simultaneously.
This requires study of the 3D distributions of events,
which takes into account both separation and
reduction of kinematical smearing automatically.

3. We have computed the 3D oscillograms of the $\nu_\mu$ events  
with  the kinematical smearing (for this the kinematics of the 
$\nu_\mu$ CC-interactions has been taken into account precisely).  
We then found the 2D asymmetry distribution in the 
$E_\nu - \cos \theta_\mu$ plane for different intervals of $y$. 
The main contribution to the identification of 
the hierarchy follows from the intermediate range $y = 0.3 - 0.7$, 
and the contributions from intervals $y = 0.8 - 1.0$  and $y = 0 - 0.2$ 
are  small.  The  inelasticity
enhances the total significance of determination of the mass hierarchy by
about $20 \%$, which is consistent with our semi-qualitative 
analysis provided that  a slight decrease of $\gamma$ is achieved. 

4. We then performed smearing of the distributions  
over the observables: the energies of muon and cascade 
as well as the angle of muon. We used the Gaussian smearing 
functions assuming different widths and their dependences on energy. 
The experimental smearing further diminishes  
the total significance by factor 1.5 - 2.4 depending on 
the energy and angular resolutions. 
The  inclusion of the   
inelasticity leads to an increase of 
the total significance by $(8 - 11)\%$ after application 
of the experimental smearing: 
The stronger the smearing, 
the weaker the significance increase. 

5. Inversion of the  mass hierarchy  and variations of other parameters
have different  effects on the $y-$distribution of events. 
This means that inelasticity measurements will alleviate the
degeneracy of the hierarchy with  
$\theta_{23}$ and $\Delta m_{32}^2$. 
Without $y$ distribution the degeneracy with $\Delta m_{32}^2$ 
reduces the significance  
by factor $\sim 1.7$. The inelasticity measurements  
increase the total significance by  $43\%$ before the experimental smearing 
and by 25\%  with a specific reasonable experimental smearing scheme. 

6. The mass hierarchy and the systematic errors 
affect the $y$ distribution differently. 
Therefore measurements of inelasticity will likely help to   reduce the impact
of systematic uncertainties. 

7. The contamination of the $\nu_\mu$  event sample with other flavors leads to a 
suppression of the oscillation effects. The selection of events with  not too large $y$ 
will help discriminate $\nu_\mu$ CC events from event  of other types and therefore 
mitigate the loss of features in the oscillatory pattern.

8. All in all,  we expect that the inclusion 
of the inelasticity of the interaction in the analysis will increase the significance  by 
$(20 - 50)\%$, which is equivalent to an increase of the exposure time 
or  effective volume by factor 1.5 - 2. 

9. It is not excluded that more sophisticated analysis 
will lead to even stronger enhancement  effect. 

The next step in enhancement of the discovery potential  of the 
Multi-megaton scale detectors  
could be related to some information 
about the direction of the cascade using detailed time 
information about development of event. Also  
the inclusion of other type of events (cascades without muons) 
in the analysis  will reinforce the discovery potential.

The inclusion of the inelasticity as an ingredient in the data analyses 
may become necessary in order to unambiguously conclude 
on the mass hierarchy in the near future.

\bigskip

\section*{Acknowledgments}
M. Ribordy is supported by the Swiss National Research Foundation under the grant PP002--114800. 
We thank S. Razzaque for providing us the raw neutrino 
oscillograms used in~\cite{ARS} and for discussions 
in the initial phase of this project and 
R. Bruijn for his help with the GENIE simulation 
in order to understand the light output and the point-like nature of the low energy cascades.


\appendix

\section{Variable change $\phi \rightarrow c_\nu$}\label{appendix:A}

Consider a muon with vector ${\bm\mu}=(s_\mu,0,c_\mu)$.
The matrix associated to the rotation $\phi$ around the $\mu$ axis is
\[R_\mu(\phi)=\left(
\begin{array}{ccc}
c_\phi c_\mu^2+s_\mu^2 & - c_\mu s_\phi & s_{\phi/2}^2 s_{2\mu} \\
c_\mu s_\phi &c_\phi & - s_\phi s_\mu \\
s_{\phi/2}^2 s_{2\mu} & s_\phi s_\mu & c_\mu^2+ c_\phi s_\mu^2 \\
\end{array}
\right).
\]
We consider a neutrino vector ${\bm\nu_0}$ at an angle $\beta$ also 
in the plane $x-z$. A possible vector is ${\bm\nu_0}=(s_{\nu_0},0,c_{\nu_0})$, 
with $\theta_\nu=\theta_\mu+\beta$. We use the matrix to generate revolution 
vector set $\{{\bm\nu}(\phi)\}_\phi$ around the muon trajectory, 
\begin{eqnarray}
{\bm\nu}(\phi)&=&R_\mu(\phi){\bm\nu_0} 
=
\left(\begin{array}{c}
c_\phi c_\mu s_{\beta}+s_\mu c_{\beta}\\
s_\phi s_{\beta}\\
-c_\phi s_\mu s_{\beta}+c_\mu c_{\beta}\\
\end{array}
\right).
\nonumber
\end{eqnarray}
The $z$-component of the vector $\bm\nu$ can be associated to $c_\nu$,
$c_\nu = -c_\phi s_\mu s_{\beta}+c_\mu c_{\beta}$
therefore
${\rm d}\phi= {\rm d}c_\nu/s_\beta s_\mu s_\phi,$
with $s_\beta s_\mu s_\phi=\sqrt{s_\mu^2 s_\beta^2-(c_\beta c_\mu - c_\nu)^2}$.

\section{Smearing function for $y$}\label{appendix:B}

%
\begin{figure}[h]
\centering  
 \includegraphics*[width=0.35\textwidth]{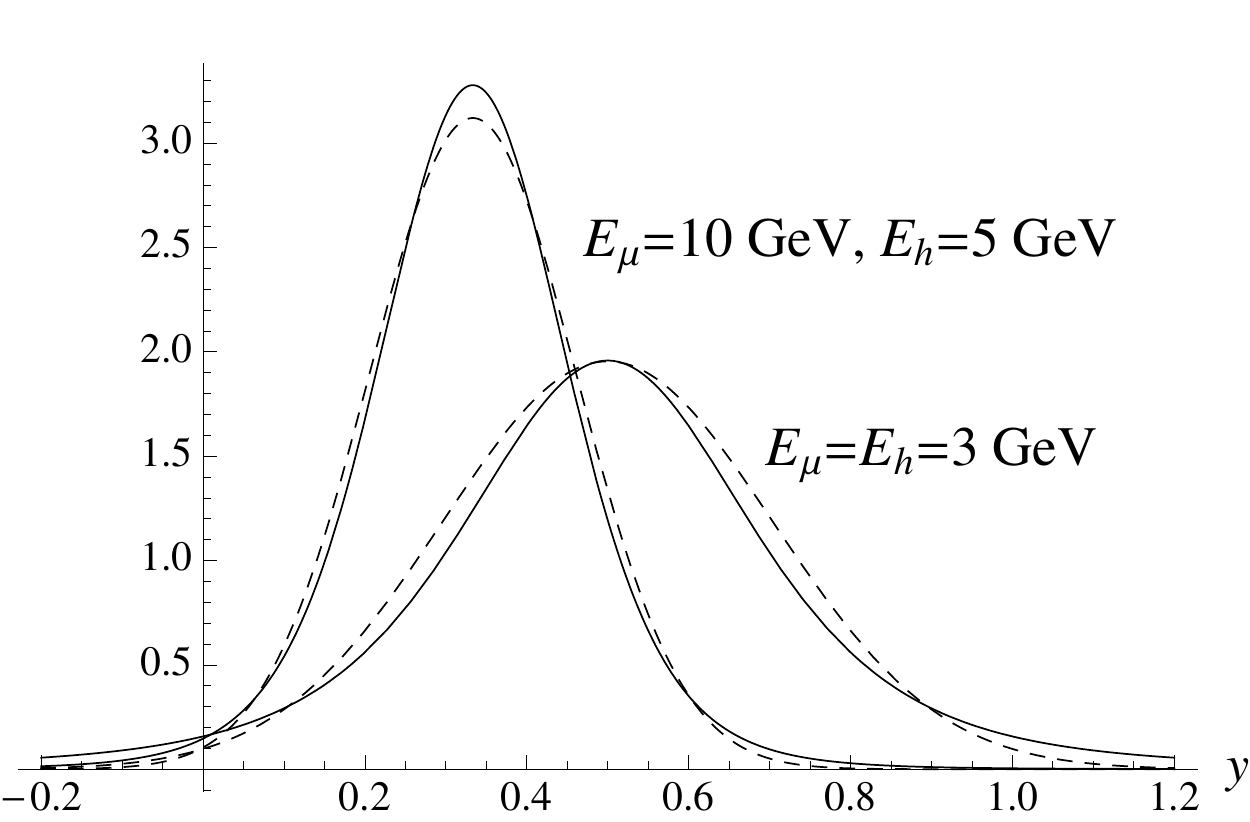}
\caption{
The reconstructed $y$-distribution for two different 
sets of $\{E_\mu, E_h\}$  (numbers at the curves)
and  $\sigma_{\mu, h} = \surd{E_{\mu, h}}$. 
The  dashed curves correspond to the Gaussian approximation.
\label{fig:illYdistr}}
\end{figure}
%

We use the notation 
$\Delta_\mu \equiv E_\mu - \tilde{E}_\mu$, $\Delta_h \equiv E_h - \tilde{E}_h$  
and  the simplified relation $y=E_h/(E_\mu+E_h)$.
In order to obtain the $y-$distribution,
we introduce $z \equiv E_\mu/E_h$, so that  $y = 1/(1 + z)$. 
Consequently, $z = 1/y - 1$ and  ${\rm d}z = - {\rm d}y/y^2$.  
Let us denote by $P_z(\tilde{z})$ the distribution of $\tilde{z}$.
Then the $y-$distribution is given by  
\be
P_y(\tilde{y})  = 
\frac{1}{\tilde{y}^2}P_{z}({1/\tilde{y}} - 1). 
\nonumber
\label{eq:gyyy}
\ee
In turn,  the distribution of ratio $z$ can be found from 
\[
P_{z}(\tilde{z}) =
\int g_h(\tilde{E}_h, E_h) \, g_\mu(\tilde{E}_\mu, E_\mu) \,
\delta \left(\frac{E_\mu}{E_h} - \tilde{z} \right){\rm d}E_h\, {\rm d}E_\mu. 
\]
The integration gives 
\begin{eqnarray}
\nonumber
P_{z}(\tilde{z}) &=&\frac{e^{-\frac{\Delta_h^2}{2\sigma_h^2} - 
\frac{\Delta_\mu^2}{2\sigma_\mu^2}}}{2\pi\,(\sigma_h^2+\sigma_\mu^2 \tilde{z}^2)^2 \,|\Delta_\mu \sigma_h^2 + 
\Delta_h \sigma_\mu^2 \tilde{z}|}\\
\times&&
\Biggl(2 \sigma_h \sigma_\mu (\sigma_h^2+\sigma_\mu^2 \tilde{z}^2)\, |\Delta_\mu \sigma_h^2 + 
\Delta_h \sigma_\mu^2 \tilde{z}| \Biggr.
\nonumber\\
+&& \sqrt{2\pi} \, e^{\frac{(\Delta_\mu \sigma_h^2+ \Delta_h \sigma_\mu^2 \tilde{z})^2}{2 \sigma_h^2 \sigma_\mu^2 
(\sigma_h^2+\sigma_\mu^2 \tilde{z}^2)}}  \,(\Delta_\mu \sigma_h^2+ \Delta_h \sigma_\mu^2 \tilde{z})^2 
\nonumber\\ \times&& \Biggl. 
\sqrt{\sigma_h^2+\sigma_\mu^2 \tilde{z}^2}\, \text{erf}\Bigl(\frac{|\Delta_\mu \sigma_h^2 + 
\Delta_h \sigma_\mu^2 \tilde{z}|}{\sqrt{2} \sigma_h \sigma_\mu \sqrt{\sigma_h^2 + 
\sigma_\mu^2 \tilde{z}^2}}\Bigr)\Biggr). 
\nonumber
\end{eqnarray}
The $y$ distribution obtained in this way is nearly Gaussian in most cases 
of interests. It  starts to deviate from Gaussian, showing  
enhanced tails, when $E_\mu$ and $E_h$ are both small. 
This is illustrated Fig.~\ref{fig:illYdistr} for the energy resolution 
widths $\sigma_{\mu, h} = \surd{E_{\mu, h}}$. 

\end{document}